\documentclass[fleqn,usenatbib,useAMS]{mnras}

\usepackage{graphicx}	
\usepackage{amsmath}	
\usepackage{amssymb}	
\usepackage{multicol}        
\usepackage{bm}		
\usepackage{pdflscape}	
\usepackage[nameinlink,capitalise]{cleveref}

\newcommand{\Msun}{\rm M_{\sun}}

\newcommand{\kpc}{\rm kpc}
\newcommand{\pc}{\rm pc}

\newcommand{\cMpc}{\rm cMpc}
\newcommand{\Mpc}{\rm Mpc}

\newcommand{\lsim}{\mathrel{\hbox{\rlap{\lower.55ex\hbox{$\sim$}} \kern-.3em\raise.4ex\hbox{$<$}}}}
\newcommand{\gsim}{\mathrel{\hbox{\rlap{\lower.55ex\hbox{$\sim$}} \kern-.3em\raise.4ex\hbox{$>$}}}}

\newcommand{\Ato}{A_{2} }

\newcommand{\Amax}{A_{\rm 2,max} }

\newcommand{\rbar}{r_{\rm bar} }
\newcommand{\rhalf}{r_{\rm 50,*} }
\newcommand{\hdisc}{h_{\rm disc, *}}

\newcommand{\tnorm}{t_{\rm norm} }
\newcommand{\tbar}{t_{\rm bar} }

\newcommand{\TNGF}{TNG50}

\usepackage[T1]{fontenc}
\usepackage{ae,aecompl}

\usepackage{newtxtext,newtxmath}
\usepackage{ulem}
\defcitealias{rosasguevara2020}{RG20}

\title[The evolution of barred galaxy population]{The evolution of the barred galaxy population in the TNG50 simulation}

\author[Rosas-Guevara et. al.]{Yetli Rosas-Guevara$^{1}$ \thanks{E-mail: yetli.rosas@dipc.org},  Silvia Bonoli$^{1,2}$, Massimo Dotti$^{3,4,6}$, David Izquierdo-Villalba$^{3,4}$,
\newauthor Alessandro Lupi$^{3,4}$, Tommaso Zana$^{5}$,
Matteo Bonetti$^{3,4,6}$, Dylan Nelson$^{7}$, Volker Springel$^{8}$,
\newauthor Lars Hernquist$^{9}$,  Mark Vogelsberger$^{10}$ \\
$^{1}$ Donostia International Physics Centre (DIPC), Paseo Manuel de Lardizabal 4, 20018 Donostia-San Sebastian, Spain\\
$^{2}$IKERBASQUE, Basque Foundation for Science, E-48013, Bilbao, Spain\\
$^{3}$Dipartimento di Fisica G. Occhialini, Universit\`{a} di Milano-Bicocca, Piazza della Scienza 3, IT-20126 Milano, Italy\\
$^{4}$INFN, Sezione di Milano-Bicocca, Piazza della Scienza 3, IT-20126 Milano, Italy\\
$^{5}$Scuola Normale Superiore, Piazza dei Cavalieri 7, I-56126 Pisa, Italy \\
$^{6}$INAF, Osservatorio Astronomico di Brera, Via E. Bianchi 46, I-23807, Merate, Italy\\
$^{7}$Max-Planck-Institut fur Astronomie, Konigstuhl 17, D-69117, Heidelberg, Germany\\
$^{8}$Max-Planck-Institut fur Astrophysik, Karl-Schwarzschild-Str. 1, 85748 Garching, Germany\\
$^{9}$Harvard-Smithsonian Center for Astrophysics, 60 Garden Street, Cambridge, MA, 02138, USA\\
$^{10}$Department of Physics, Kavli Institute for Astrophysics and Space Research, Massachusetts Institute of Technology, Cambridge, MA 02139, USA\\}
\date{Last updated 2020 June 10; in original form 2013 September 5}

\pubyear{2020}

\begin{document}
\label{firstpage}
\pagerange{\pageref{firstpage}--\pageref{lastpage}}
\maketitle

\begin{abstract}
We use the magnetic-hydrodynamical simulation TNG50 to study the evolution of barred massive disc galaxies.  Massive spiral galaxies are already present as early as $z=4$, and bar formation takes place already at those early times. The bars grow longer and stronger as the host galaxies evolve, with the bar sizes increasing at a pace similar to that of the disc scale lengths. The bar fraction mildly evolves  with redshift for galaxies with $M_{*}\geq10^{10}\Msun$,  being greater than $\sim40\%$ at $0.5<z<3$  and $\sim30\%$  at $z=0$. When bars larger than a given physical size ($\geq 2\,\rm kpc$) or the angular resolution limit of twice the I-band angular PSF FWHM of the HST are considered, the bar fraction dramatically decreases with increasing redshift, reconciling the theoretical predictions with observational data. We find that barred galaxies have an older stellar population, lower gas fractions and star formation rates than unbarred galaxies. In most cases, the discs of barred galaxies assembled earlier and faster than the discs of unbarred galaxies. We also find that barred galaxies are typical in haloes with larger concentrations and smaller spin parameters than unbarred galaxies. Furthermore, the inner regions of barred galaxies are more baryon-dominated than those of unbarred galaxies but have comparable global stellar mass fractions. Our findings suggest that the bar population could be used as a potential tracer of the buildup of disc galaxies and their host haloes. With this paper, we release a catalogue of barred galaxies in TNG50 at $6$ redshifts between $z=4$ and $z=0$.

\end{abstract}

\begin{keywords}
galaxies:structure -- galaxies:evolution -- methods:numerical
\end{keywords}


\section{Introduction}
Stellar bars are non-axysimmmetric structures in the inner parts of disc galaxies. They populate more than  30 per cent  of massive disc galaxies ($ M_{*}\geq10^{10} M_{\odot}$) in the local Universe (e.g., \citealt{sellwood1993,masters2011,gavazzi2015}). Stellar bars are believed to be a crucial ingredient in the secular evolution of disc galaxies (e.g., \citealt{debattista2004, athanassoula2005}), since they can efficiently redistribute gas, stars, and dark matter in the central parts of galaxies (e.g., \citealt{athanassoula2002,athanassoula2003,debattista2006,sellwood2012}).  The clearest prediction of this is gas inflows funnelled into the centres of galaxies, producing rapid star formation (e.g., \citealt{spinoso2017, donohoe2019,george2019}) or feeding central black holes (e.g., \citealt{shlosman1989,fanali2015}). The presence of a bar has been associated with the rapid consumption of star forming gas in galaxies \citep{fraser2020}. 

In addition to the effects of bars on shaping their host galaxies, the existence (or absence) of a bar could provide information on the galaxy assembly history. One powerful way to investigate the evolution of bars and their hosts in a cosmological context is to determine the fraction of barred galaxies as a function of various galaxy properties, such as stellar mass, gas fractions, or disc structural properties. 
Observational studies in the local Universe have shown that bars are more frequently found in massive red and gas-poor galaxies than in blue, star-forming and gas-rich galaxies (e.g., \citealt{barazza2008,masters2012,gavazzi2015,consolandi2016,cervantes2017}). 
The dependence of the bar fraction on redshift is also a diagnostic of the evolution of bars and their host discs. 

Despite significant challenges, some observers have been able to determine the bar frequency up to $z=1$.
Earlier studies, based on small samples of disc galaxies, find contrasting trends in the evolution of the bar fraction.  For instance,  \cite{abraham1999} find a decreasing bar fraction as a function of increasing redshift using HST observations, while \cite{elmegreen2004} and \cite{jogee2005}, infer a constant fraction.
Later, \cite{sheth2008}  using a larger disc sample from the COSMOS field, find that the fraction of barred disc galaxies  declines rapidly with increasing redshift from $\sim 0.65$  at $z\sim 0.2$ to $\sim 0.20$ at $z\sim 0.8$. Recently,  \cite{melvin2014} observe a similar decreasing trend of the bar fraction using a different  selection of disc galaxies from COSMOS and with a different bar identification provided by the Galaxy Zoo Hubble (GZH) project. The apparent discrepancies between earlier and later studies have been attributed to differences in the sample selection and observational biases. In particular, \cite{erwin2018}, who studies a  spiral galaxy sample from the Spitzer Survey of Stellar Structure in Galaxies  (S$^4$G), indicates that  bar fractions at high redshift might be underestimated due to a combination of poor angular resolution and the connection between the bar size and the galaxy stellar mass.

The existence of high redshift bars could give some insights into the formation of these structures and, generally, of high redshift disc galaxies and their haloes. At high redshift, galaxy evolution is more rapid and violent and galaxies are expected to be fed by frequent mergers and rapid gas accretion (e.g., \citealt{pillepich2019,dekel2020a}). Consequently, high redshift disc galaxies should be severely disturbed, as filamentary dense and cold gas is efficiently accreted. This fresh material accreted into the disc  leads to high gas densities, causing extreme star formation rates (SFRs) and  strong supernova (SN) feedback (e.g., \citealt{keres2005,keres2009}), thus making dynamically cool disks difficult to sustain. On the other hand, cold streams can favour the growth of
extended discs if the accreted material is mostly corotating and coplanar for a sufficiently long time (e.g., \citealt{sales2012}).   Recent observations suggest the existence of
massive ($M_{*}> 10^{10}\Msun$) rotation-dominated discs at high redshift
($z>4$) with high rotation velocity to velocity dispersion ratios: $V_{\rm rot}/\sigma\sim 10$   (e.g \citealt{rizzo2020,neeleman2020,fraternali2021,lelli2021}).
While these objects may represent outliers of star-forming galaxies at high redshifts \citep{genzel2020},
they may also be potential hosts of stable bar structures.

At lower redshift, \cite{posti2019} and \cite{marasco2020} use Spitzer combined with HI interferometry to estimate the halo and stellar mass of spiral galaxies that are more massive than the Milky Way ($M_{*}\sim 8\times 10^{10}\Msun$). The authors find that these galaxies are dominated by baryons with high stellar mass fractions ($M_{*}/M_{\rm tot}>0.8$), at global and disc scales. These massive spiral galaxies are expected to have lower stellar mass fractions than observed,  based on  abundance matching models \citep{moster2018} that predict the stellar-to-halo mass relation, as AGN feedback is efficient at decreasing star formation, creating some tension between observations and abundance matching models.

An important role in the formation and evolution of a bar is also played by the dark matter halo in which galaxies are embedded.  Since the early 70s,  numerical studies of N-body discs indicate that the presence of a dominant spherical component is needed to prevent bar instability in discs \citep{ostriker1973}. The dark matter component is now thought to be a crucial factor in the formation (or not) of a bar, although the specifics of its role are still being debated. Several numerical simulations of idealised galaxies have indicated that, under certain conditions, the exchange of angular momentum between the dark matter and disc component can induce and even speed-up bar formation (e.g.,\citealt{saha2013,sellwood2016}). 
For instance, \cite*{saha2013} show that bar formation is favourable in haloes corotating with the disc with a dark matter spin parameter ($\lambda_{\rm DM}$) in the range of  0 and 0.07. 
However, \cite{long2014} show that the bar growth is halted in haloes with larger spins ($\lambda_{\rm DM}\gsim0.03$). \cite{collier2018} suggested that not only the spin parameter but also the shape of the dark matter halo is important to the evolution of bars. \cite{yurin2015} consider the stability of discs by inserting already formed stellar discs in haloes using the Aquarius simulation project that consists of zoom-in, dark matter-only simulations. These authors found that the 3-D shape of the dark matter halo could affect the stability of the disc. Independently of the results, all these studies show the importance of the angular momentum exchange between the dark matter and the stellar components in the formation and  evolution of a bar (see also \citealt{sheth2012,athanassoula2013}).
The drawback of these previous works, though they are very useful, is that they do not take into account the assembly history of a galaxy, and/or they do not include other baryonic processes such as star formation and feedback that are important for the buildup of a galaxy.

Barred galaxies have started to be explored in a fully cosmological context recently, thanks to high-resolution zoom-in simulations (e.g., \citealt{kraljic2012,scannapieco2012,bonoli2016}).  In particular, \cite*{bonoli2016} present a zoom-in simulation of a Milky Way-type galaxy,  \textit{ErisBH}, a sibling of the \textit{Eris} simulation \citep{guedes2011}, but with  black hole (BH) subgrid physics included. \cite{bonoli2016} and \cite{spinoso2017} find that the simulated galaxy forms a strong bar below $z\sim1$, and the authors point out that the disc in the simulation is more prone to instabilities compared to the original \textit{Eris}, possibly because of early AGN feedback affecting the central part of the galaxy. \cite{zana2018c}, studying an enhanced suite of \textit{Eris}, highlight the effects of the feedback processes on the formation time and final properties of the bar. Recently, \cite{fragkoudi2020} have studied barred galaxies which form a boxy/peanut (b/p) bulge in the Auriga suite, which consists of high resolution, magnetohydrodynamical cosmological zoom-in simulations of galaxy formation \citep{grand2017}. The authors find that the simulated galaxies that can reproduce many chemodynamical properties of the stellar populations seen in the Milky-way bulge have quiet merger histories. They  also find that in some galaxies of their sample, a star-forming ring forms around the bar, affecting the metallicity of the inner regions of the galaxy,
emphasising the possible role of the bar on the surrounding disc. Moreover, \cite{fragkoudi2021} find that in barred galaxies the stellar component dominates the dynamics of the central region, with the dark matter component being subdominant. They also find that barred galaxies have generally larger stellar-to-halo fractions with respect to unbarred ones.

The study of the evolution of barred galaxies and their properties in a statistical framework is now possible also thanks to a new generation of cosmological hydrodynamic simulations (\citealt{vogelsberger2014a,schaye2015,pillepich2018b,nelson2018}, see also the recent review of \citealt{vogelsberger2020a}). Analysing  the EAGLE simulation, \cite{algorry2017} discover that bars slow down very quickly as they evolve, expanding the inner parts of the dark matter halo.  \cite{rosasguevara2020} study massive barred disc galaxies at $z=0$ in the TNG100 simulation (see also \citealt{peschken2019,zhao2020,zhou2020} for Illustris and IllustrisTNG), finding that barred galaxies are less star-forming  and more gas poor than unbarred galaxies. Following the evolution of barred galaxies back in time, the authors find that these objects assembled most of their disc component and black holes before bar formation, and earlier than unbarred galaxies.

Our main goal in this paper is to extend our analysis using the TNG100 and  investigate the theoretical predictions for the bar population and their host galaxies across time. In this work, we use the simulation TNG50 \citep{pillepich2019,nelson2019b}, which is the highest resolution run of the TNG project and allows to study with greater details the dynamics of galaxies at all redshfits. The TNG50 simulation encompasses a 51.7  Mpc region, simultaneously resolving the formation of individual galaxies with a resolution of hundreds of pcs, so that it is possible to study the structural properties of galaxies in some detail.
Combining the volume statistics and the high resolution, the TNG50 simulation has been able to reproduce the diversity of disc galaxies since early epochs, such as the fraction of discs as a function of time and stellar mass \citep{pillepich2019}, the high redshift galaxy luminosity functions compatible to current observations and the predictions  from the James Webb Space Telescope (JWST) \citep{vogelsberger2020b,shen2020,shen2021} making the simulation ideal for also studying the barred population across time.

We explore the properties of large-scale and well-resolved bars from $z=4$ down $z=0$ in galaxies with $M_{*}\geq10^{10}\Msun$.  We also investigate the barred galaxy population in a cosmological context, to gain insights into the conditions for a galaxy to contain a bar.  We base our analysis on comparing the properties of barred galaxies and their host haloes against unbarred galaxies. 
 
The paper is structured as follows. In section~\ref{sec:method},  we give a brief overview of the IllustrisTNG project, of our selection of the disc galaxies, and of our methodology for identifying a bar and provide the definition of bar. In section~\ref{sec:results}, we present the theoretical predictions of bars and their discs in the TNG50 simulation. In section~\ref{sec:galaxyprop}, we describe the properties of barred galaxies by comparing them to unbarred galaxies $z=4$ down to $z=0$. We also explore the properties of the host haloes of barred galaxies and compare them to the host haloes of unbarred galaxies. Finally, in section~\ref{sec:summary},  we summarise and discuss our findings.

\section{Methodology}
\label{sec:method}
\subsection{TNG Simulations}
The IllustrisTNG (The Next Generation) project\footnote{\citep{nelson2019a}; http://www.tng-project.org}  (\citealt{nelson2018,naiman2018,pillepich2018b,marinacci2018,springel2018})
includes three main cosmological, gravo-magneto-hydrodynamical simulations of galaxy formation with volumes ranging from $(50)^3$ to $(300)^3\,\cMpc^3$  with different spatial  and mass resolutions.
The IllustrisTNG simulations have been performed with the moving-mesh \textsc{AREPO} code \citep{springel2010},
adopting the Planck cosmology parameters with constraints  from \cite{planck2016}: $\Omega_\Lambda=0.6911$, $\Omega_{\rm m}=0.3089$, $\Omega_{\rm b}=0.0486$, $\sigma_8=0.8159$, $h=0.6774$, and $n_{s}=0.9667$  where  $\Omega_\Lambda$, $\Omega_{\rm m}$, and  $\Omega_{\rm b}$ are the average densities of matter, dark energy, and baryonic matter in units  of the critical density at $z=0$,  $\sigma_8$ is the square root of the linear variance,  $h$ is the Hubble parameter ($H_{0}\equiv h \,100 \rm km \, s^{-1}$) and, $n_{s}$  is the scalar power-law index of the power spectrum of primordial adiabatic perturbations. 

The subgrid physics of IllustrisTNG hinges on its predecessor, Illustris \citep{vogelsberger2013,vogelsberger2014a,vogelsberger2014b,genel2014,nelson2015,sijacki2015} with substantial modifications to star formation feedback (winds), the growth of supermassive black holes, Active  Galactic Nuclei (AGN) feedback, and stellar evolution and chemical enrichment. A complete description of the improvements in the subgrid physics and calibration process can be found in \cite{pillepich2018a} and \cite{weinberger2017}.   A summary of the improvements concerning Illustris is shown in Table~1 of \cite{pillepich2018a}.

In this work, we focus on the \TNGF~ simulation \citep{pillepich2019,nelson2019b}, which is the highest resolution simulation that is part of the TNG suite and, at the same time, provides a large enough cosmological volume for studying the statistical properties of galaxies at intermediate masses. The simulation evolves $2160^3$ dark matter particles and initial gas cells in a 51.7 comoving Mpc region from $z=127$ down to $z=0$. The mass resolution is $4.5\times 10^5 \Msun$ for dark matter particles, whereas the mean gas mass resolution is $8.5\times10^4\Msun$. A comparable initial mass is passed down to stellar particles, which subsequently lose mass through stellar evolution.
The spatial resolution for collisionless particles (dark matter, stellar, and wind particles) is 575 comoving pc down to $z=1$, after which it remains constant at 288 pc in physical units  down to $z=0$. In the case of the gas component, the gravitational softening  is adaptive and based on the effective cell radius,
down to a minimum value of 72 pc in physical units which is imposed at all times. The details of the simulation are given in Table~\ref{table:simulations}.

Galaxies and their haloes  are identified as bound substructures  using  an \textsc{FoF} and  then a \textsc{SUBFIND} algorithm \citep{springel2001} and tracked over time  by the \textsc{Sublink} merger tree algorithm \citep{rodriguezgomez2015}. 
Halo masses ($M_{200}$) are defined as  all matter within the radius $R_{200}$ for which
the inner mean density is $200$ times the critical density.
In each FoF halo, the `central’ galaxy (subhalo) is the first (most massive) subhalo of each FoF group. The remaining galaxies within the
FoF halo are its satellites.  The stellar mass of a galaxy is defined as all stellar matter assigned to the subhalo.

\begin{table}
\caption{Main features of the TNG50 simulation. From top to bottom:  box side-length,  number of initial resolution elements including gas cells and  dark matter (DM) particles,  initial mass of gas cells and of  DM particles, the minimum proper softening length allowed for gas cells, the proper softening length for the collisionless particles at $z=0$, and their comoving softening length.}
\begin{center}
\begin{tabular}{|l|l|l|} 
\hline
$L$    &  [$\Mpc$]  &   $51.7$   \\ 

$N$    &            &   $2\times2160^3$ \\ 
$m_{\rm g}$  &[$M_\odot$]&  $8.5\times10^4 $\\ 
$m_{\rm DM}$ & [$M_\odot$]& $4.5\times10^5$ \\ 
$\epsilon_{\rm gas,min}$ &  $[\pc]$  & $72$  \\
$\epsilon_{\rm DM,stars,0}$ & $[\kpc]$ & $0.288$\\
$\epsilon_{\rm DM,stars,z}$ & $[\kpc]$  & $0.58$ to $0.29$ \\ 
\hline
\end{tabular}
\end{center}
\label{table:simulations}
\end{table}

\subsection{ Disc galaxy sample}
\label{sub:discsample}
In this work we focus on the analysis of disc galaxies with a stellar mass $\geq10^{10}\Msun$, to ensure that the galaxies analysed are well-resolved. To identify  the disc and bulge components of the galaxies we use the kinematic decomposition  computed as in \cite{genel2015}, which is based on \cite{marinacci2014} and \cite{abadi2003} and consistent with the selection of the disc sample in \cite{rosasguevara2020} (hereafter \citetalias{rosasguevara2020}) for the TNG100 simulation. Galaxies are first rotated such that the z-axis is located along the direction of the total angular momentum of the stellar component. For each stellar particle within $10 \times \rhalf$, where $\rhalf$ is the radius within which 50 per cent of the total stellar mass is contained, the circularity parameter, $\epsilon=J_z/J(E)$, is calculated. $J_{z}$ is the specific angular momentum of the particle around the symmetry axis, and  $J(E)$ is the maximum specific angular momentum possible at the specific binding energy of each stellar particle.
The mass of the stellar disc comprises the stellar particles with $\epsilon\geq 0.7$, while the bulge mass is defined as twice the mass of stellar particles with a circularity parameter $\epsilon<0$.
The disc-to-total ($D/T$) ratio is defined as the ratio between the disc stellar mass and the stellar mass enclosed in $10\times r_{50,*}$. We define disc-dominated galaxies as those galaxies  with $D/T\geq0.5$. We look for $M_{*}\geq10^{10}\Msun$ disc-galaxies at the six discrete redshifts $z=0,0.5,1,2,3,4$.  Our final sample consists of  1062 galaxies (see Table~\ref{table:bars} for further details). We note that we do not separate between central and satellites throughout the paper, except for the host halo properties  which include a part of the section \ref{subsec:buildup} and section \ref{subsec:prophaloes}.

\subsection{The sample of barred galaxies}
\label{subsec:barredsamples}

The disc sample defined above is split into barred and unbarred galaxies, depending on the presence or absence of a clear and persistent bar structure in the disc.
Non-axisymmetric structures are identified by Fourier decomposing the face-on stellar surface density  (e.g., \citealt{athanassoula2002,zana2018a,rosasguevara2020}). We focus on  $\Ato(R)$, which is sensitive to non-axisymmetric structures and  is defined by the ratio between the second and zero terms of the Fourier expansion and its phase $\Phi(R)$:
\begin{eqnarray}
A_{2}(R)= \frac{|\Sigma_{j}m_je^{2i\theta_j}|}{\Sigma_{j}m_j}, \\  
\label{eq:a2}
\Phi(R)= \frac{1}{2}{\rm arctan}\bigg[\frac{\Sigma_{j}m_j{\rm sin}(2\theta_j)}{\Sigma_{j}m_j\rm{cos}(2\theta_j) }  \bigg],
\label{eq:phase}
\end{eqnarray}
where  $m_j$ is the mass of the jth particle and $\theta_j$ is the angular coordinate on the galactic plane.
To compute $\Ato(R)$ and  $\Phi(R)$,  the summations are done over all the stellar particles within cylindrical shells of radius $R$, coaxial to the centre of the galaxy and centred in the middle plane. The cylindrical shells have a  width of $0.1\,\kpc$ and height of $1\,\kpc$. Both quantities $\Ato(R)$ and  $\Phi(R)$, have been used to characterise a bar structure, where the bar strength is defined as the value of the peak of $\Ato(R)$, $\Amax$. The phase should be constant inside the bar, however, there is noise in simulated and observed bars. To define the constancy of the phase, we calculate the standard deviation ($\sigma$) of $\Phi(R)$, including each time a new cylinder shell and consider the phase constant when $\sigma\leq 0.1$.  Then, we define the bar extent ($\rbar$) as the maximum radius where $\sigma\leq 0.1$, and the A2 profile first dips at  $0.15$ or the minimum value of $\Ato(R)$ before $\Ato(R)$ is going up again after reaching its peak ($\rm max(0.15, min(\Ato))$)\footnote{This is a different proxy for the bar extent from the one  used in \citetalias{rosasguevara2020}, where the bar extent is defined as the radius at which the maximum of $\Ato$ is reached. We do not adopt this definition since upon visual inspection, it appears to underestimate the bar length.}. The last condition allows us to calculate the bar length in galaxies where there is not a sharp break in A2  between the end of the bar and the beginning of the spiral arms. The value of 0.15 is arbitrary, but we found that it gives the best results when visually comparing the calculated bar radius with the bar extent from the  stellar density maps. We have also compared our catalogue with the catalogue from Zana et al. (submitted), finding good agreement at all redshifts.




 As large values of  $\Ato(R)$  could also be due to transient events, such as mergers and interactions, we conservatively assume a bar to be real only if:
\begin{enumerate}
\item The maximum of $\Ato$, $\Amax$, is greater than $0.2$,
\item $\rbar>r_{\rm min}$ where $r_{\rm min}=1.38\times\epsilon_{*,z}$ is a  minimum radius imposed and $\epsilon_{*,z}$ corresponds to the proper softening length for stellar particles, and $r_{\rm min}$ spans from $0.16$ to $0.4\,\kpc$,
\item The estimated age of the bar is larger than the time between the analysed output and 2 previous simulation outputs ($0.33$ Gyrs at $z=0$ and $0.17$ Gyrs at $z=4$).
\end{enumerate}
The last condition is particularly important at high redshifts ($z=[2,4]$), where galaxies could be extremely disturbed,  experiencing  frequent mergers and stream-fed accretion \citep{zolotov2015,capelo2017,pillepich2019,dekel2020a}. At low redshifts, this condition may reject newly formed bars, but keeps consistency with the bar selection of \citetalias{rosasguevara2020} where the bar structure was present in previous outputs.
The formation time of a  bar, $\tbar$, is computed by tracing back the evolution of $\Amax$ up to the lookback time  when $\Amax \leq 0.2$  and  $\rbar(\tbar)\leq r_{\rm min}$ for more than 2 snapshots.  We additionally require  that, within this time, the relative difference between the $\Amax$ at a given snapshot and the $\Amax$ at two snapshots before does not surpass the value of 0.45. This further ensures that the bars we detect are stable structures.


Of the entire sample of disc galaxies, we find that $431$ are barred galaxies and, of these, $190$ are strong bars ($A_{2,\rm max} \geq 0.4$).  Table~\ref{table:bars} offers further details on the disc and bar samples, including the number of  galaxies that are centrals and the bar fractions at the different redshifts considered.  Note that given our minimum mass cut-off ($M_{*}>10^{10} M_{\sun}$), the number of galaxies at high-z is  relatively small, as the number density of galaxies with stellar masses above this value drops significantly at $z>2$ (e.g.~\citealt{ilbert2013,davidzon2017}). All disc galaxies that do not satisfy the previous criteria are considered ``unbarred''.



\begin{figure*}
\begin{tabular}{cc}
\includegraphics[width=1\columnwidth]{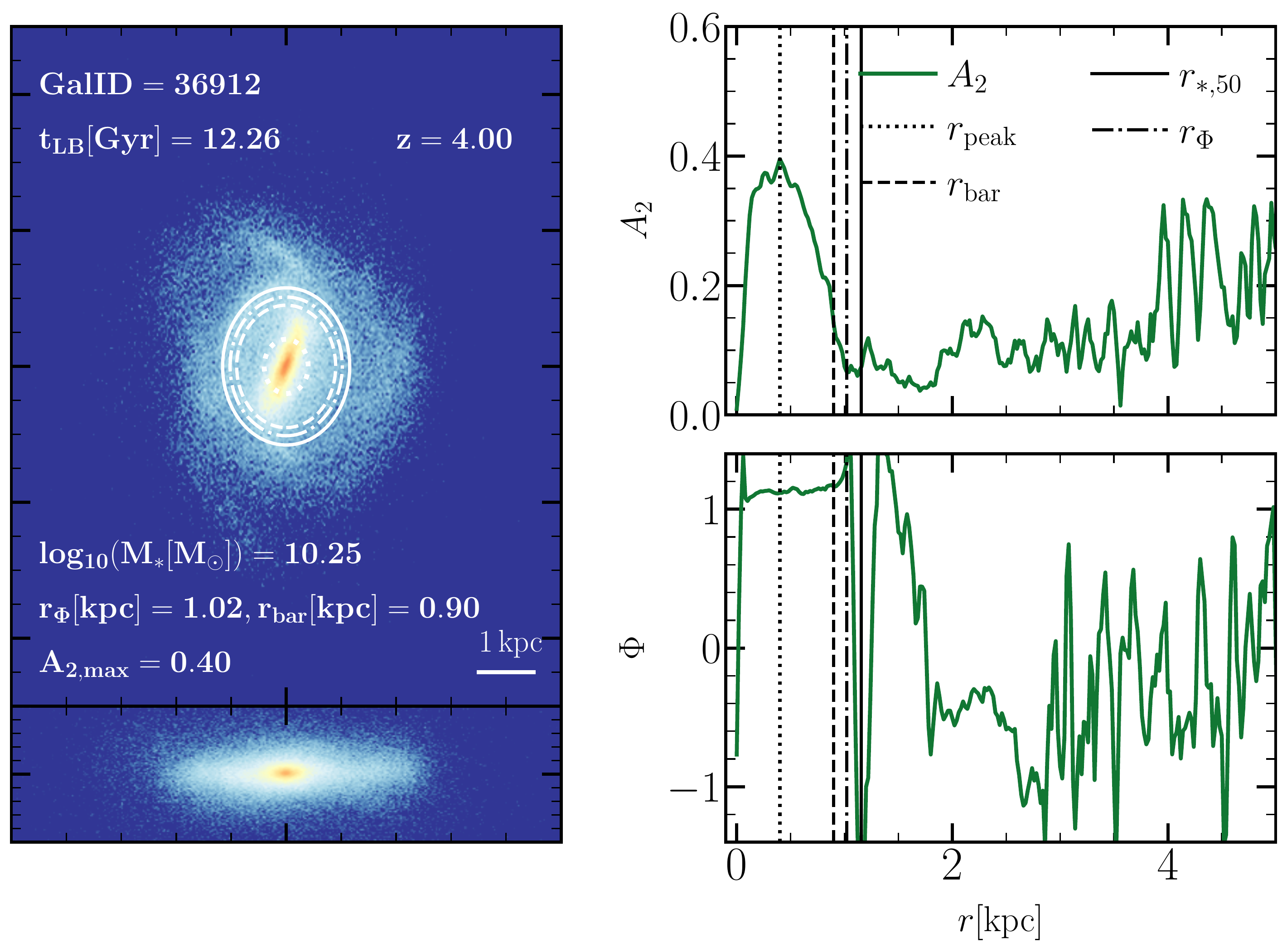} &
\includegraphics[width=1\columnwidth]{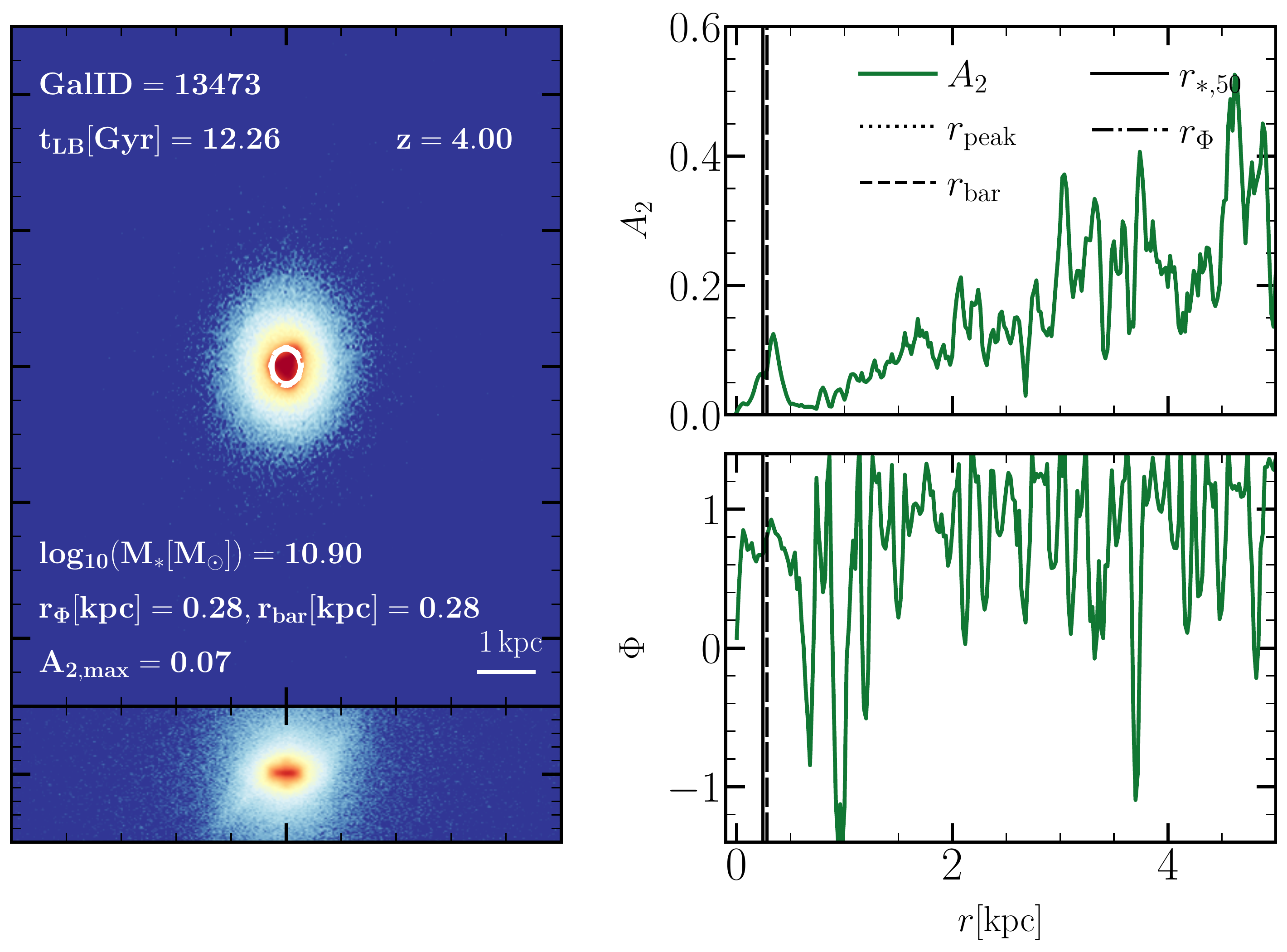}  \\
\includegraphics[width=\columnwidth]{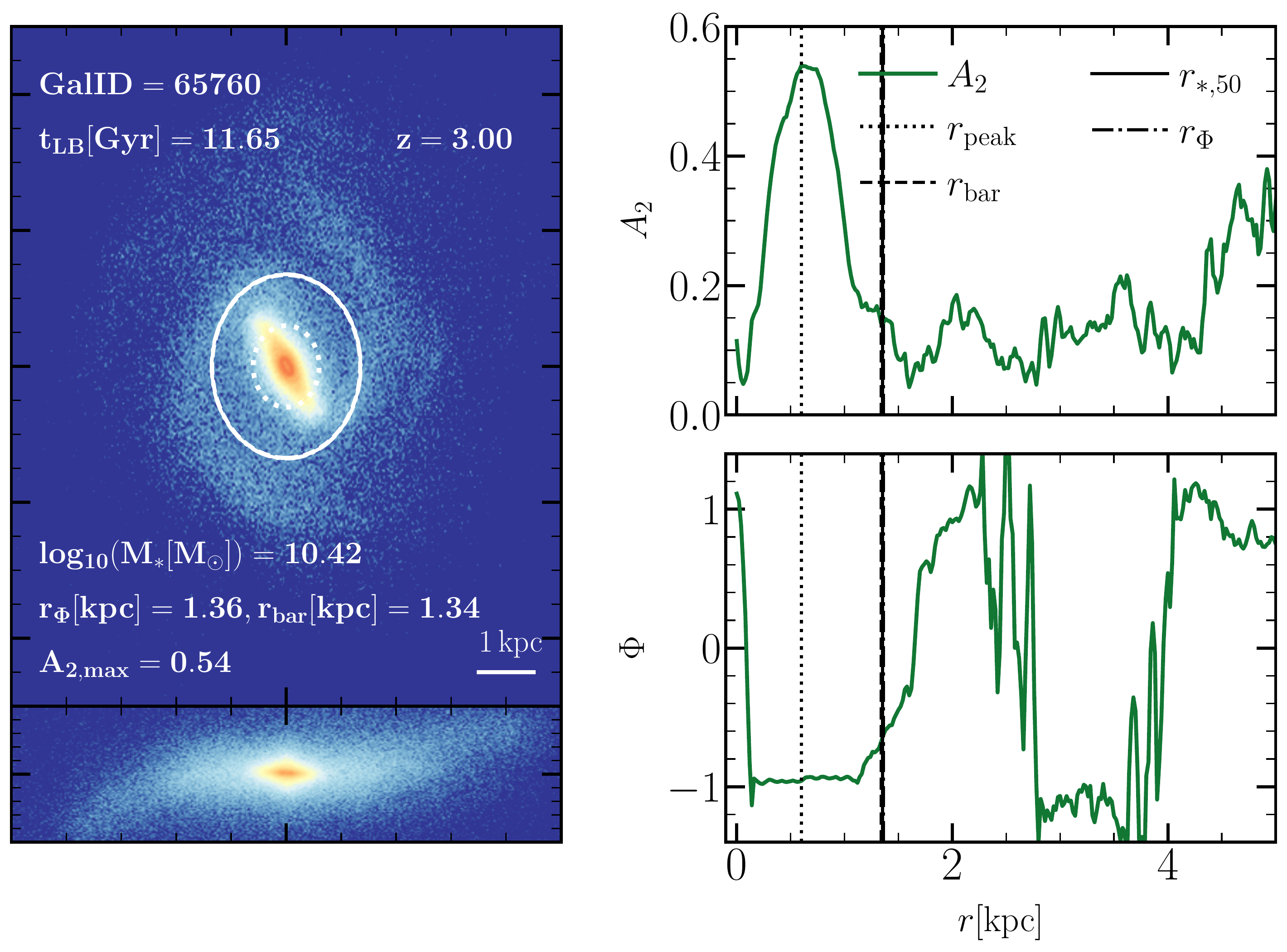}&
\includegraphics[width=\columnwidth]{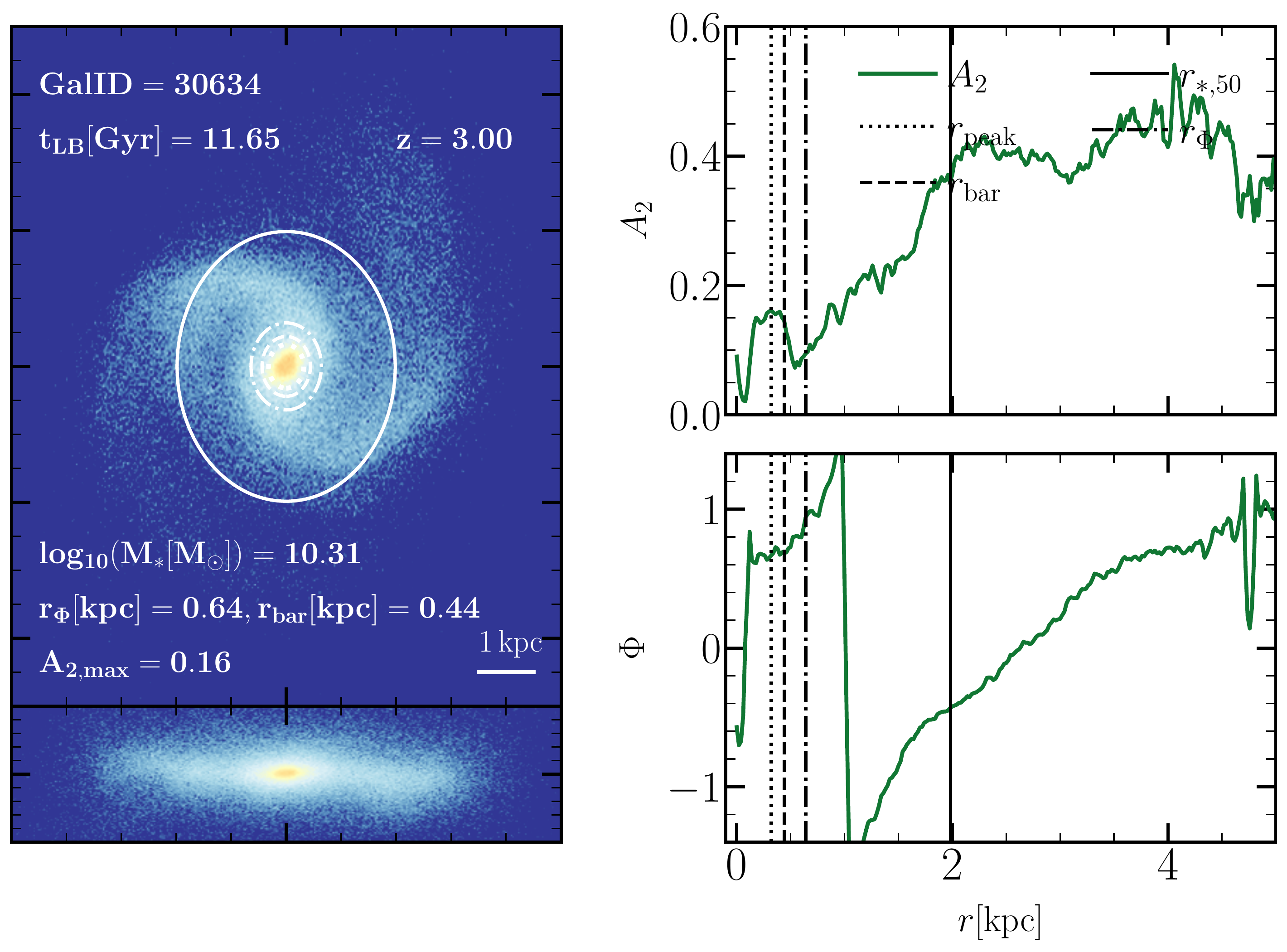}  \\
\includegraphics[width=\columnwidth]{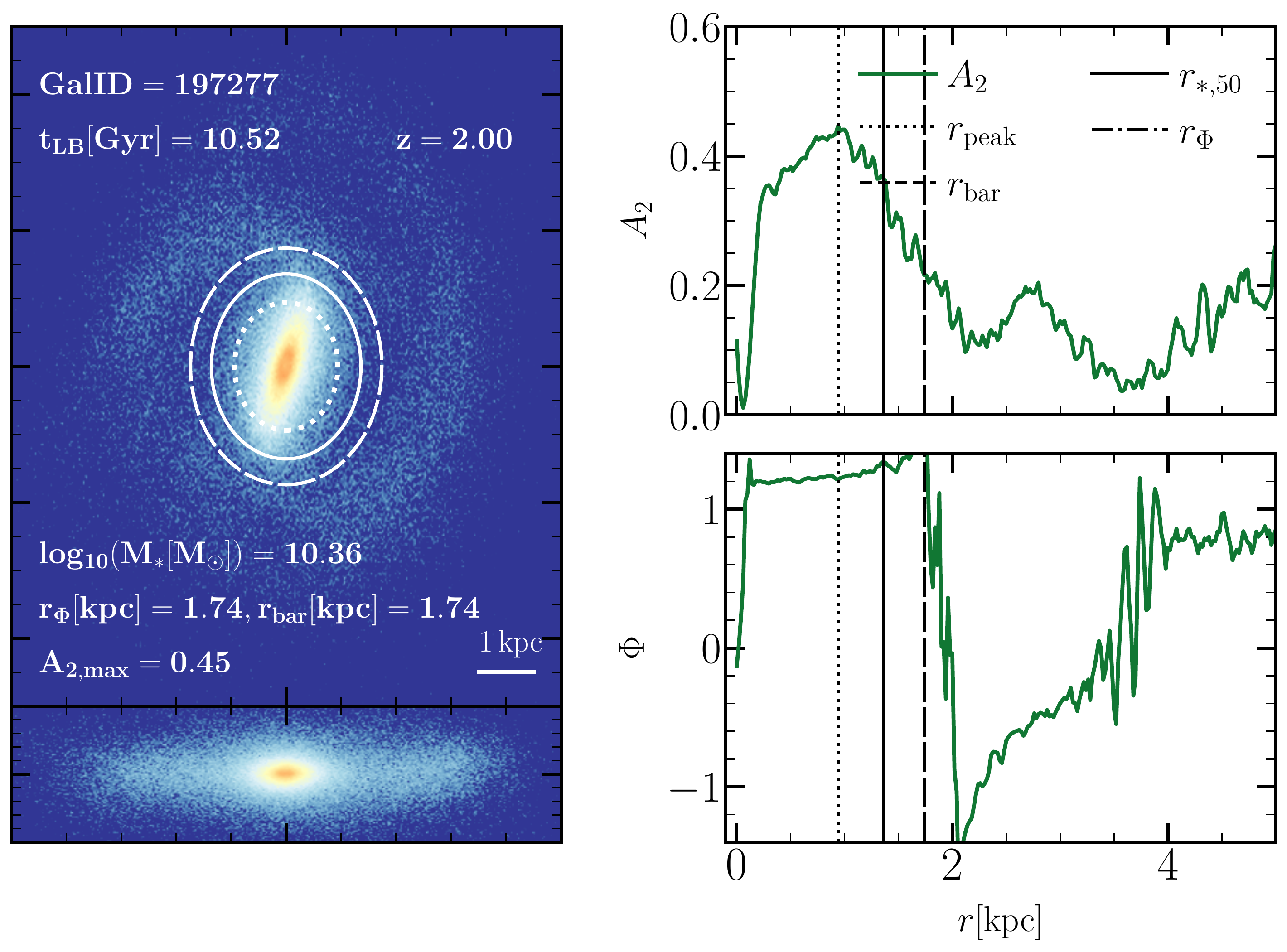} &
\includegraphics[width=\columnwidth]{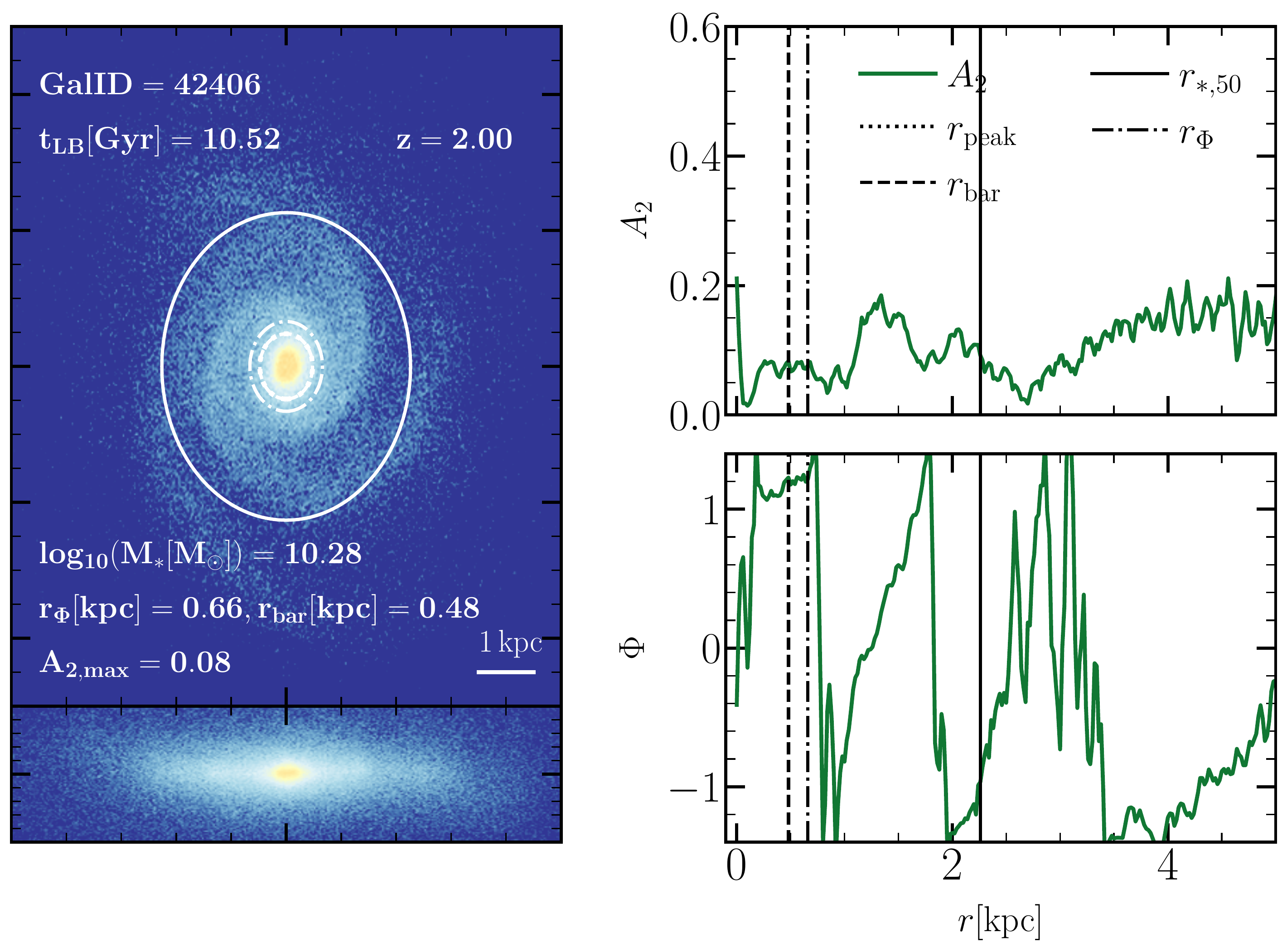}  \\
\end{tabular}
\caption{ Examples of barred galaxies (left panels) and unbarred galaxies (right panels) in  3 snapshots at: $z=4$ (top row), $z=3$ (middle row), and $z=2$ (bottom row). The left column of each panel includes the face-on (top) stellar density maps of $10\times10\times1\, \rm kpc^{3}$ and edge-on (bottom) views of each galaxy. In the right column of each panel, the $\Ato$ profile of the Fourier decomposition of the face-on stellar surface density (green curve) and the phase are plotted. The values of $\Ato$ and the phase are smoothed with a Savitzky-Golay filter. Dotted vertical lines indicate the radius corresponding to the peak of $\Ato$, $r_{\rm peak}$, and vertical dashed and dotted-dashed lines correspond to the bar size, $\rbar$, and the maximum radius at which the phase is constant, $r_\Phi$, respectively. We include the stellar half-mass radius, $\rhalf$, of the galaxy as a solid line. Note that typically $\rbar$ is smaller than or equal to $r_{\Phi}$.} \label{fig:Barexamples}
\end{figure*}
\begin{figure*}
\begin{tabular}{cc}
\includegraphics[width=1\columnwidth]{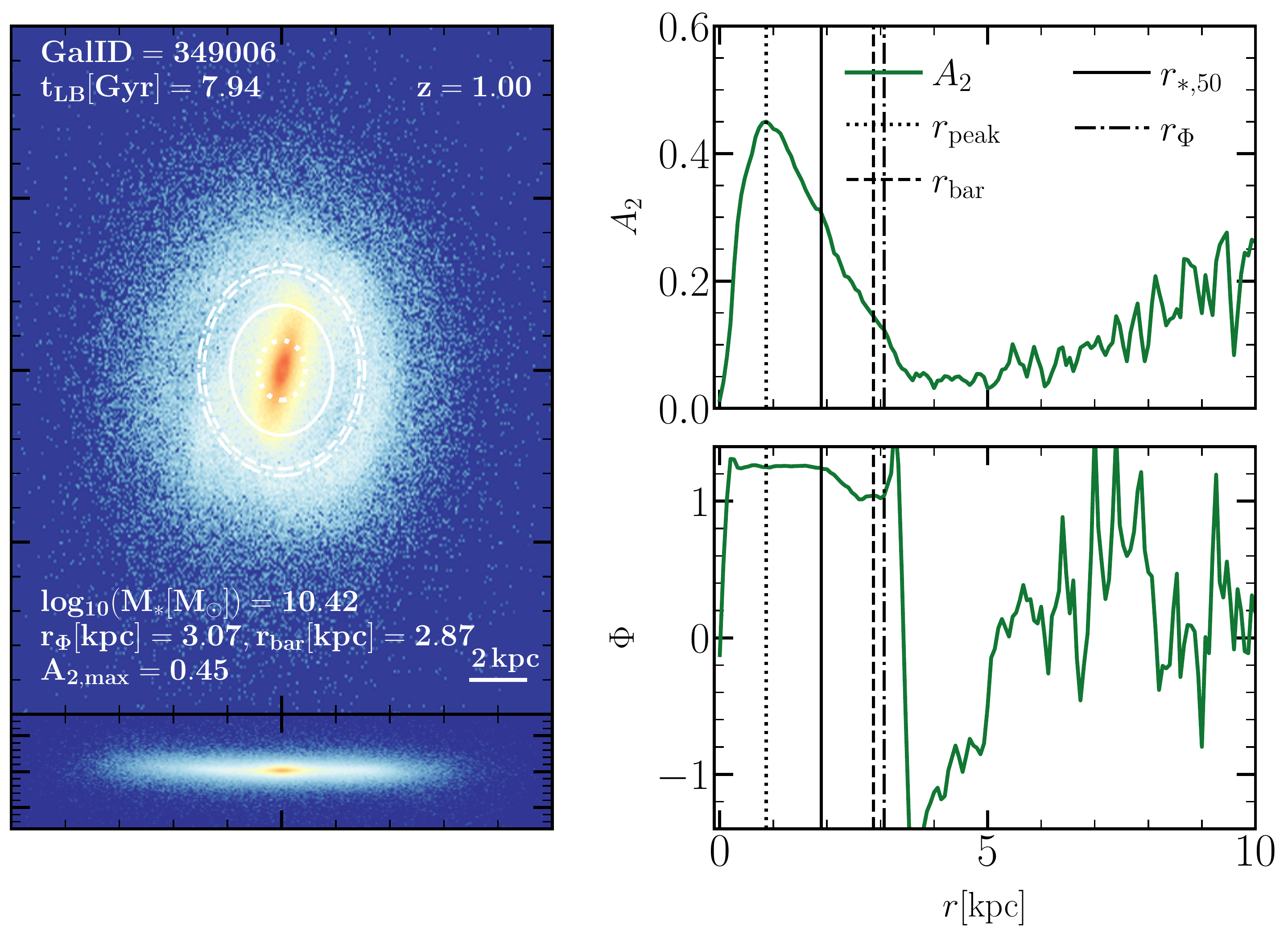} &
\includegraphics[width=1\columnwidth]{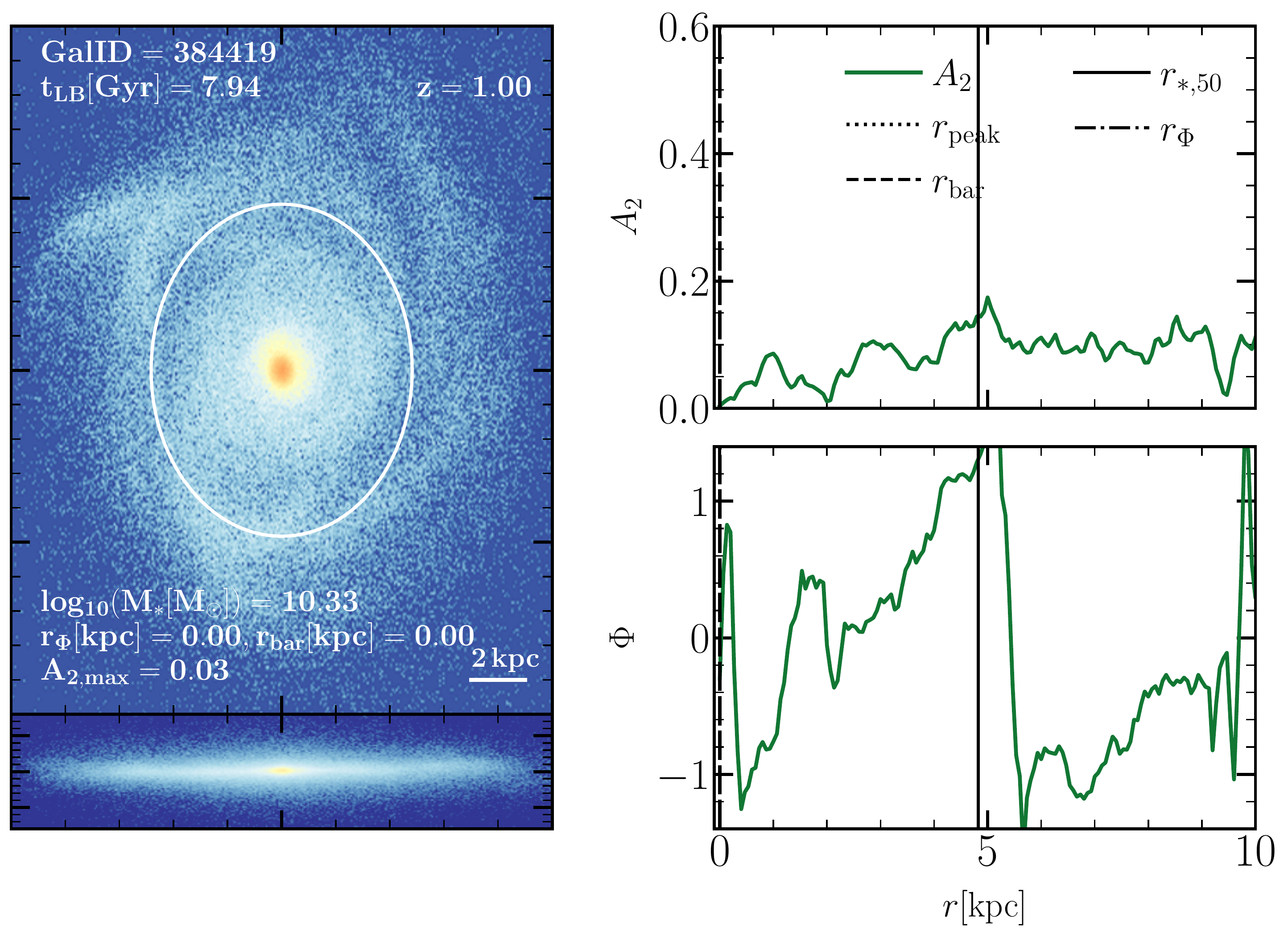}  \\
\includegraphics[width=1\columnwidth]{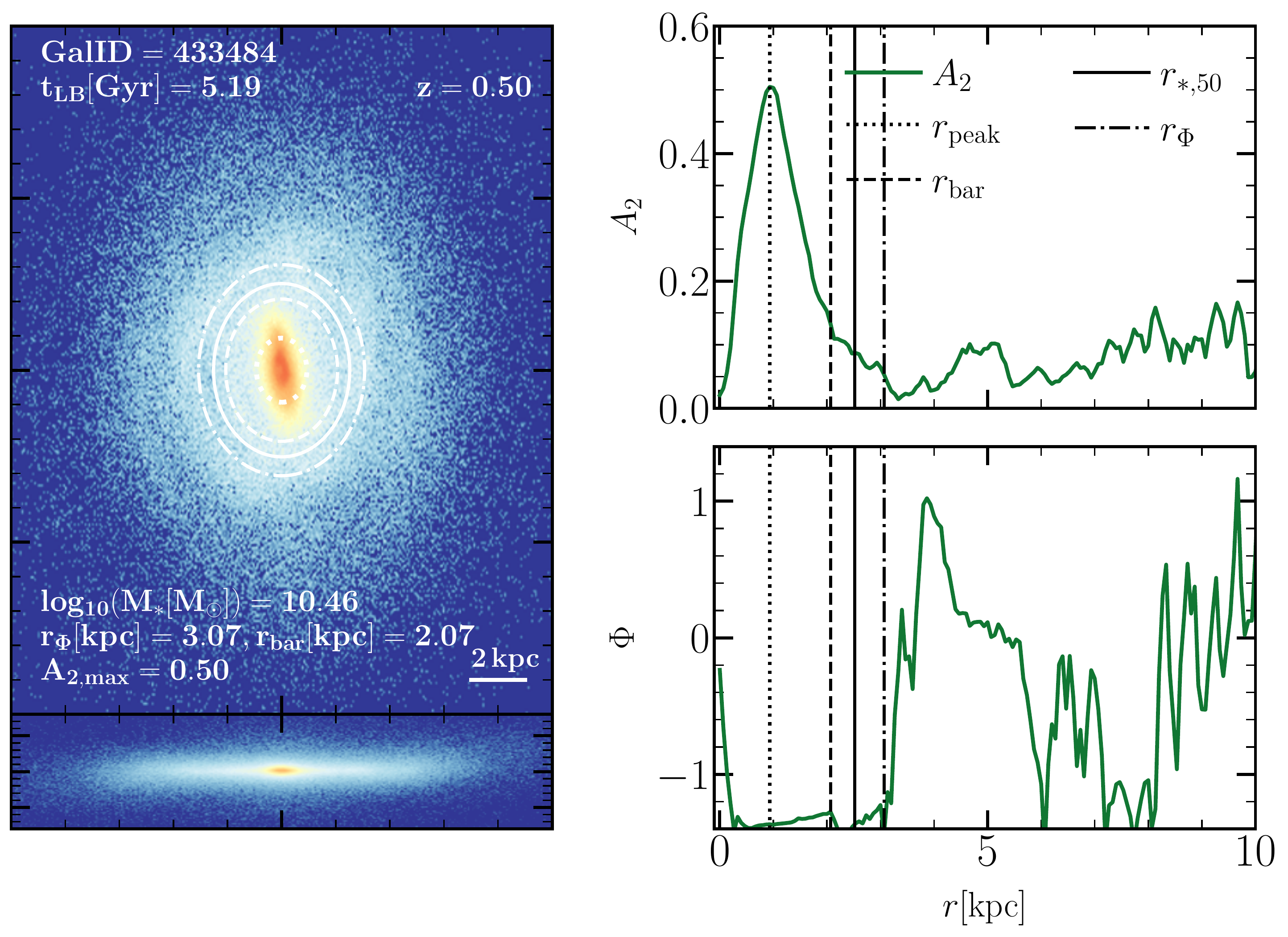} &
\includegraphics[width=1\columnwidth]{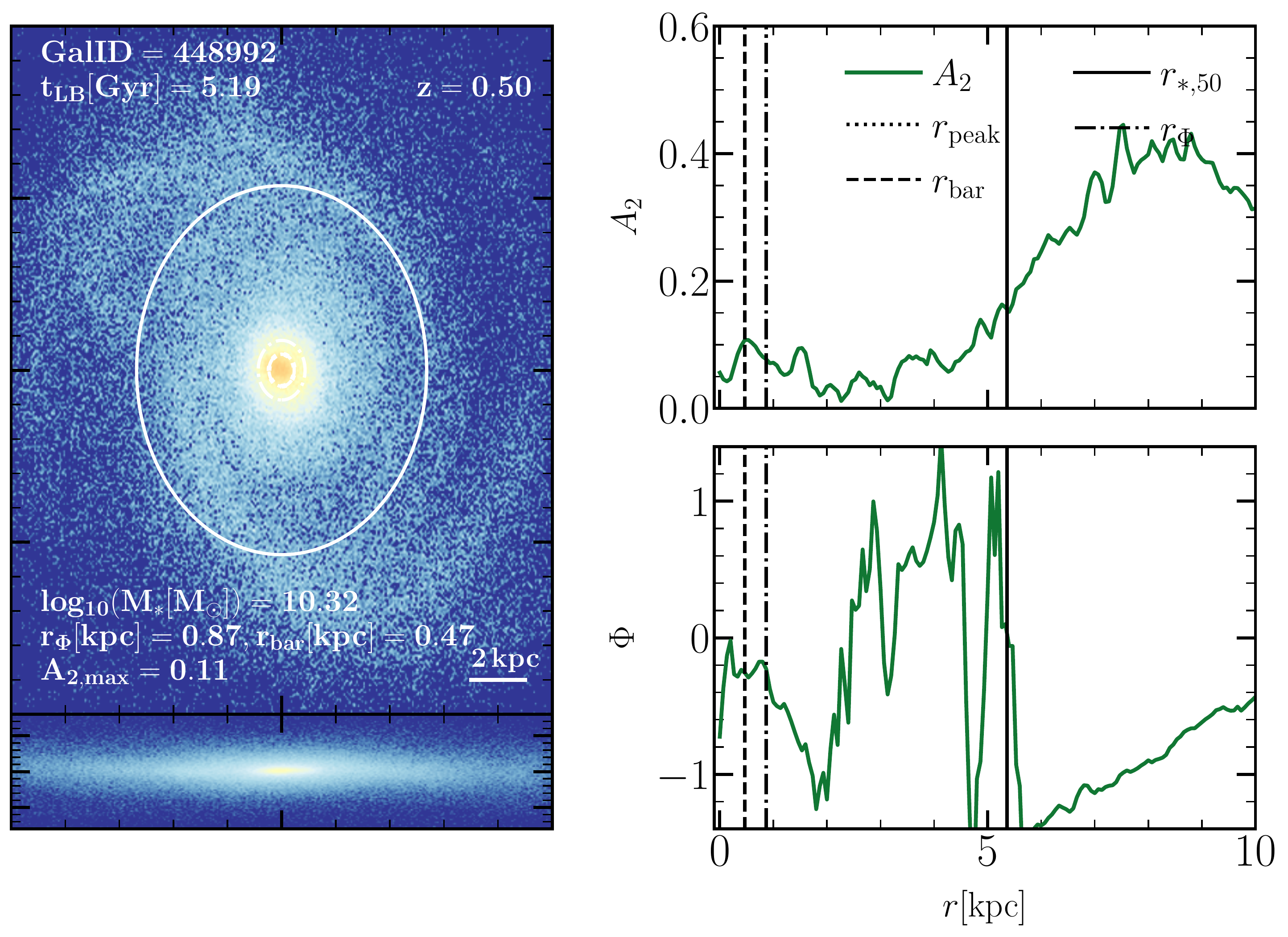} \\
\includegraphics[width=1\columnwidth]{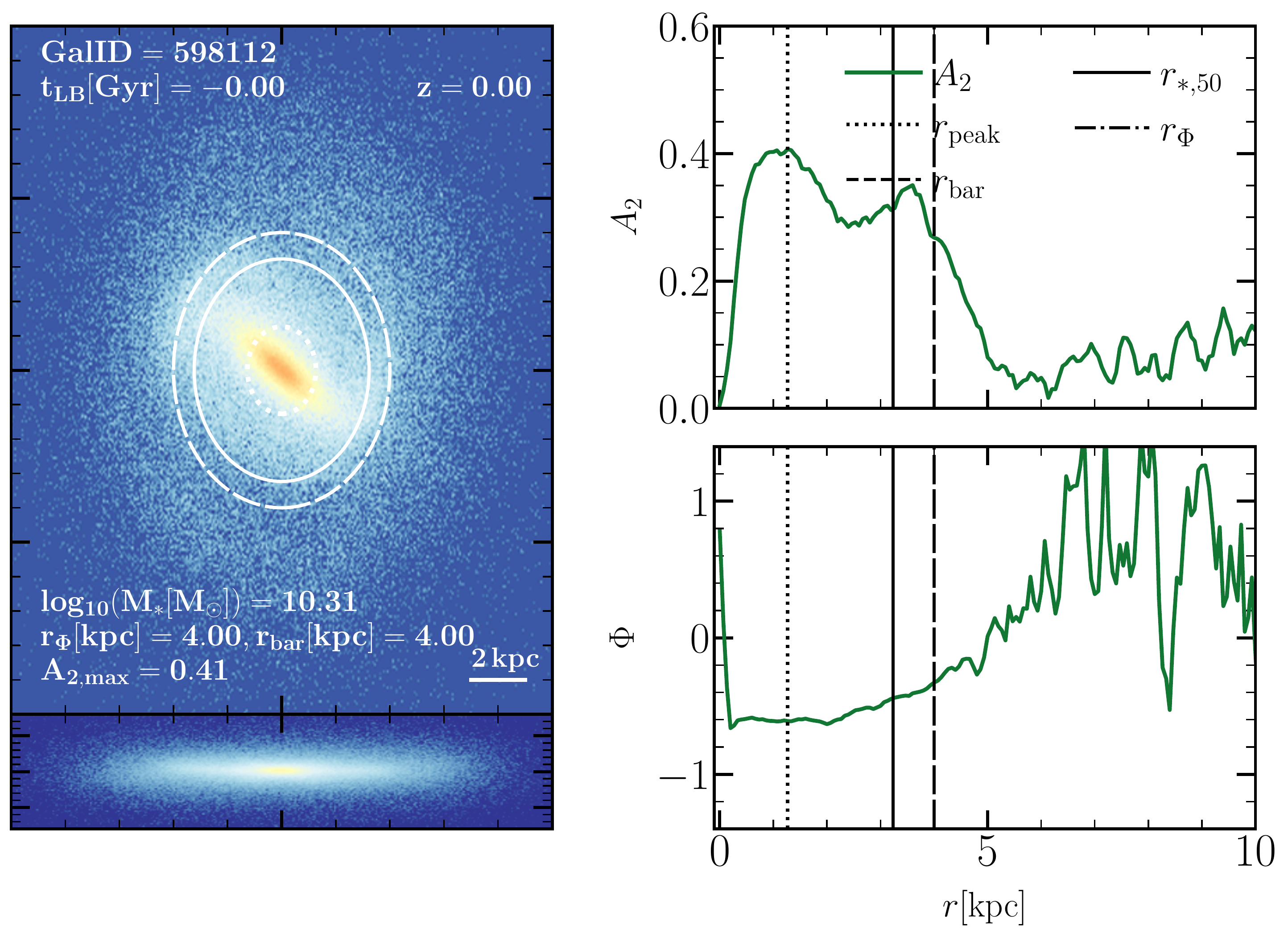} &
\includegraphics[width=1\columnwidth]{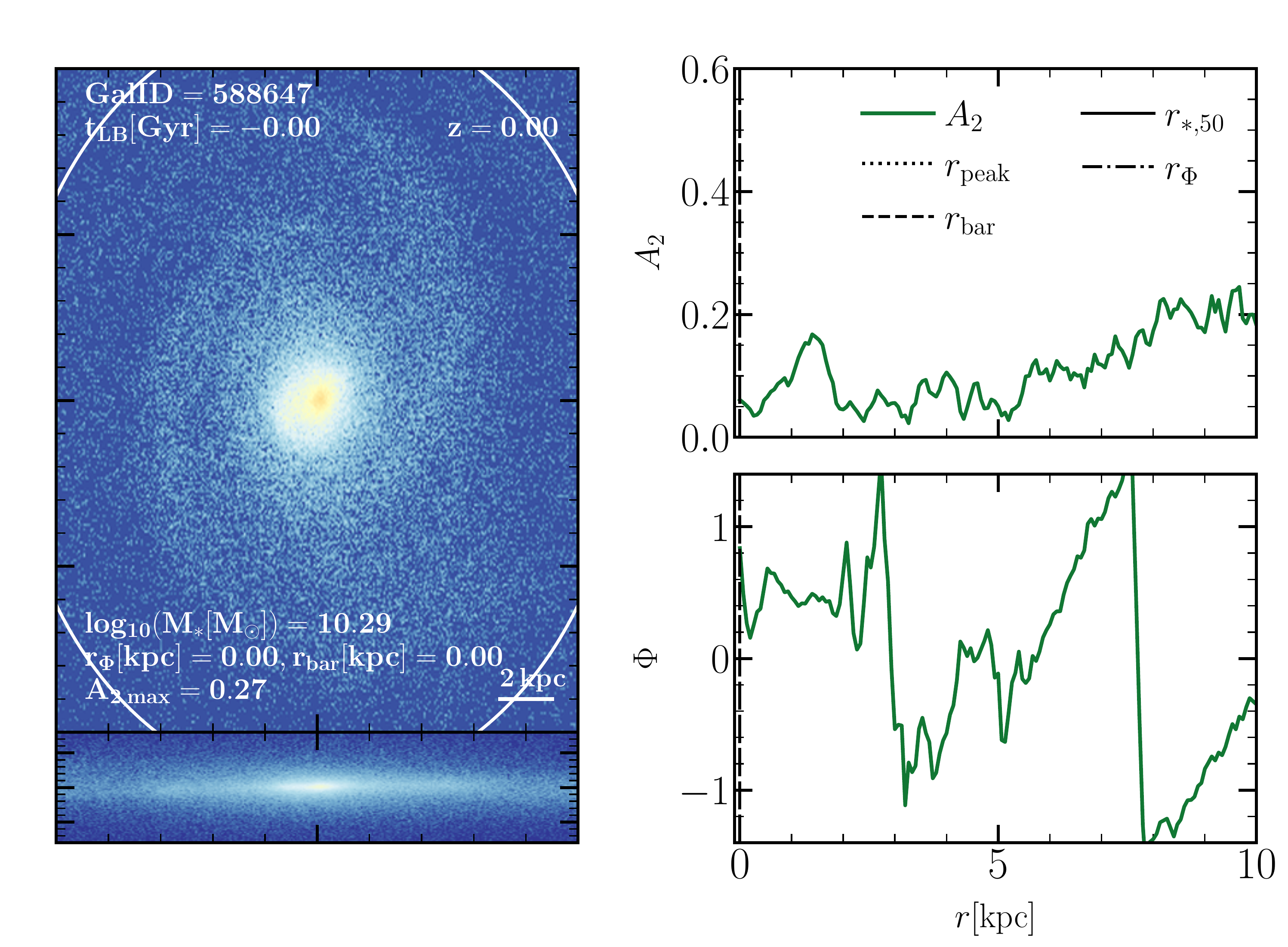}  \\

\end{tabular}
\caption{ Examples of barred galaxies (left panel) and unbarred galaxies (right panel) at  $z=1$ (top row), $z=0.5$ (middle row), and $z=0$ (bottom row). The left column of each panel includes face-on (top) stellar density maps of $20\times 20\times 1\, \rm kpc^{-3}$ and edge-on (bottom) views of each galaxy. As in Fig~\ref{fig:Barexamples}, the right column of each panel shows the $\Ato$ profile of the Fourier decomposition of the face-on stellar surface density (green curve) and the phase.The values of $\Ato$ and the phase are smoothed with a Savitzky-Golay filter. Dotted vertical lines represent the radius corresponding to the peak of $\Ato$, $r_{\rm peak}$, and vertical dashed and dotted-dashed lines correspond to the bar size, $\rbar$, and the maximum radius at which the phase is constant, $r_\Phi$, respectively.} \label{fig:Barexamples2}
\end{figure*}

\cref{fig:Barexamples,fig:Barexamples2}  present examples of barred and unbarred galaxies  at all the redshifts considered.  The face-on (top) and edge-on (bottom) stellar density maps show prominent, well-developed discs with a  bar in the inner part of the disc as early as $z=4$. The bar is characterised by the shape of the $\Ato$ profile (right plots in each panel) where the peak of $\Ato$ is the strength of the bar and its phase is constant in the central parts. In contrast, the right column of  \cref{fig:Barexamples,fig:Barexamples2}  corresponds to disc galaxies that have not formed a bar at the same observed redshift. 

\begin{table}
\caption{From left to right columns: redshift, number of disc galaxies, number of barred galaxies, number of strong bars ($\Amax\geq0.4$), number of disc galaxies that are centrals, number of barred galaxies that are centrals, and bar fractions. The binomials errors on the bar fractions are calculated using
the number of galaxies and bars in each redshift bin as $\sigma =(f_{\rm bars}(1- f_{\rm bars})/n_{\rm discs})^{0.5}$}
\begin{center}
\begin{tabular}{|l|l|l|l|l|l|l|} 
\hline
$z$    &     $n_{\rm discs}$  &   $n_{\rm bars}$  & $n_{\rm sb}$&   $n_{\rm central}$ & $n_{\rm bars,central}$  & $f_{\rm bars}$  \\
\hline
$4$    &         7  &   2     &  1  &  7  &  2  & $0.28\pm 0.17$   \\
$3$    &         24 &   10    &  6  &  22 &  9  & $0.41\pm 0.10$  \\
$2$    &         104 &  48    &  23 &  84  & 36 & $0.46\pm 0.04$  \\
$1$    &         260 &  125   &  48 &  205 & 98 & $0.48\pm 0.03$\\
$0.5$  &         318 &  141   &  59  & 247 & 103& $0.44\pm 0.03$ \\
$0$    &         349 &  105   &  53  & 258 & 76 & $0.30\pm 0.02$   \\

\hline
\end{tabular}
\end{center}
\label{table:bars}
\end{table}

\section{Bar properties connected to their host galaxies}
\label{sec:results}
In this section, we present the structural properties of the TNG50 bars  and their host galaxies. 
To study the redshift evolution of bar and galaxy properties, we show the results at 4 redshifts: the bin $z=[2,4]$, $z=1$, $0.5$, and $0$. We join the samples of galaxies at $z=2,3$ and $4$ given the low number of objects available at $z \geq 2$.



\begin{figure*}	
\begin{tabular}{cc}

\includegraphics[width=1\columnwidth]{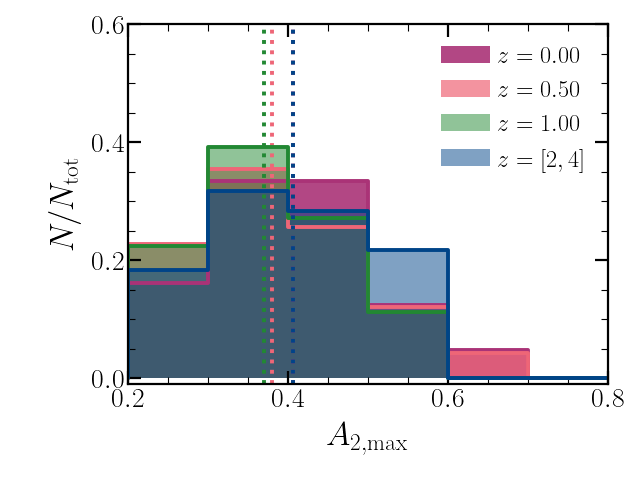}  &
\includegraphics[width=1\columnwidth]{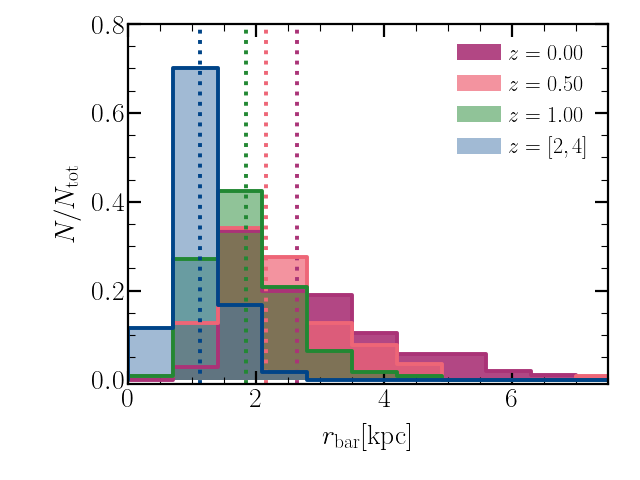}  \\
\end{tabular}
\caption{The distribution of the strength (left panel) and extent (right panel) of bars  at different redshifts. The vertical lines show the median bar strengths and the median bar lengths at each redshift. On average, there is no evolution in the bar strength, whereas the bar extent increases with time.}
\label{fig:distributionbars}
\end{figure*}


Fig~\ref{fig:distributionbars} presents the distributions of the strength (left panel) and length (right panel) of the bars at $z=[2,4]$ and  at $z=1,0.5,0$ separately.  Vertical dotted lines represent the median strength and median extent of the bars at a given redshift.
Overall, we find no significant evolution in the strength of bars. 
Bar extents, on the other hand,  are smaller at high redshift with a median of  $ 1.1\, \kpc$, increasing systematically with decreasing redshift, reaching a median value of $2.6\, \kpc$ at $z=0$. The bar sizes at $z=0$ are consistent with the typical bar sizes found in  late-type galaxies in the local Universe \citep{erwin2005}. For instance, \cite{erwin2019}, using a subset of the disc galaxy sample from S$^4$G \citep{diaz2016a}, find that the typical size of galaxies with $M_{*}=10^{10.5}\Msun$ (which is the median stellar mass of our barred galaxies) is $\sim 2.5\,\kpc $. High redshift bars are more difficult to detect and measure, and fewer data are available. \cite*{sheth2008} using a sample of spiral galaxies from the COSMOS survey, find no significant evolution of the bar size up to $z=1$. However, their sample are limited by the angular resolution of $0.05''$ pixels, not able to resolve small ($<2$ kpc) bars.  As we will show later, limitations in the  spatial resolution might also be the reason why the bar fraction is observed to decrease with increasing redshift.
 
\begin{figure*}
\begin{tabular}{cc}
\includegraphics[width=\columnwidth]{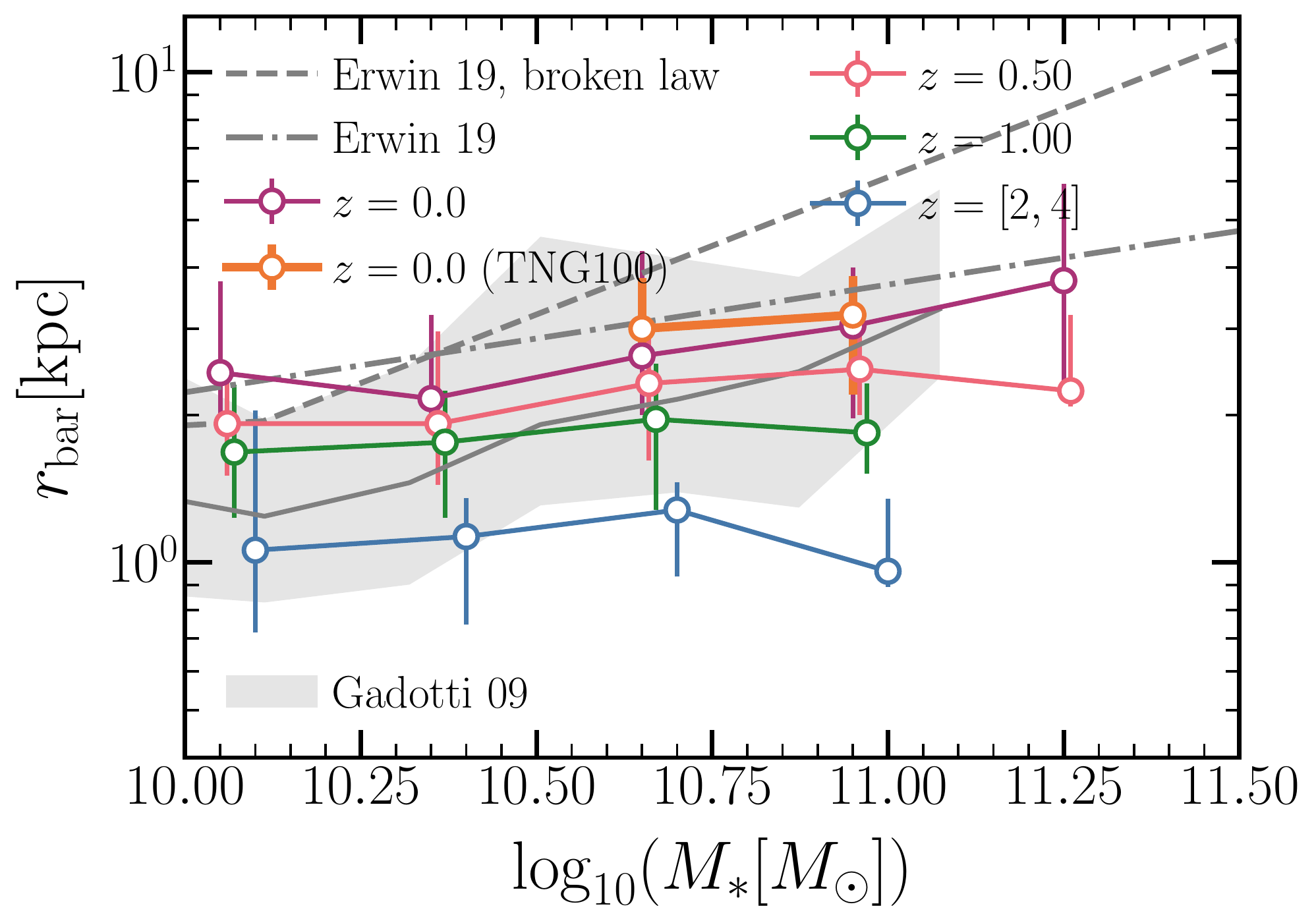} &
\includegraphics[width=\columnwidth]{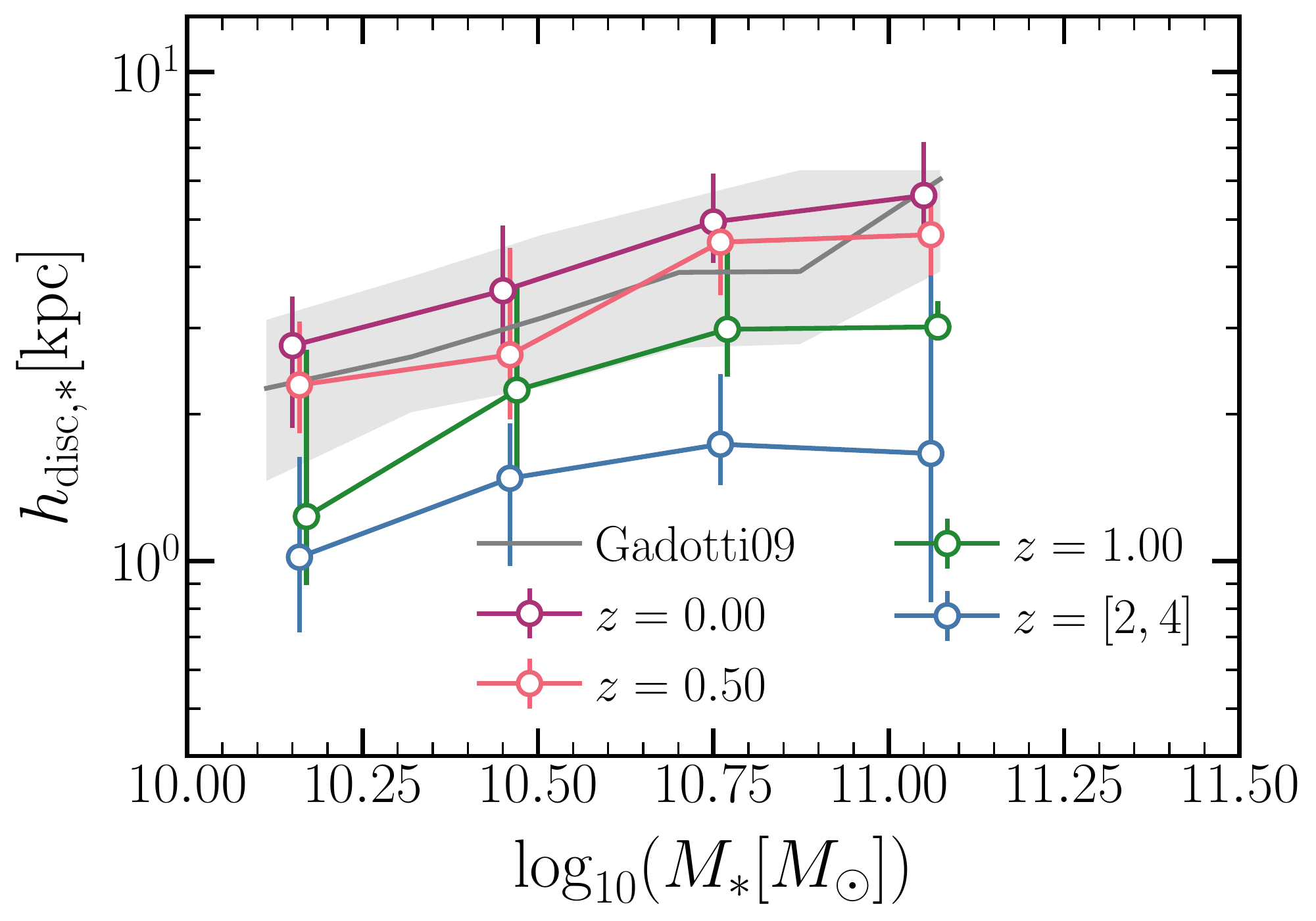}   \\
\end{tabular}
\caption{Bar extents (left panel)  and disc scale lengths (right panel) as a function of stellar mass. Solid lines with circles correspond to the median relation for barred galaxies. Error bars correspond to the 20$^{\rm th}$ and $80^{\rm th}$ percentiles of the distribution in each sample.  The median observed relation in the local Universe by \protect\cite{gadotti2009}  and the linear fit and the broken linear fit estimated by \protect\cite{erwin2019} using S$^4$G are also included. 
}  \label{fig:rbarhdiscrelations}
\end{figure*}

\begin{figure}	
\includegraphics[width=1.\columnwidth]{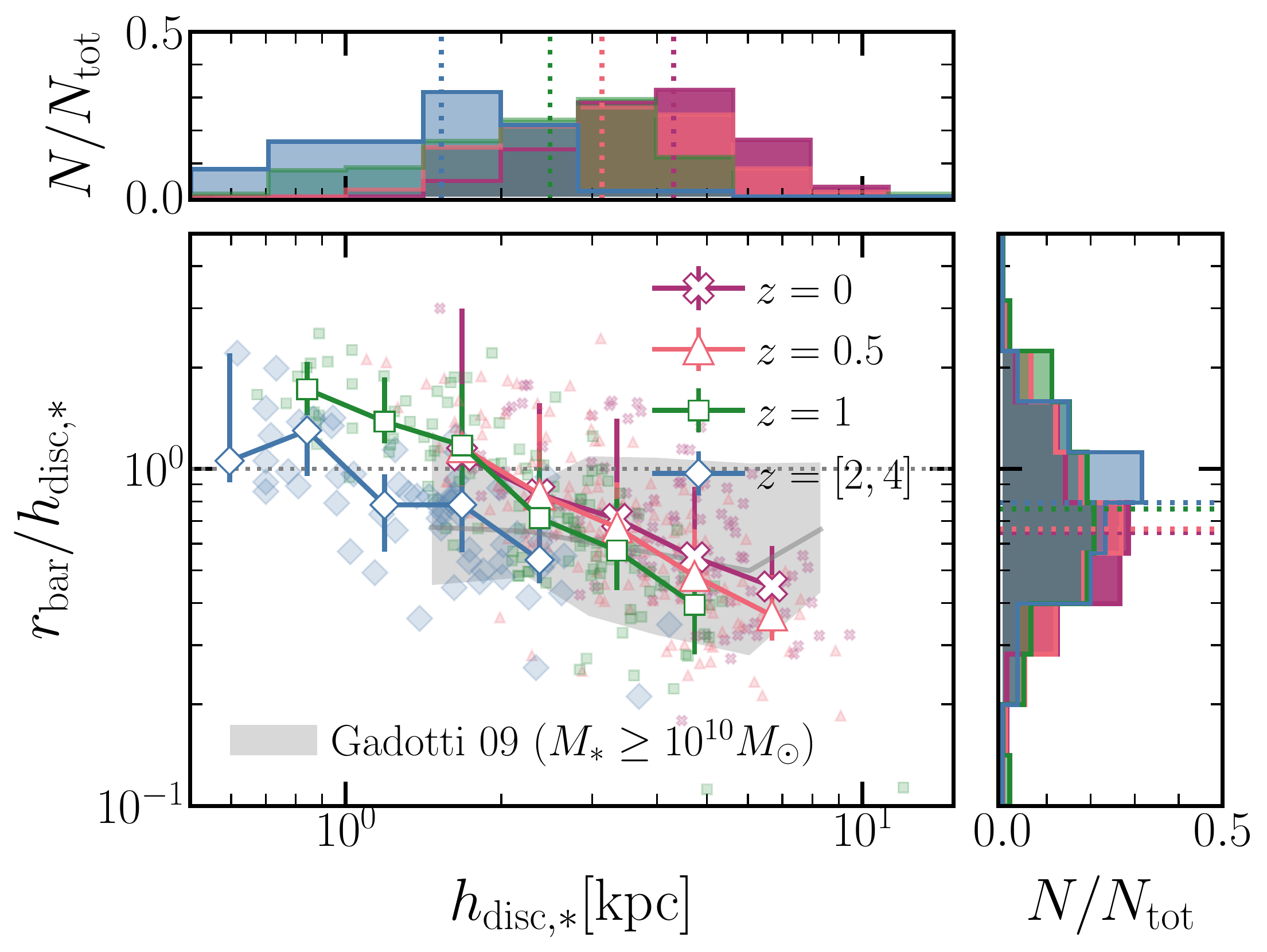}  
\caption{The bar extent relative to the disc scale length, $ \rbar/\hdisc$, in the barred TNG50 galaxies as a function of $\hdisc$ where $\hdisc$ is the scale length of the disc. Different markers and colours represent different redshifts, as indicated in the legend. The grey shaded region and the grey solid line correspond to the observational estimates from  \protect\cite{gadotti2009} for disc galaxies at $z=0$ . The median relation is shown as solid lines with markers for each redshift and only in disc length bins with more than 5 galaxies, and error bars represent the $20^{\rm th}$ and $80^{\rm th}$ percentiles of the distribution.
Panels along the margins show the distributions of disc scale lengths and  $ \rbar/\hdisc$. Dotted lines represent the median of each distribution.}
\label{fig:rbarhdisc}
\end{figure}


We now present, in the left panel of Fig~\ref{fig:rbarhdiscrelations}, the median bar sizes as a function of the galaxy stellar mass at high redshifts ($z=[2,4]$) and $z=1,0.5,0$. Close inspection suggests that the bar sizes roughly increase with galaxy stellar mass larger than $10^{10.4}\Msun$, in particular for $z<1$, and roughly flatten for more massive galaxies. Fixing the stellar mass, we find the median bar sizes grow with decreasing redshift, by more than a factor of 2 from  $z=[2,4]$ to $z=0$, with most of the evolution happening before $z=1$.

The figure also includes  the observed relation  by \cite{gadotti2009,gadotti2011}, who analysed a sample of 300 barred galaxies in the local Universe, and the relation estimated by \cite{erwin2019} from the Spitzer Survey of Stellar Structure in Galaxies (S$^4$G).  Both studies have shown a positive relation  between bar sizes and  galaxy stellar masses, which is in close accord with our barred sample. The predicted relation of our $z=0$ barred sample seems to suggest a flatter correlation than that seen  globally in observations.  In particular, \cite{erwin2019} find that the correlation between bar sizes and galaxy stellar mass could be bimodal, with a steeper relation for galaxies with $M_{*}\geq10^{10.1}\Msun$ (dashed line) than the typical unimodal correlation (dotted-dashed line) and also steeper than our sample.    

Furthermore, we  incorporate the barred sample at  $z=0$ from \citetalias{rosasguevara2020} who use the TNG100 simulation (faint magenta line with circles),  which  contains more massive galaxies ($M_{*}\ge10^{10.5}\Msun$), embedded in a larger volume although at a lower resolution than the  TNG50 simulation. The relation is flatter than observations. We remark that our aim is not to carry out a direct apples-to-apples comparison between observations, other simulations and our barred disc samples, since each study uses different proxies for the bar length and different bias selection. We only want to guide the reader on the global trends.

The extent of bars is also connected to the size of the disc in which they are embedded.  We consider as a measurement of disc  “size”, the disc scale length, $\hdisc$, calculated from the stellar surface density profiles, which are simultaneously fitted to the sum of a Sersic and an exponential profile using a particle swarm optimisation that minimises the $\chi^2$. The procedure is described in Appendix~\ref{app:decomp}, where examples are also provided. 

 In the right panel of Fig.\ref{fig:rbarhdiscrelations} we show how the disc scale length for barred galaxies is related to the stellar mass and how the relation evolves with redshift.  As expected, the disc scale length increases with the galaxy stellar mass at all redshifts.
  For a given stellar mass,  high-redshift discs   are smaller than their analogues at lower redshift ($\hdisc\sim 1.2 \, \kpc$ at $z=[2,4]$ vs.  $\hdisc\sim 4.5 \, \kpc$ at $z=0$). These values are  consistent with the estimates of the sizes of $z=2$ massive disc galaxies ($\sim 1.5\,\kpc$) found in the observational study of \cite{vanderwel2011} using HST observations. This is compatible with the evolution of galaxy size-mass relation found by \cite{vanderwel2014} for late-type galaxies using the CANDELS survey since $z=3$. Besides, it is also in agreement with the evolution of the complete TNG50 galaxy population in \cite*{pillepich2019} (see also the size evolution of the TNG100 galaxies in \citealt{genel2018}). The median relation of barred galaxies in TNG50 is slightly higher than that estimated by \cite{gadotti2009}, albeit within the distribution scatter. We also obtained agreement in comparison to  the linear broken fit provided by \cite{erwin2019}.
 It is worth noting that the relation is similar in unbarred galaxies. 
 To ensure that our results do not depend on the method used to estimate the disc scale length, we present the evolution of this relation using the half stellar-mass radius, $\rhalf$, of the galaxies in Appendix \ref{app:sizes}, finding similar trends. 

While both the bar size and disc scale length increase with decreasing redshift (in particular for smaller galaxies), the dependence with mass is steeper for the disc scale length.
We explicitly look at how bar sizes and disc sizes relate with each other in Fig. \ref{fig:rbarhdisc}, where we show the ratio between the bar extent and the disc scale length for the barred sample as a function of $\hdisc$.
The median is calculated only for bins with more than 5 galaxies. The panels along the margins how the $\hdisc$ and $\rbar/\hdisc$ distributions, and dotted lines represent the median of each distribution.  First of all, as discussed above, we see a clear evolution of the disc scale lengths for the entire sample, increasing from $\approx 1.2\, \rm kpc $ at $z=[2,4]$ to $\approx 4.5\,\kpc$  at $z=0$.  The $\rbar/\hdisc$ ratio, instead, barely evolves with time and in our sample is not  a flat function of $\hdisc$. This is consistent with the small growth of the bar lengths with redshift.

To guide the reader, we compare these values with estimates from the local Universe by \cite{gadotti2009,gadotti2011} (grey shaded region). In the figure we only include the barred massive spiral galaxies  ($M_{*}\geq 10^{10}\Msun$) studied from the authors to make a more consistent comparison, since as we showed previously, both bar and galaxy sizes depend on stellar mass. We find a rough agreement between the predicted relation and observations at $z=0$, although our dependence with the disc scale length is stronger.
This is because the correlation between bar sizes and stellar masses is steeper in \cite{gadotti2009}
than the one predicted  by our barred sample, as we have seen in Fig. \ref{fig:rbarhdiscrelations}.
There is currently no data on how the relation evolves with redshift, given the difficulties of properly detecting and characterising a bar. It would be interesting, though, to observe how this relation evolves with time.



\begin{figure*}
\includegraphics[width=2.1\columnwidth]{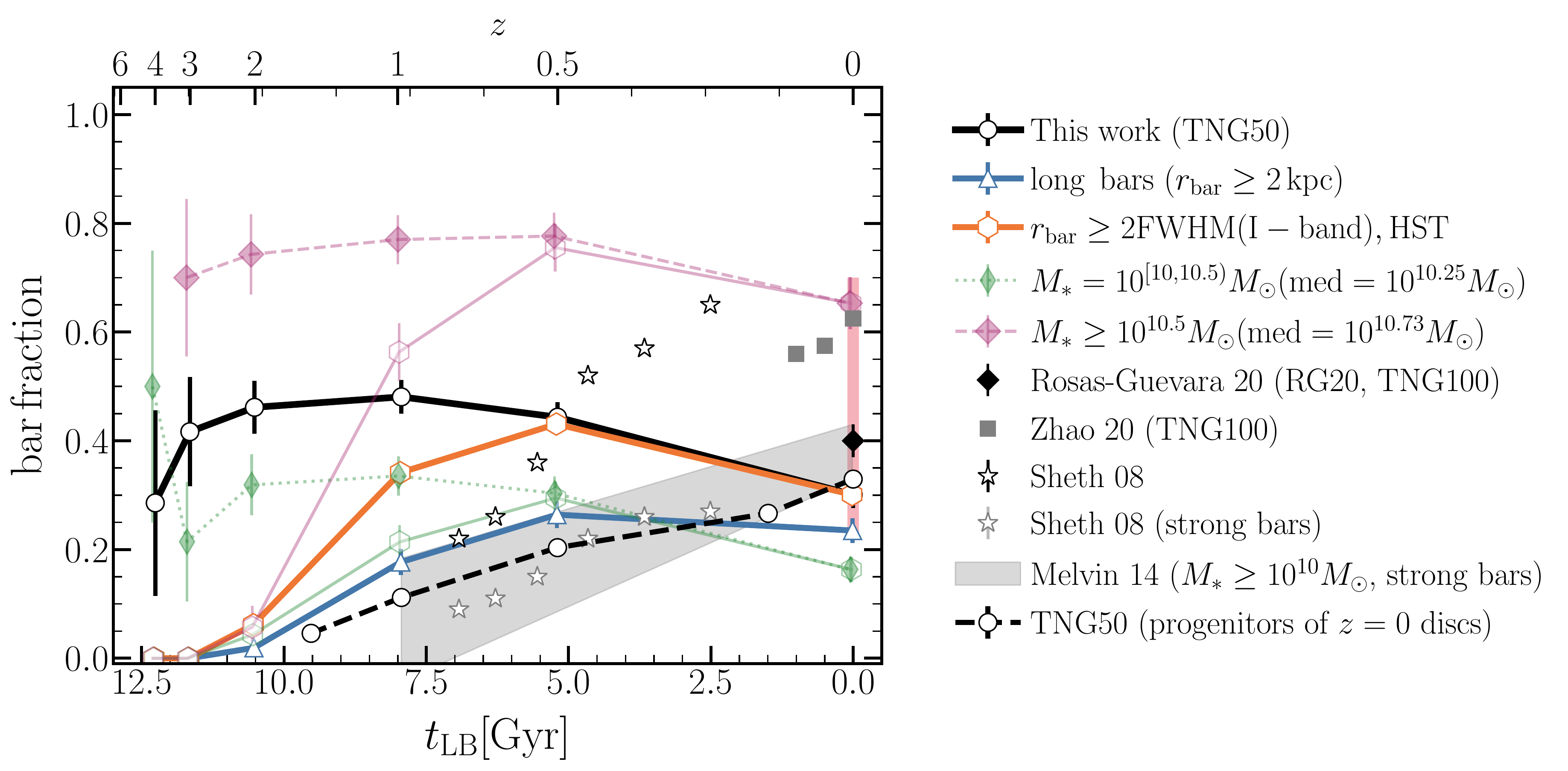}

\caption{
The evolution with redshift (look back time) of the total bar fraction (black solid lines with circles) for our parent disc sample and two stellar mass bins: $\geq 10^{10.5}\Msun$ (dashed lines with purple diamonds) and $< 10^{10.5}\Msun$ (dotted lines with green diamonds). Error bars correspond to binomial errors that are calculated using the number of galaxies and bars in each redshift bin as $\sigma =(f_{\rm bars}(1- f_{\rm bars})/n_{\rm discs})^{0.5}$. The total bar fraction is nearly constant with redshift  up to $z =0.5$, then decreases at $z=0$ and it is consistent with observational estimates (pink shaded area). The bar fractions of \protect\cite {melvin2014} (grey shaded area) and \protect\cite {sheth2008} (black and grey stars)  decline with increasing redshift. This trend is recovered when we consider only long bars ($\rbar > 2\kpc $, blue line with triangles) or bars whose sizes are above the spatial resolution of the HST images (orange lines with hexagons). The bar fraction limited by the spatial resolution of the HST images for the two stellar mass bins is also included (faint lines with purple and green hexagons). Grey squares \protect\citep{zhao2020} and black diamonds \protect\citepalias{rosasguevara2020} show the predicted values from the TNG100 simulation for galaxies with $M_{*}\geq10^{10.6}\Msun$ and, $M_{*}\geq10^{10.4}\Msun$ respectively The dashed line present the bar fraction estimated only considering the progenitors of $z=0$ disc galaxies} \label{fig:barfractionz}
\end{figure*}

To conclude this section, in Fig. \ref{fig:barfractionz}, we show the predicted bar fraction as a function of redshift, where we define as bar fraction the number of disc galaxies with a bar over the total disc galaxy number at any given redshift (black line, see also Table~\ref{table:bars}). At $z=0$ the predicted bar fraction is, $0.30$ which is broadly consistent with observational estimates that range between 25 and 70 per cent (\citealt{barazza2008,aguerri2009,masters2011,cheung2013} and more recently \citealt{melvin2014,diaz2016a,erwin2018}).  The large discrepancy between different observational results is attributed to several selection effects, such as the method of identification of the bars \citep{erwin2018}, and how galaxies are selected \citep{melvin2014}. Note, however, that the predicted fraction of bars in the local Universe is also roughly consistent with the theoretical works of \citetalias{rosasguevara2020} and \cite{zhao2020} who analysed the TNG100 simulation using different methods for the disc galaxy selection and bar identification and derived bar fractions of  $0.40$ and $0.60$, respectively. These values are also consistent with the results of  \cite*{algorry2017} who analysed the EAGLE simulation and obtained a bar fraction of $0.4$. \cite*{peschken2019} instead, analysed the Illustris simulation and found a lower bar fraction ($\sim 20$ per cent).

We inspect the redshift evolution of the bar fraction in Fig.~\ref{fig:barfractionz}. We find  a mild  evolution of the bar fraction  with redshift (solid black line), increasing from $0.28$  at  $z=4$ and reaching the highest value of $0.48$ at $z=1$.  Then, it smoothly declines to $0.30$ at $z=0$.
To make a qualitative comparison between the predicted and observed bar fractions,  we have included the estimated bar fraction by \cite{sheth2008} (black and grey stars) and \cite{melvin2014} (shaded grey region) who both used the COSMOS survey. Although both observational estimates indicate that the bar fraction decreases as redshift increases,  they differ from each other in their galaxy selection and bar identification methods. For instance, \cite{melvin2014}  used an identification method from the Galaxy Zoo project and concentrated exclusively on strong bars in massive galaxies ($\geq10^{10}\Msun$), whereas  \cite{sheth2008} focused on both  strong and weaker bars, and their selection is complete for galaxies with $M_{*}\gsim10^{10.5}\Msun$ at $z=0.9$.  These differences are consistent with the fact that the discrepancies between  \cite{sheth2008} and   \cite{melvin2014} are smaller at redshifts higher than $z>0.5$.

Additionally, the galaxy samples of both \cite{sheth2008} and \cite{melvin2014}  have  limited spatial resolution,  allowing only structures $>2$ kpc to be resolved. Motivated by this, in Fig. \ref{fig:barfractionz}, we also show the bar fraction when only long bars  are considered (blue solid line), where we define  long bars  as those with $\rbar\geq 2\kpc$.  With the exception of a slight increase at $z=0.5$, we observe a gradual decrease in the bar fraction with increasing redshift in this case, which is broadly consistent with the trend observed in the data of  \cite{melvin2014} and  \cite{sheth2008}. We note, however, that the predicted bar fraction is slightly lower than that reported in \cite{melvin2014}. This could be  due to two factors. The first one is that TNG50 has a small volume with low statistics of massive disc galaxies (mostly because of cosmic variance). By including the data from \citetalias{rosasguevara2020} who uses TNG100, which has a larger volume but a lower resolution, the bar fraction is higher and in agreement with \cite{melvin2014}. The second possible reason
could be that our selection criteria identify too few bars. While indeed our criteria are conservative, our bar sample is consistent with the one derived by Zana et al., who define disc galaxies and bar structures with a different approach and reach similar conclusions on the evolution of the bar fraction. Along the same line, we estimate the bar fraction of the TNG50 galaxies, selecting only bar sizes larger than twice the PSF FHWM of the HST $I$--band  as  described in \cite{erwin2018}. This limit may represent a lower bound on the angular size of detectable bars, and it is equal to $0.09''$. The results are depicted in the figure as an orange line, and show a slight increase in the bar fraction at $z=0.5$, followed by a constant decrease at higher redshifts.  These two exercises reconcile the observed and predicted evolutions of the general bar fraction. As  pointed out by \cite{erwin2018},  the bar fractions at higher redshifts may be underestimated if selection effects due to the limited angular resolution of the survey are not properly accounted for. Bars smaller than the  available angular resolution would be omitted. Indeed, as illustrated in Fig. \ref{fig:distributionbars}, at higher redshift the TNG50 bars are generally smaller than in the local Universe.

To conduct a fair comparison of observed and expected bar fraction evolution, all potential selection biases and the restricted angular resolution of each observational survey and simulation must be taken into account. This, however, is beyond the scope of the paper, which seeks to familiarise the reader with the spectrum of observations.

We also explore how the bar fraction of the TNG50 discs varies with stellar mass by dividing the sample into two stellar-mass bins: $\geq 10^{10.5}\Msun$ and  $<10^{10.5}\Msun$ (see the pink and green lines in Fig. \ref{fig:distributionbars}). For the high mass bin, the shape of the bar fraction as a function of redshift is almost flat with larger values than $0.7$, with the highest values ($\sim 0.8$) at $z=0.5$ and followed by a decrease ($0.68$) at $z=0$. This is roughly comparable with the findings of  \cite{zhao2020} (grey squares), who reported an almost constant bar fraction of 0.6 from $z=1$ at the current epoch, analysing disc galaxies with $M_{*}>10^{10.6}\Msun$ in the TNG100 simulation.
For the low mass bin, the bar fractions exhibit a similar dependence on redshift as the total bar fraction, but with lower values at all times (always below 0.35 at $z\leq3$).

The predicted redshift evolution of the bar fraction has implications in the formation and evolution of bars and the assembly of the host galaxies. Since early times, the bar fraction has increased up to $z=1$ consistent with the evolution of the cosmic star formation history and the epoch of bar formation \citep{simmons2014}. At lower redshifts, the probability of a disc forming  a bar decreases as redshift decreases. This small decrease in the bar fraction at low redshift could be for a variety of  reasons: (1) Despite the fact that the number of discs increases  over time \citep{pillepich2019,vanderwel2014},  they do not efficiently form a bar.  Indeed, this is consistent with our findings in section \ref{subsec:buildup} (see Fig. \ref{fig:agespassembly})  that unbarred galaxies form later in time than barred galaxies even though they have similar massive discs and smaller bulges (see Fig. \ref{fig:evolutionmorph}). (2) Another possibility is that the rate of disc destruction in barred galaxies could be different from unbarred galaxies. For instance, we have seen that disc galaxies at $z=1$ undergo a morphological transformation.  Many of these discs are destroyed before they reach $z=0$, but others are capable of developing a bar. If the  merger histories of barred and unbarred galaxies were different, this would affect the bar fraction.  As an illustration, in Fig. \ref{fig:barfractionz}, we show the bar fraction estimated using only the progenitors of the $z=0$ disc galaxies (dashed lines), which exhibit a decreasing trend with increasing redshift in remarkable agreement with \cite{melvin2014}.  This is also the case in \cite{fragkoudi2020} who find a similar redshift evolution of the bar fraction with the Auriga suite. Nonetheless, the authors obtain a higher bar fraction than observed and our predicted bar fraction at lower redshift. The third possibility (3) is  that bars could be destroyed but not the discs, as a result of bar buckling, which causes  the bar to  evolve along  the perpendicular axis, via vertical instabilities \citep{combes1981,debattista2004}  or flybys that may destroy or delay the bar formation \citep{zana2018b,peschken2019}. Which of these processes (formation/destruction of new discs, dissolution/formation of bars) dominates the evolution of the bar fraction at different cosmological times is something to explore in the future.


\section{Properties of barred and unbarred galaxies}
\label{sec:galaxyprop}


\begin{figure*}
\begin{tabular}{c}
\includegraphics[width=2\columnwidth]{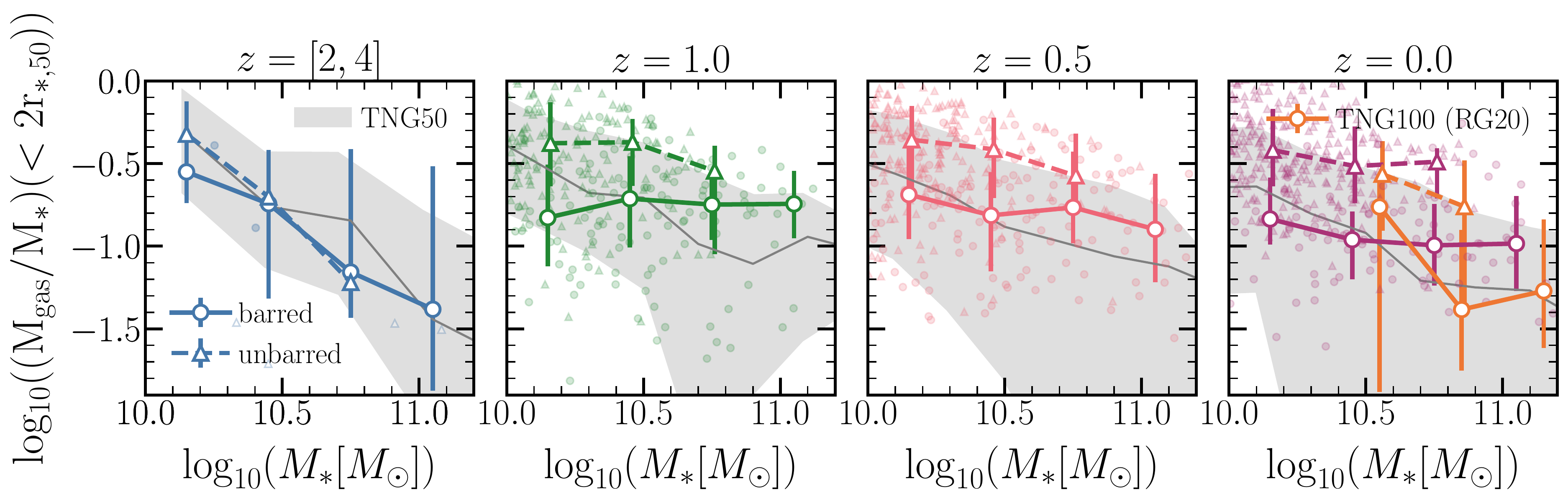} \\
\includegraphics[width=2\columnwidth]{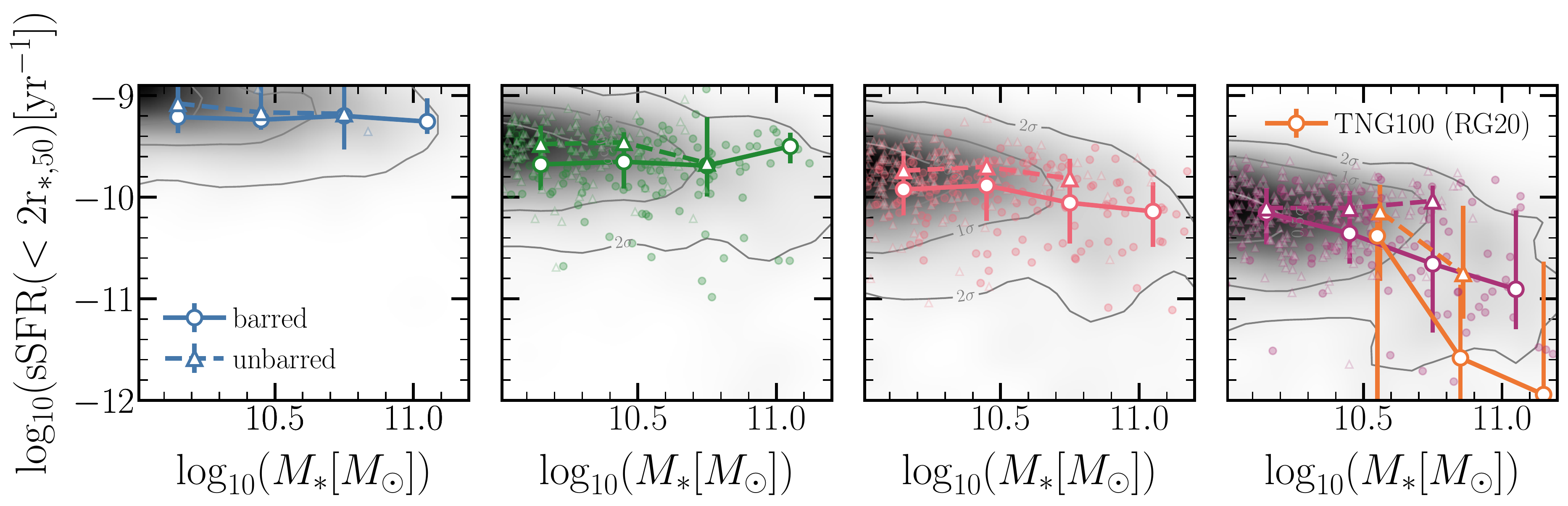} \\
\end{tabular}
\caption{Gas fractions (top row) and specific star formation rates (bottom row) as a function of stellar mass at redshift $[2,4]$ and $z=1,0.5,0$.
Circles correspond to the medians  for barred galaxies and triangles for unbarred galaxies, and the error bars represent the 20$^{\rm th}$ and 80$^{\rm th}$ percentiles. Grey lines and shaded regions correspond to the median gas fractions and the 20$^{\rm th}$ and 80$^{\rm th}$ percentiles for all the TNG50 galaxies. The diffuse density maps and contours show the $\rm sSFR$ distributions for all the TNG50 galaxies at a given redshift. Faint purple symbols and lines correspond to the barred and unbarred galaxies from \protect\citetalias{rosasguevara2020}. At all redshifts, barred galaxies display a lower gas fraction. The SFRs of the two populations are generally comparable except for galaxies with $M_{*}\leq 10^{10.5}\Msun$ at low redshifts ($z<0.5$), that present less SF activity, as also seen in the TNG100 galaxies.}  \label{fig:gasfracmstar}
\end{figure*}

\begin{figure*}

\begin{tabular}{c}
\includegraphics[width=2\columnwidth]{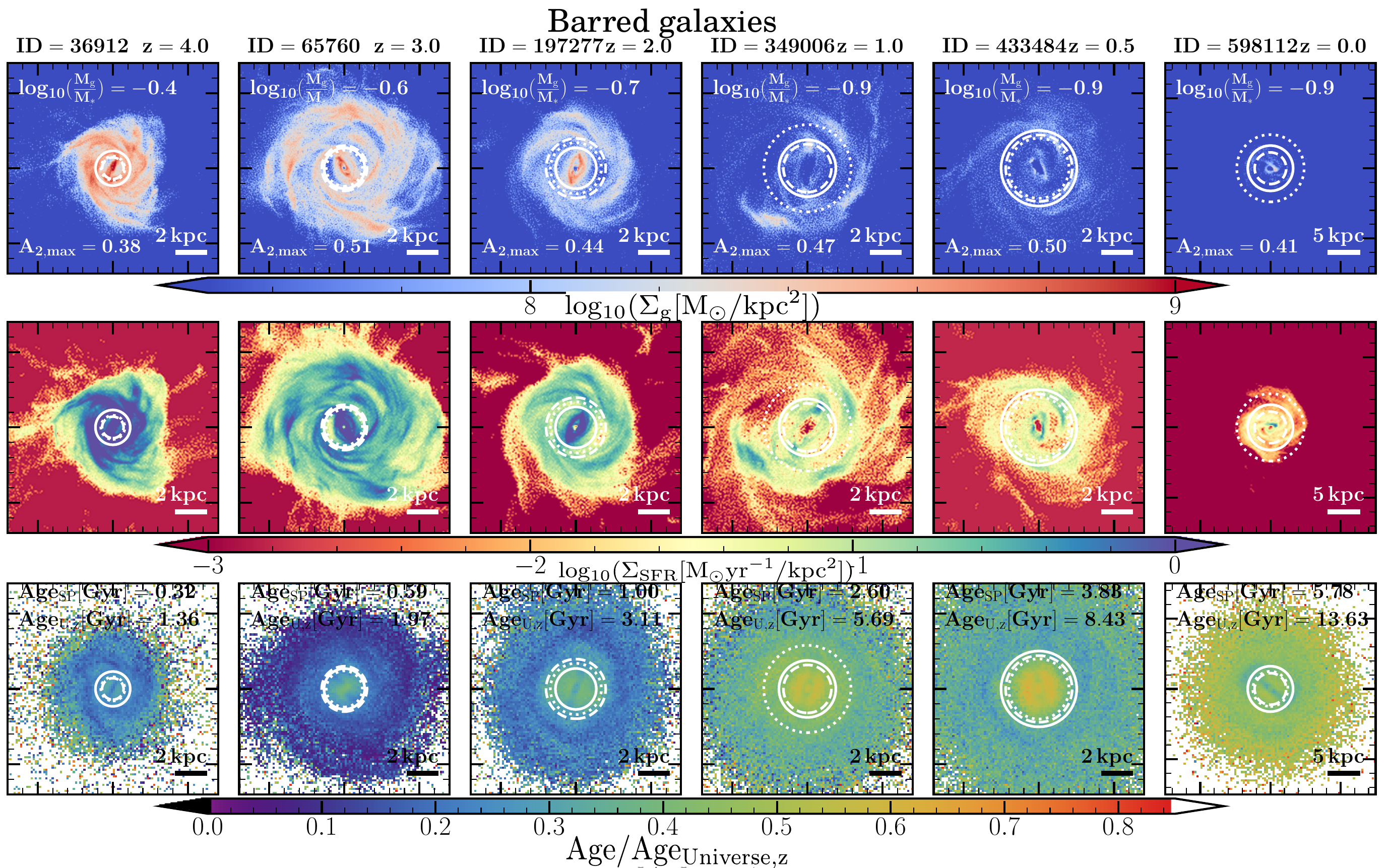} \\
\includegraphics[width=2\columnwidth]{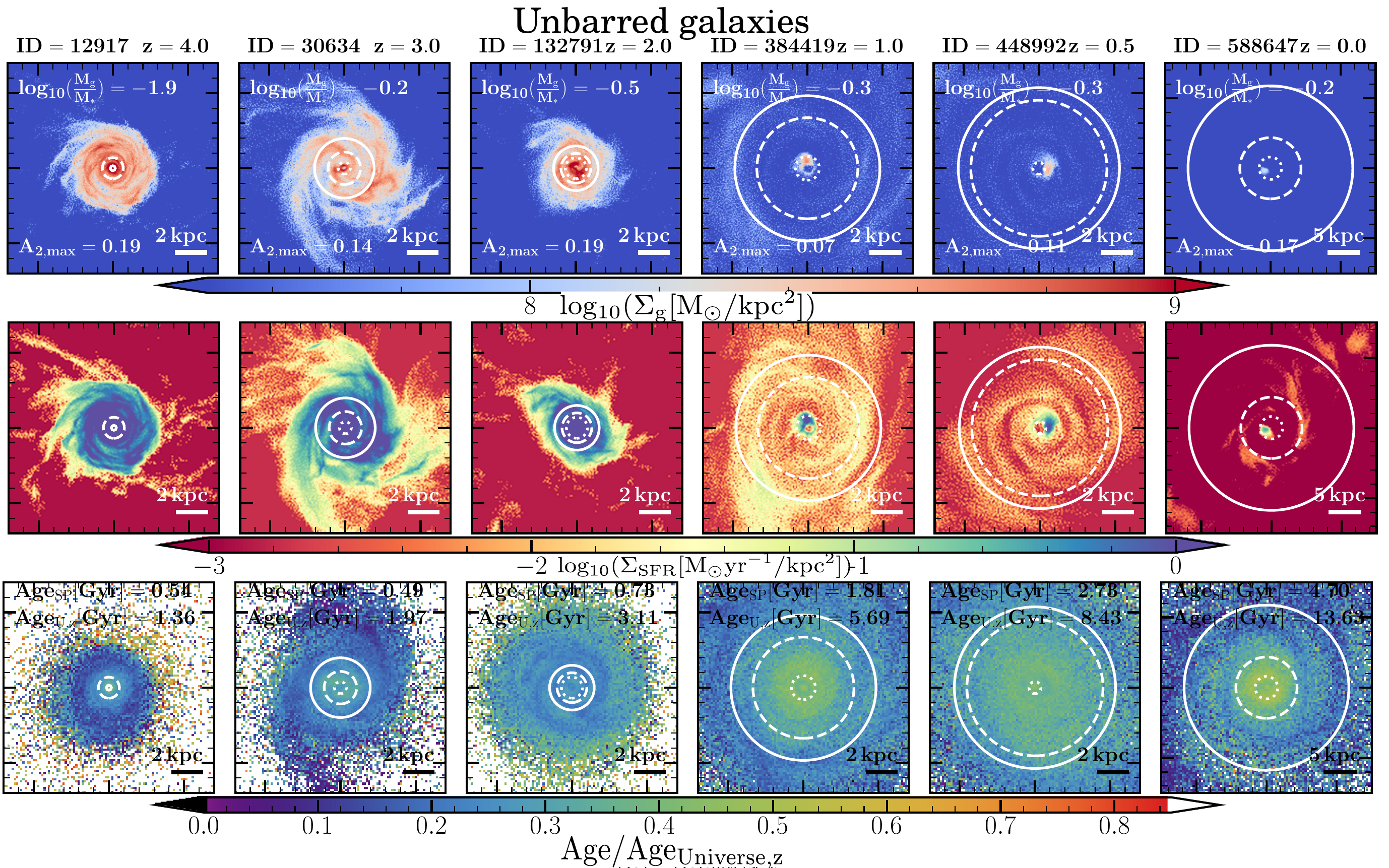} \\
\end{tabular}

\caption{ Gas (top rows), star formation (middle rows) face-on density maps and the average  age over the Universe age  maps  of the stellar component (bottom rows) for the barred and unbarred galaxies shown in Figs~\ref{fig:Barexamples} \& \ref{fig:Barexamples2}. Their  stellar masses are all in the range  $10^{10.25-10.5}\Msun$. Circles with white solid, dashed and dotted lines indicate the stellar half mass radius, the stellar disc scale length and the bar size, respectively. Maps are obtained from slices of $14\times14\times1$ kpc in $z\geq0.5$ and $30\times30\times1 \,\rm kpc^3$ in $z=0$.  At $z\geq2$, barred and unbarred galaxies are  gas rich with high star formation rates.
Going down to $z=0$, overall, barred galaxies present older  stellar  populations than unbarred galaxies.}
\label{fig:gasdensitymaps}
\end{figure*}

 Many studies have explored the properties of barred and unbarred galaxies in the local Universe, and there is growing evidence that
 bars are more likely to be found in disc galaxies that are  red, massive and gas-poor than those that are in blue, low-mass, and gas-rich (e.g., \citealt{masters2011,masters2012,cheung2013,gavazzi2015,consolandi2016,cervantes2017,geron2021}). In this section, we compare barred galaxies to unbarred galaxies without a transient bar structure and with the same stellar masses/halo masses as the barred sample. In subsection \ref{subsec:galprop}, we present the properties of the barred and unbarred galaxies at stacked high redshifts $z=[2,4]$ and redshifts $z=1,0.5,0$. We also investigate the buildup of galaxies in subsection \ref{subsec:buildup}. In subsection \ref{subsec:prophaloes}, we present the properties of the host haloes of barred and unbarred galaxies and the relation between stellar and halo mass.
\subsection{Galaxy Properties}
\label{subsec:galprop}
We first analyse the gas fraction and the specific star formation rates for unbarred and barred galaxies. The top panel of Fig. \ref{fig:gasfracmstar} shows the median evolution of the gas fraction-stellar mass relation where we define the gas fraction, $M_{\rm g}/M_{*}$, as the ratio between gas mass and stellar mass enclosed in $2 r_{*,50}$. The median gas fraction of barred galaxies  (solid lines with circles) is systematically lower than the median of unbarred galaxies for a given stellar mass, but with a large scatter at high redshift. This difference is as large as 0.5 dex at $z=0$.  Note, however, that for the highest mass bins ($M_{*}>10 ^{10.5}\Msun$) there are not enough unbarred galaxies for a proper statistical comparison. However, this is consistent with the difference in gas fraction found in \citetalias{rosasguevara2020} for barred and unbarred galaxies with $M_{*}\geq 10^{10.4}\Msun$ at $z=0$ in the TNG100 simulation (faint lines with symbols in the figure). \citetalias{rosasguevara2020} found a difference up to $0.6$ dex in galaxies with  $M_{*}>10 ^{10.7}\Msun$.

Along the same lines, we characterise the specific star formation rate ($\rm sSFR$) as the ratio between the star formation rate and the stellar mass enclosed in twice $r_{*,50}$. The evolution of the sSFRs for barred and unbarred disc galaxies is shown in the bottom panel of Fig. \ref{fig:gasfracmstar}.  Diffuse density maps and grey contours show the location of all the TNG50 galaxies within this diagram.
From the figure, it is clear that both barred and unbarred galaxies follow the median evolution of all the TNG50 galaxies.  The median values of sSFR of the TNG50 galaxies steadily decrease with decreasing redshift. Focusing on the difference in the SF activity between barred and unbarred galaxies, both high redshift galaxies with and without bars present typical values of high star formation rates, following the ballpark of the galaxy population at these high redshifts.   Towards lower redshifts, barred massive galaxies ($M_{*}>10^{10.5}\Msun$) tend to be less star-forming active,  with values close to the already formed quenched galaxy population (sSFR$\lsim10^{-11}\rm yr$). To expand the stellar mass range of unbarred galaxies at $z=0$,  we  have included the sSFRs in massive barred and unbarred galaxies from \citetalias{rosasguevara2020} from the  TNG100 simulations in a higher stellar mass bin.  They find a more pronounced difference, although with a higher scatter, in the sSFRs in massive spiral galaxies with and without a bar (faint lines at $z=0$).

\begin{figure*}
\begin{tabular}{cc}
\includegraphics[width=1.\columnwidth]{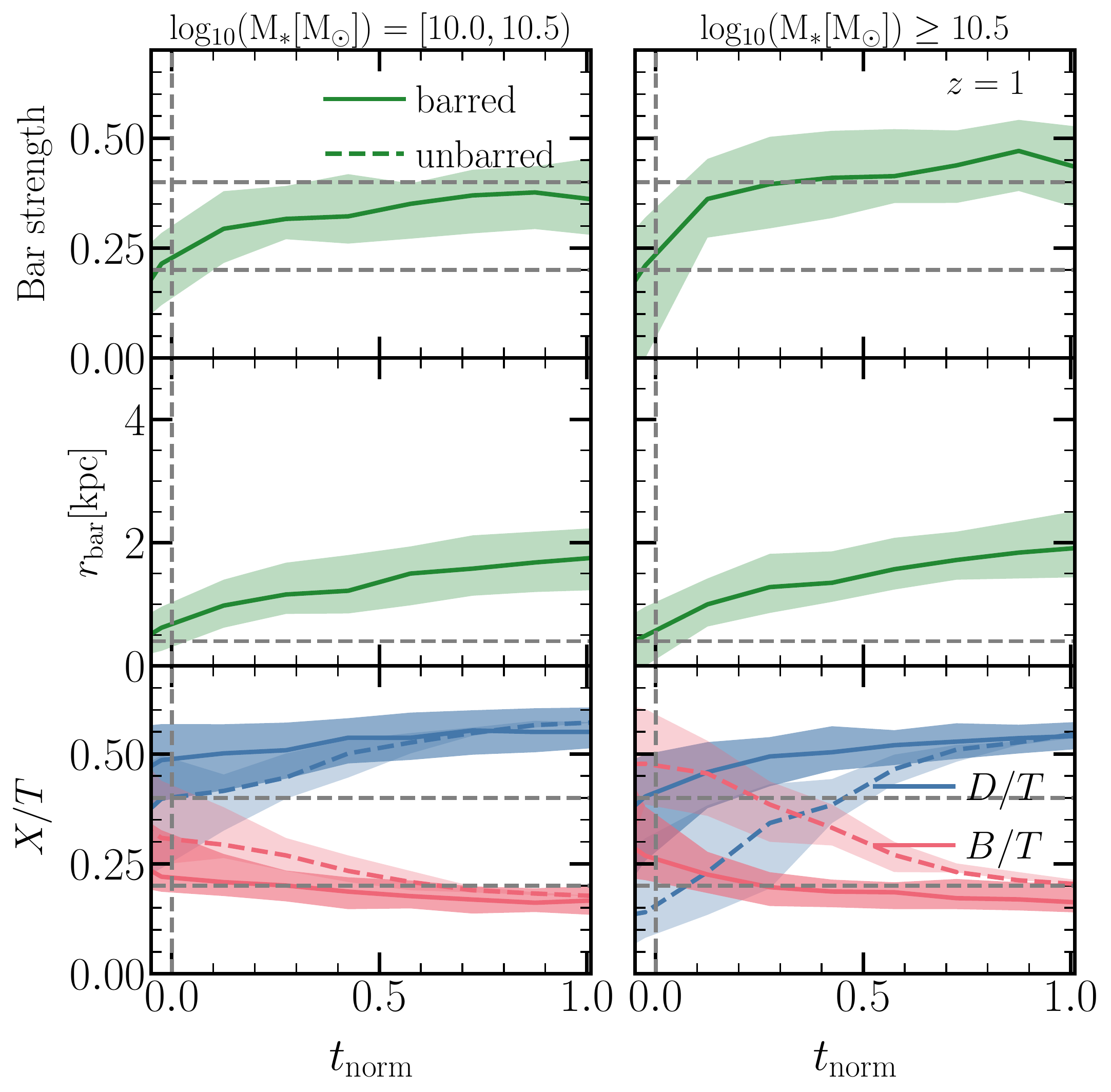} &
\includegraphics[width=1.\columnwidth]{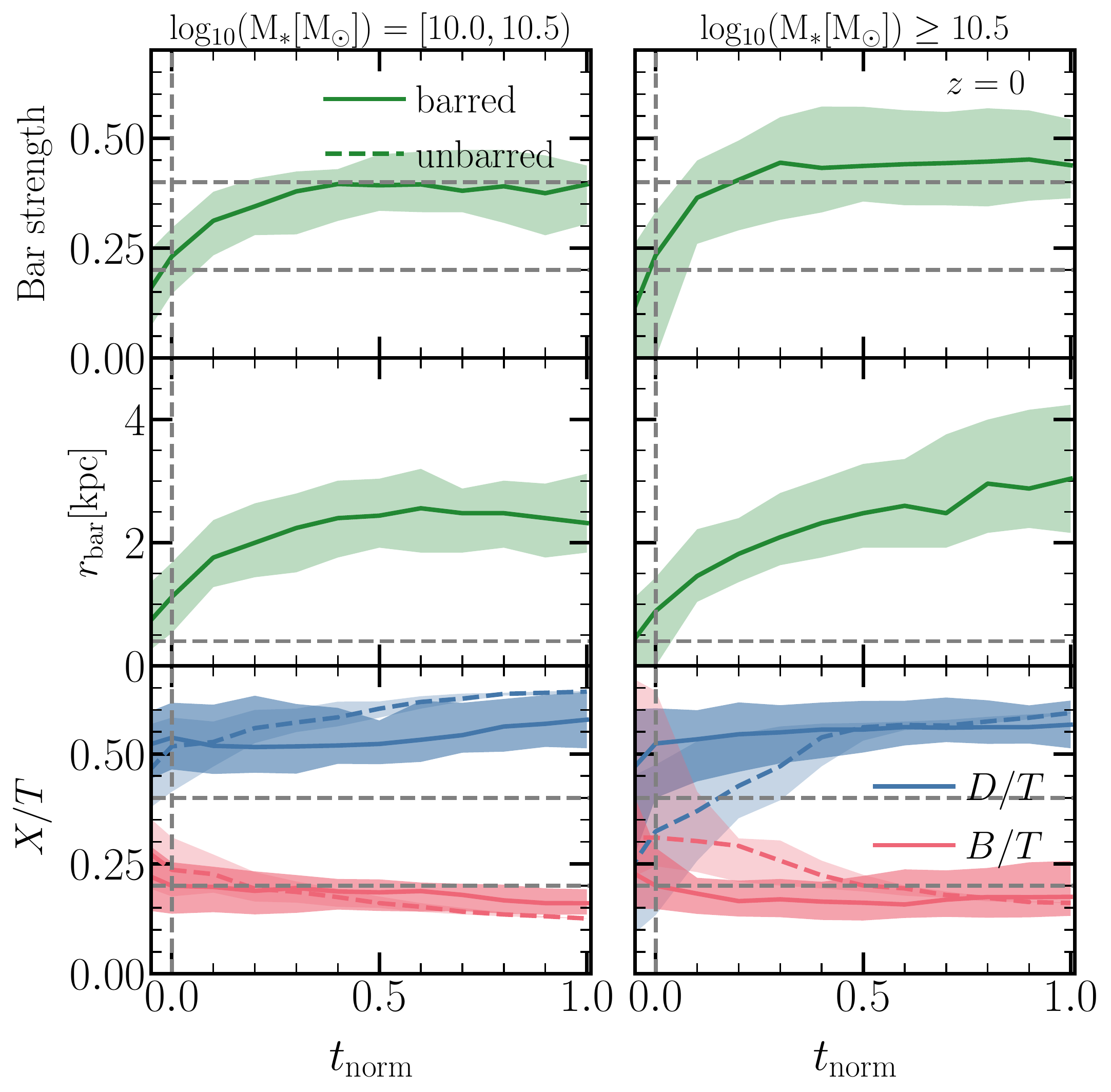} \\
\end{tabular}
\caption{ The  evolution of bar properties and  morphology of barred galaxies at $z=1$ (left panel) \& $z=0$ (right panel) and  for two mass bins, as stated in each column. The evolution is given in terms of $t_{\rm norm}$, where $t_{\rm norm}=(\tbar-t_{\rm LB})/(\tbar-t_{\rm LB,z})$, $\tbar$ and $t_{\rm LB}$ denote the bar formation time and the lookback time respectively and $t_{\rm LB,z}$ the lookback time at $z=1$ or $z=0$. The evolution of morphology of each barred galaxy is compared to the median morphology of unbarred galaxies at each time (see details in the text). Horizontal dashed lines represent values of 0.2 and 0.4 and vertical dashed lines represent the time of bar formation. \textit{Top panels}: The evolution of the median bar strength, which rises with time. \textit{Middle panels}: The evolution of the median bar sizes follow a similar increasing trend as the bar strength.  \textit{Bottom panels}: The median evolution of the $D/T$ and $B/T$ ratios for barred (solid lines) and unbarred galaxies (dashed lines).  Discs in barred galaxies formed fast, with a dominating disc component at the time of bar formation, whereas those in unbarred galaxies formed later and at a slower rate.}
\label{fig:evolutionmorph}
\end{figure*}

To give a visual impression of the spatial distribution of gas and star-forming gas in barred and unbarred galaxies at different redshifts, we show density maps in Fig. \ref{fig:gasdensitymaps}. The figure shows examples of barred (top figure) and unbarred galaxies (bottom figure) from Fig. \ref{fig:Barexamples} \& \ref{fig:Barexamples2} at $z=[4,0]$ . All the galaxies shown have similar stellar masses ($M_{*}=10^{10.25-10.50}\Msun$).
 
High-redshift barred galaxies ($z\geq2$) are as rich in gas and star-forming gas as unbarred galaxies. At these high redshifts, barred galaxies display high levels of star formation, primarily in spiral arms and along the stellar bar whereas  unbarred galaxies exhibit a higher rate of star formation in their spiral arms and central regions. The surface density of the total and star-forming gas decreases with decreasing redshifts ($z\leq1$) for both unbarred and barred galaxies, as seen in Fig. \ref{fig:gasfracmstar}. At these high redshifts, barred galaxies display high levels of star formation, primarily in spiral arms and along the stellar bar, whereas unbarred galaxies exhibit a higher rate of star formation in their spiral arms and central regions. The surface density of the total and star-forming gas decreases with decreasing redshifts ($z\leq1$) for both unbarred and barred galaxies, as seen in Fig. \ref{fig:gasfracmstar}. At lower redshifts, it appears that there is no clear variation in the distribution of the star-forming gas between barred and unbarred galaxies.


The bottom row of each figure shows, for the same galaxies, face-on maps of the age distribution of the stellar population at the observed redshift over the Universe age, $\rm Age/Age_{\rm Universe,z}$. 
 
The figure depicts the assembly of barred and unbarred galaxies at different redshifts, showing that the inner parts are built up first,  consistent with an inside-out formation scenario (e.g., \citealt{fallefstathiou1980}).   On average, there is no substantial difference in the median age of the stellar population between high redshift barred and unbarred galaxies, since both assembled in the previous Gyr. We caution the reader that we are normalising by the age of the Universe, and  this is not the real age of the stellar population.
However, inspecting closely the central parts of the barred galaxies, it is possible to see a young stellar population component along the major axis of the bar. 

By $z=0.5$, most of the assembly of barred and unbarred discs has already taken place, and the stellar populations of the galaxies are globally older. Note that age variations between barred and unbarred galaxies become more prevalent. The interior components of the barred galaxies are older than their unbarred analogues. While the star formation activity is lower or null in the barred galaxies at $z<1$,  it is  still possible to locate a younger population along the large axis of the bars. In the outer regions of unbarred galaxies, the stellar population is younger than the one of barred galaxies. 
At  $z=0$, the  stellar population formed $\sim 6$ Gyr ago ($\rm Age/Age_{\rm Universe,z}\sim 0.4$) in the barred galaxy in our example, whereas the unbarred galaxy  stellar population is formed $\sim 5$ Gyr ago ($\rm Age/Age_{\rm Universe,z}\sim 0.3$). As we will see in the next section, more massive galaxies present larger age differences in the stellar population.

\subsection{The buildup of barred and unbarred galaxies}
\label{subsec:buildup}

\begin{figure}

\includegraphics[width=\columnwidth]{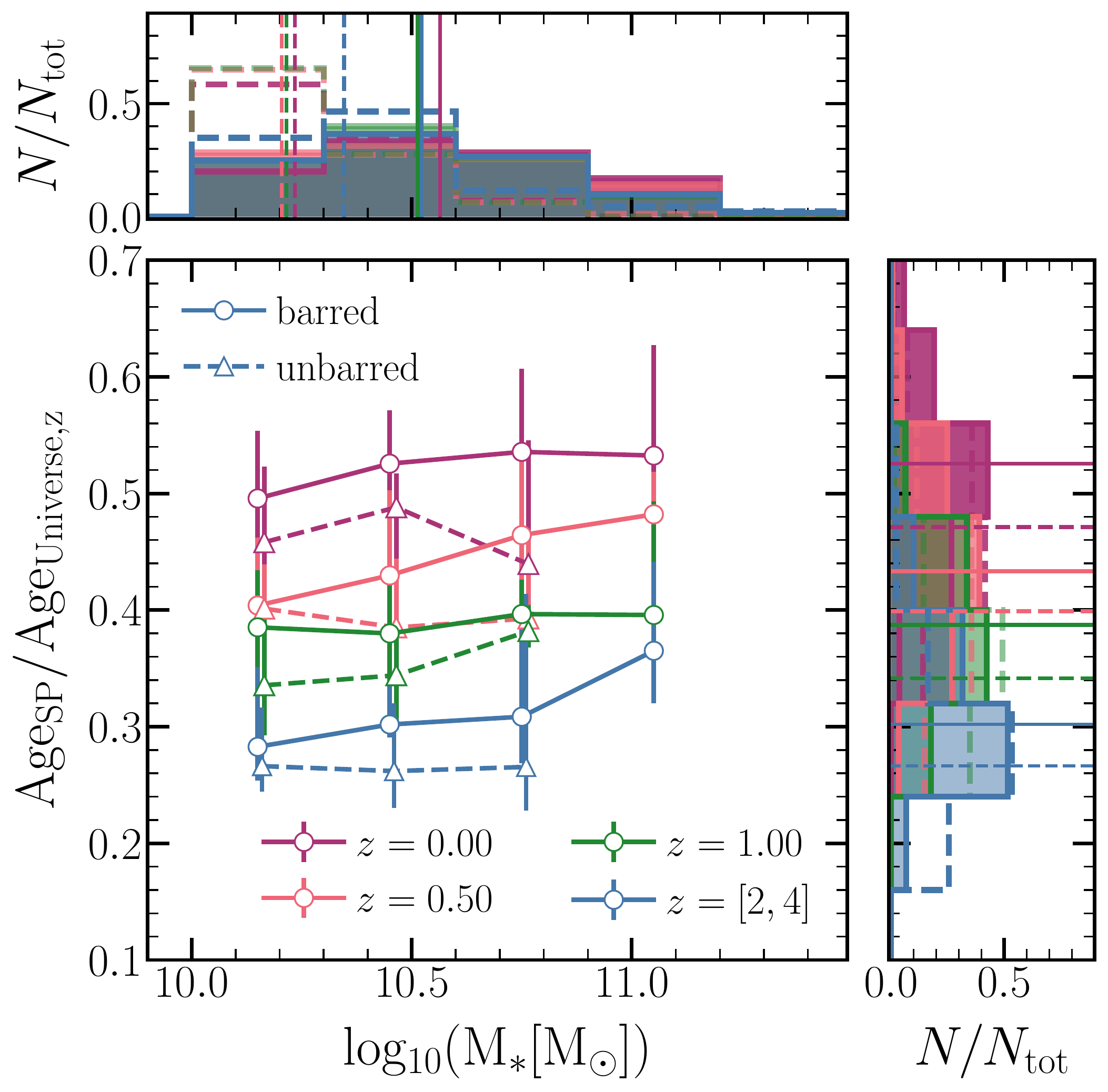}  \\
\caption{ The median age $(\rm Age_{\rm SP}/ Age_{\rm Universe,z})$ as a function of the logarithmic stellar mass at several redshifts where $\rm Age_{\rm SP}$ represents the mass-weighted age of the stellar population inside $2r_{50,*}$ and $\rm Age_{\rm Universe,z}$ is the Age of the Universe at the given redshift.  Solid lines with circles correspond to barred galaxies, and dashed lines with triangles to unbarred galaxies.  The medians are calculated on;y for stellar mass bins with more than 5 galaxies. Error bars  represent the $20^{\rm th}$ and $80^{\rm th}$ percentiles of each distribution. The normalised histograms of  the stellar mass  and  $(\rm Age_{\rm SP}/ Age_{\rm Universe,z})$ distributions are shown in panels along the margins. The solid and dashed  lines represent the median of the barred and unbarred samples distributions, respectively. Barred galaxies overall show older stellar populations than those in unbarred galaxies.}
\label{fig:agespassembly}
\end{figure}

\begin{figure}

\includegraphics[width=\columnwidth]{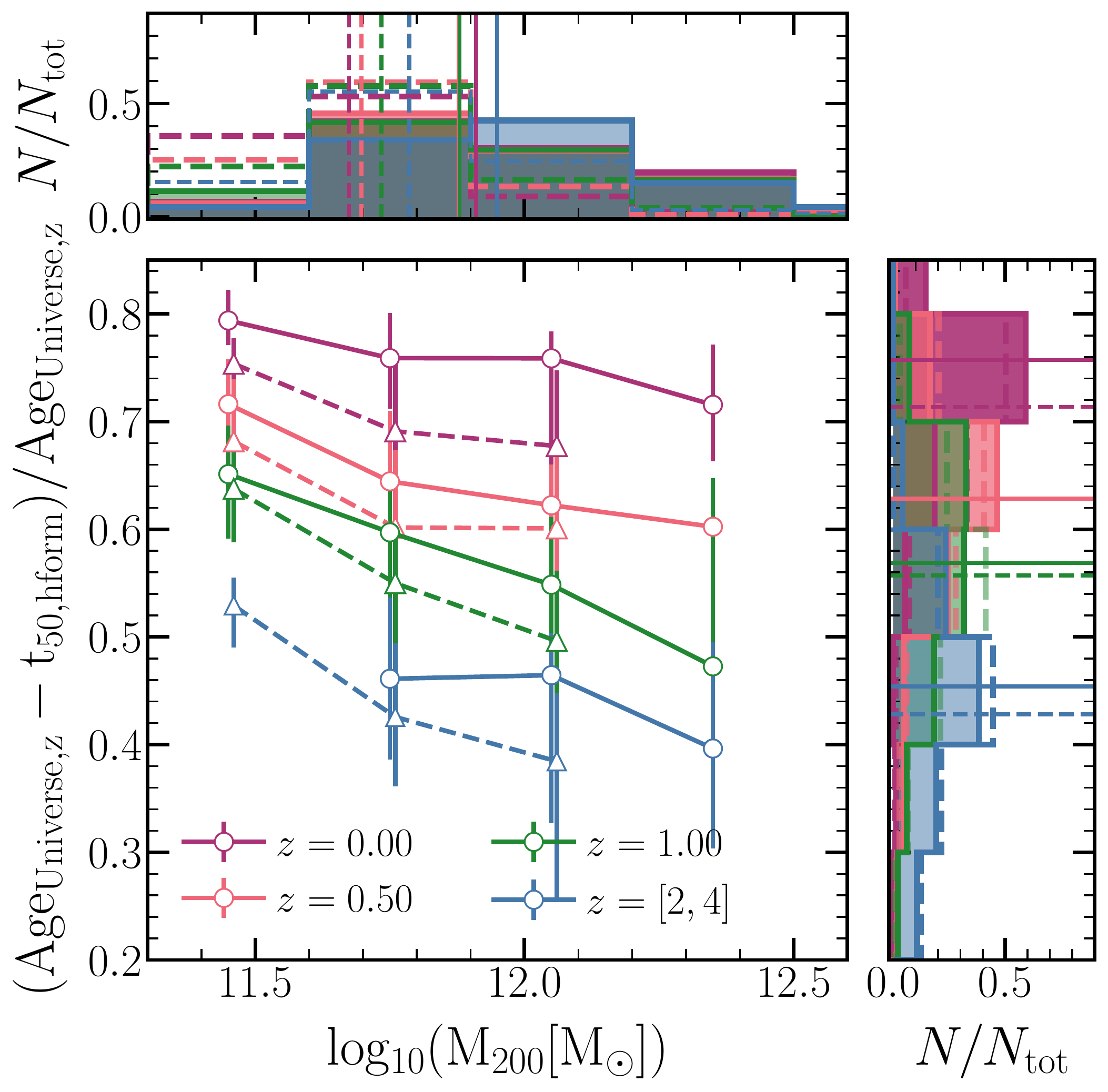}

\caption{The halo formation time, $t_{50,\rm hform}$ relative to and normalised by the Age of the Universe at a given redshift, as a function of the halo mass. Solid lines with circles correspond to the host haloes of barred galaxies, and dashed lines with triangles to the host haloes of unbarred galaxies.  The medians are only calculated for halo mass bins with more than 5 galaxies. Error bars  represent the $20^{\rm th}$ and $80^{\rm th}$ percentiles of each distribution.   Halo mass and halo formation time distributions are shown in the panels along the margins. Host haloes of barred galaxies formed earlier that those of unbarred galaxies since $z=1$,  with the greatest difference in halo formation time at $z=0$.}    
\label{fig:timeassembly}
\end{figure}
We now look at the evolution of bar properties and how this is related to the assembly of the disc and bulge components of the galaxy.  The top and middle panels of Fig. \ref{fig:evolutionmorph} show the  median evolution of the  strength ($\Amax$) and size ($\rbar$) of bars at $z=1$ and $z=0$ , respectively. The evolution of bars is shown in galaxies divided into two stellar-mass bins:  $M_{*}<10^{10.5}\Msun$ and $M_{*}\geq 10^{10.5}\Msun, $ and it is in terms of  $\tnorm$  where $\tnorm=(\tbar-t_{\rm LB})/(\tbar-t_{\rm LB,z} )$
and $t_{\rm LB}$ corresponds to the lookback time,  $\tbar$ to the formation time of each bar defined in subsection~\ref{sub:discsample}, and $t_{\rm LB,z}$ is the lookback time at $z=1$ or $z=0$.  The time  $\tnorm=0$ corresponds to the time of bar formation, whereas $\tnorm=1$ indicates $z=0$ or $z=1$. The top and middle panels of Fig. \ref{fig:evolutionmorph} show that the bar strength and the bar size grow significantly over time. While the bar strength and size of $z=1$ bars steadily increase with time, $z=0$ bars undergo a rise during a short period of time and then remain almost constant for more than 80 per cent of the bar life. Notably, the interval in physical time ($\Delta \tnorm $) is considerably shorter in $z=1$ bars, which are most likely  younger than  $z=0$ bars. This behaviour is visible in both stellar mass bins.

The evolution of the disc-to-total mass fractions, $D/T$, and bulge-to-total mass fractions, $B/T$ as a function of $\tnorm$ is shown in the bottom panel  of Fig. \ref{fig:evolutionmorph}. The morphology of barred galaxies is compared to that of a control sample of unbarred galaxies. This control sample of unbarred galaxies  was chosen to have similar stellar masses to barred galaxies at $z=1$ and $z=0$. Following the approach used by \citetalias{rosasguevara2020} (see their Fig. 4),  we compare the $B/T$ and $D/T$ of each barred galaxy to the median $B/T$ and $D/T$ of the control unbarred galaxy sample at a given time in particular at the formation time of the bar.

The figure emphasises the evolutionary differences and similarities between barred and unbarred galaxies at various redshifts.
By closely inspecting  the evolution of $D/T$ and $B/T$ in the bottom panels of Fig.  \ref{fig:evolutionmorph}, we find that barred galaxies already have a clear disc-dominated morphology at the time of bar formation, since the disc (solid blue lines) hardly grows with time after that and the bulge component only grows mildly. Unbarred discs, on the other hand, are still evolving at the time at which the bar formed in barred galaxies ($\tnorm=0$), and they will continue to grow slowly with time. This difference in the disc assembly between barred and unbarred galaxies is greater for the highest stellar mass bins.

The bulge component of barred galaxies is already  present and small   ($B/T \sim 0.2$ at $z=0$, just above 0.2 at $z=1$) at the time that the bar formed ($\tnorm=0$).  Once the bar form, $B/T$ hardly changes with time.  In the case of unbarred galaxies, instead,  the $B/T$ is higher ($B/T\gsim 0.3 $) during the epoch of the bar formation in barred galaxies and visibly declines with time for $z=1$ galaxies. The $B/T$ of $z=0$ galaxies  does not  change substantially over time.
These patterns  confirm the results derived by  \citetalias{rosasguevara2020}  using the TNG100 simulation for  barred galaxies with $M_{*}\geq10^{10.4}\Msun$.

The difference in the assembly histories of disc galaxies with and without a bar has some implications in the formation of the bar itself. For one part, we have that the fast assembly of the discs in the barred galaxies combined with smaller bulges at early times facilitates  bar formation while  later-formed massive discs  with smaller bulges do not develop a bar. This is also shown in \cite*{izquierdo2022}, in which the authors find that  the galaxies are more compact  than  unbarred galaxies both  before and after bar formation in the TNG100 and TNG50 simulations. Similarly, the findings of  \cite*{bonoli2016} and \cite*{spinoso2017}  who investigate the twin  zoom-in simulations of the Milky Way  ErisBH and Eris point out that  ErisBH forms a bar because AGN feedback helps to reduce the bulge size in contrast to Eris, allowing for the triggering of bar instabilities. In addition, \cite*{saha2018}  show that cold stellar discs surrounding a classical bulge become strongly unstable to non-axisymmetric perturbations, using isolated simulations. However, if the bulge was denser than the internal part of the disc, there was no bar formation even though the disc was prone to develop bar instabilities (lower values of Toomre parameter) which is consistent with the fact that unbarred galaxies do not form a bar, later in time.   Furthermore, as we will show in the next section there are weak differences in the properties of the halos hosting barred and unbarred galaxies and these differences are weakly preserved even in the absence of  baryons, hinting that halo properties may play a role in the (non) formation of a bar.

The age of the stellar populations can be an interesting indicator of these differences observed in the buildup of barred and unbarred galaxies. Fig. \ref{fig:agespassembly} shows $(\rm Age_{\rm SP}/Age_{\rm Universe,z})$ as a function of stellar mass, where $\rm Age_{\rm SP}$ is the mass-weighted average age of the stellar populations in the galaxies in $2\,\rhalf$ and $\rm Age_{\rm Universe,z}$ corresponds to the Age of the Universe at the observed redshift. The marginal histograms represent the mass distribution and $(\rm Age_{\rm SP}/Age_{\rm Universe,z})$ for barred  (filled histograms) and unbarred galaxies (dashed line, empty histograms), respectively.  The vertical lines represent the median value at each redshift.
The figure confirms that barred galaxies formed earlier than unbarred galaxies at a given stellar mass and at all redshifts. The greatest difference in the mass-weighted age between barred and unbarred galaxies spans from  $(\rm Age_{\rm SP}/Age_{\rm Universe,z})=0.05$  ($\sim 67\rm Myr$) at $z\geq 2$  to 0.1 ($1.3 \, \rm Gyr$) at $z=0$.
Our findings are also compatible with the recent data from \cite{fraser2020} who, using a sample of barred galaxies from the MaNGA galaxy survey, show that the star formation histories of barred galaxies peak earlier than their unbarred analogues.  The authors have also observed that barred galaxies have built up their stellar mass before the unbarred galaxies with similar stellar mass.

To verify that the older stellar population is linked to an early assembly of the underlying halo,  we measure the formation time of haloes, $t_{50,\rm h form}$, as the lookback time when the main progenitor has assembled  50 per cent of its halo mass. To prevent complicated changes due to tidal effects in the halo formation times for satellites, we include here only the haloes of central galaxies.
Fig. \ref{fig:timeassembly} shows $( \rm Age_{\rm Universe,z}- t_{50,\rm h form})/ \rm Age_{\rm Universe,z}$, as a function of halo mass, $M_{200}$, where $(\rm Age_{\rm Universe,z}-t_{50,\rm h form})/\rm Age_{\rm Universe,z}$ is the halo formation time relative to the Universe age at a given redshift and normalised by this.

The figure illustrates the declining trend between the halo formation time and halo mass. The decreasing trend is
expected because haloes form hierarchically, as large massive haloes form later than less massive haloes.
In the relation between stellar population age and stellar mass, the trend is opposite since baryons are subjected to other mechanisms, such as feedback processes, that impact the buildup of the galaxy.

The figure also shows that host haloes of barred galaxies, overall, form earlier than the host haloes of unbarred galaxies for a specific halo mass.  This difference in the formation time increases towards lower redshifts, with the greatest formation time difference of $\sim 1$ Gyr for the most massive haloes ($M_{200}\geq 10^{11.7}\Msun$).  At higher redshifts, $z>2$, there is no difference between the formation time of the host haloes of barred to those of unbarred galaxies,
because the assembly process is rapid at this time. Besides,  bars are relatively young at this stage, which makes it more complicated to see the differences properly.

\subsection{Properties of the host haloes}
\label{subsec:prophaloes}
In this section, we focus on the properties of haloes that host barred and unbarred galaxies, motivated by the difference in the halo formation time between them. We focus primarily on $z=0$, which is where the biggest difference in halo formation time is found. The connection between halo properties and the formation and evolution of bars has been already investigated in several works which made use of isolated simulations (e.g., \citealt{debattista2000,athanassoula2002,athanassoula2003}).  These studies have pointed out that the shape of the halo  \citep{athanassoula2013} or the halo spin \citep{saha2013} could be key properties that can trigger the formation of bars and damp or enhance their subsequent growth  \citep{long2014}.

For this section,  we only study the host haloes of central galaxies to avoid complex evolutions owing to tidal effects for satellites. Note that in our disc galaxy samples, central galaxies are 75 per cent of the disc samples at $z=0$  (see Table~\ref{table:bars}).

First, we consider the ratio $V_{\rm max}/V_{200}$, where $V_{\rm max}$ is the peak value of its circular velocity curve $V_{\rm c}$  within  $r_{200}$   and $V_{200}= V_{\rm c}(r_{200})$. Here,  $V_{\rm c}= (GM(r)/r)^{1/2}$ and $r_{200}$ is the radius enclosing a mean overdensity 200 times  the critical value. This ratio is typically used as a proxy for halo concentration in cosmological simulations using only dark matter particles (e.g., \citealt{gao2005}).

\begin{figure}
\begin{tabular}{c}
\includegraphics[width=\columnwidth]{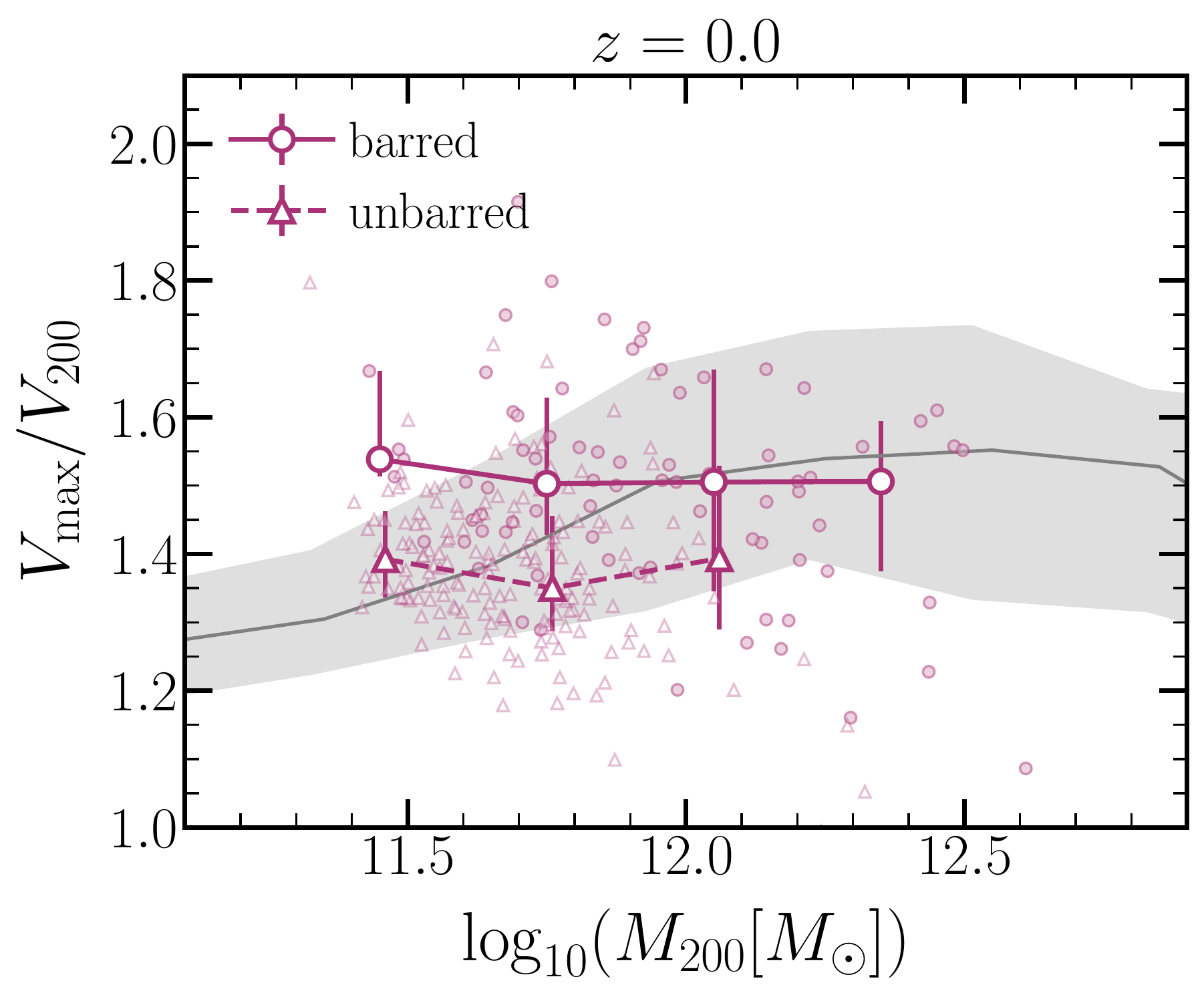} \\
\includegraphics[width=\columnwidth]{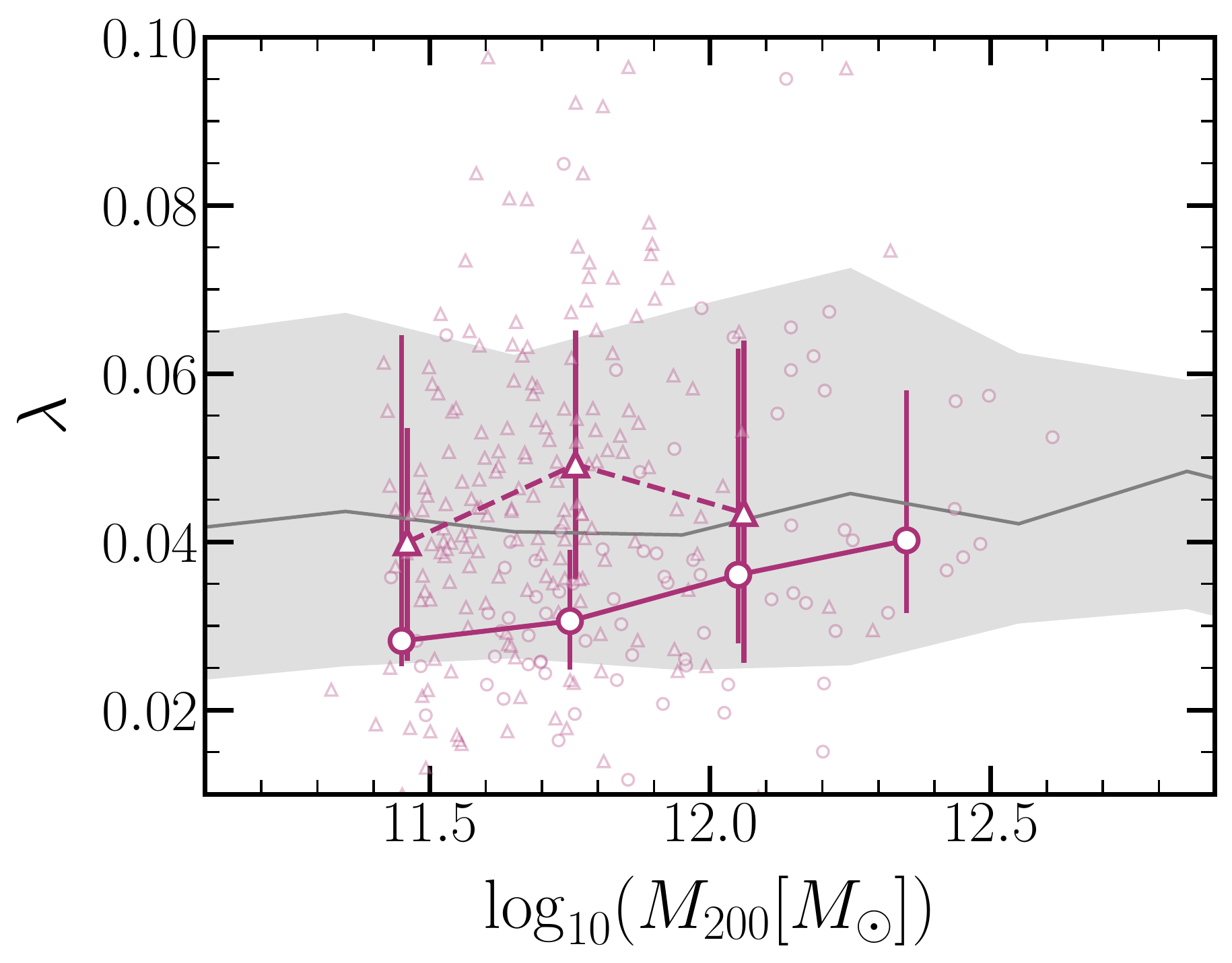} \\ 
\end{tabular}
\caption{\textit{Top panel:} $V_{\rm max}/V_{200}$-$M_{200}$ relation at $z=0$   
\textit{Bottom panel:} $\lambda$-$M_{200}$ relation where the halo spin parameter $\lambda=\mid\!\!\! \vec J\!\!\!\mid \!\!\!/( \sqrt{2} M_{200}V_{200}r_{200})$. Medians are calculated in halo mass bins with more than 5 galaxies and error bars represent the $20^{\rm th}$ and $80^{\rm th}$ percentiles of each distribution.  Grey shaded regions and solid lines represent the $20^{\rm th}$ and $80^{\rm th}$ percentiles  and the median  for all the central TNG50 galaxies, respectively.  Haloes hosting barred galaxies are likely to have higher $V_{\rm max}/V_{200}$ ratios (more concentrated) with smaller spins than those populated by unbarred galaxies.}  \label{fig:m200concentration}
\end{figure}

The top panel of  Fig. \ref{fig:m200concentration} depicts $V_{\rm max}/V_{200}$  as a function of $M_{200}$ at  $z=0$.   The median  $V_{\rm max}/V_{200}$ of haloes hosting barred galaxies (solid lines with circles) exhibits a weak indication of being above the median of unbarred galaxies (solid lines with circles). This hints that barred galaxies reside in haloes more concentrated than those hosts in unbarred galaxies  A similar variation of $V_{\rm max}/V_{200}$  is found at higher redshifts, $z=[2,4]$ with a large scatter (not shown here).

The relation between $V_{\rm max}/V_{200}$ and halo mass seems to be almost flat for both barred and unbarred galaxies (i.e., disc galaxies). This is in contrast with the trend seen for all the central galaxies in TNG50 (grey shaded region and the grey solid lines), where there is a slight increase at higher halo masses.  

Interestingly, looking into the halo spin as a function of the halo mass, the relation reverses. The bottom panel of Fig. \ref{fig:m200concentration} shows the halo spin, $\lambda$, as a function of halo mass at $z=0$. We define  $\lambda=\mid\!\! \vec J\!\!\mid \!\!/( \sqrt{2} M_{200}V_{200}r_{200})$ where  $\vec J$ is the angular momentum  evaluated at $r_{200}$ \citep{bullock2001}.  Despite a large dispersion in the samples, barred galaxies have lower halo spin on average than unbarred galaxies. This occurs at all redshifts analysed (here only shown at $z=0$) with the host haloes of barred galaxies having a median of $\lambda=0.034$ whereas the median $\lambda$ of the host haloes of  unbarred is  slightly higher (median $\lambda=0.044$).
One noteworthy aspect is that, in contrast to the overall central TNG50 population of galaxies, the halo spin of both barred and unbarred galaxies seems to be weakly increasing with increasing halo mass.

We also compare the counterpart haloes in a dark matter-only simulation, which started with the same initial conditions as in TNG50. We find that the $V_{\rm max}/V_{200}$ difference remains
weak  or null (not shown here) whereas the difference in halo spin between barred
and unbarred galaxies is preserved.

Our results are consistent with the findings of \cite{saha2018} using idealised simulations of disc galaxies. The authors find that spinning haloes facilitate the formation of a bar against static haloes. However,  \cite{long2014} point out that haloes with higher spin ($\lambda\geq 0.03$) do not fostered  the growth of the size and strength of the bar even though the disc were unstable to form a bar.

\begin{figure}
\includegraphics[width=\columnwidth]{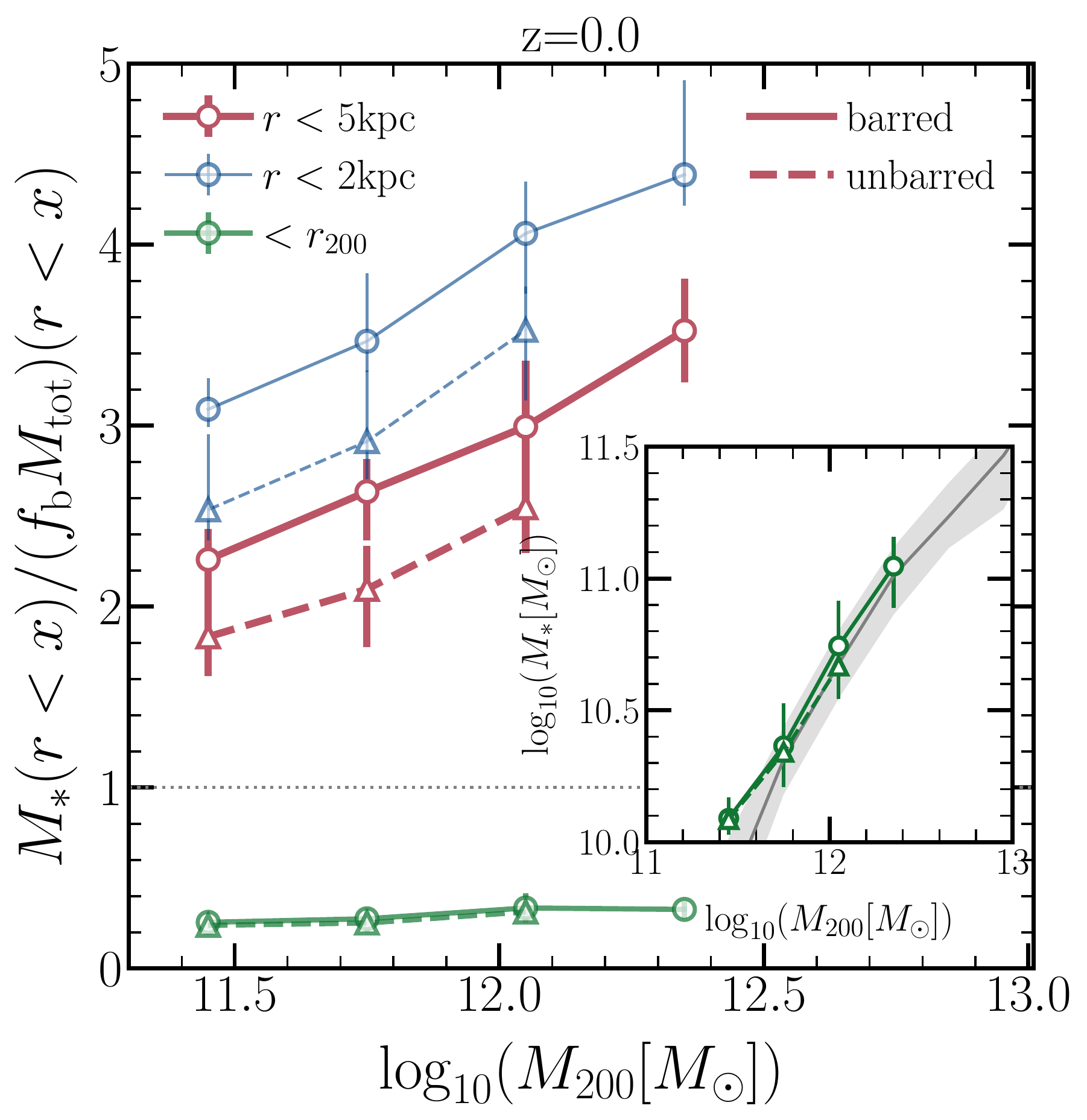} 
\includegraphics[width=\columnwidth]{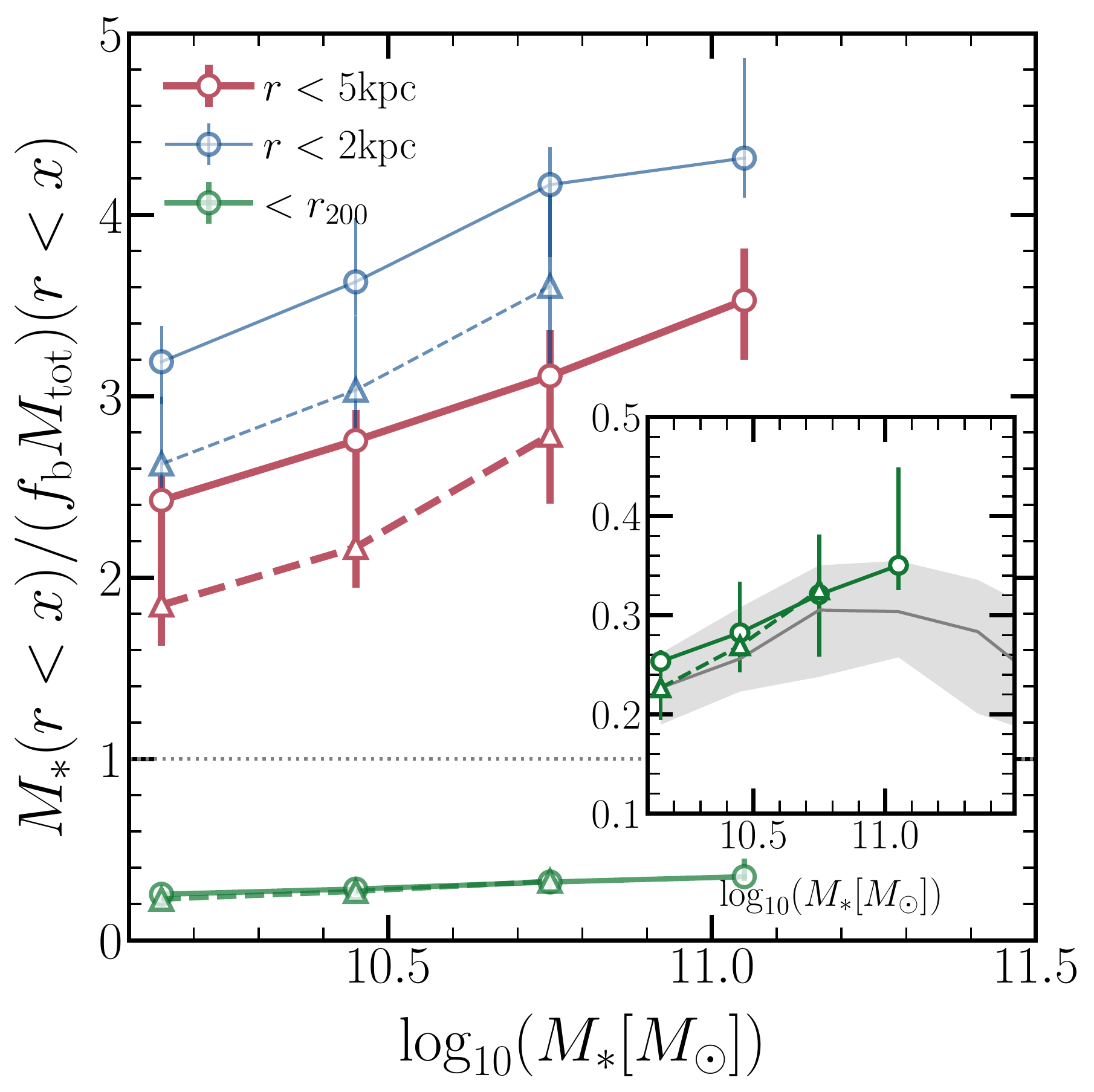}  

\caption{The $z=0$ median stellar mass fraction ($M_{*}/f_{\rm b}M_{\rm tot}$) normalised by the universal baryonic fraction,$f_{\rm b}$ as a function of $M_{200}$ (top panel) and $M_{*}$ (bottom panel), where $M_{\rm tot}$ is the mass of the total halo (stellar, gas and dark matter). $M_{*}/f_{\rm b}M_{\rm tot}$  is calculated for several apertures: $r<2\,\rm kpc$ (the thinnest blue  lines), $r<5\,\rm kpc$ (the thickest red lines), and $r<r_{200}$ (green lines). The solid lines with circles and dashed lines with triangles represent the median relation in barred and unbarred galaxies, respectively. The insets represent the median halo mass-stellar mass relation  (top panel)  and a zoom-in of the stellar mass fraction as a function of stellar mass (bottom panel) for barred, unbarred galaxies, and all the galaxies in the simulation (solid line and shaded region). The medians are calculated only for bins with more than 5 galaxies. Error bars correspond to the $20^{\rm th}$ and $80^{\rm th}$ percentiles of each sample.  Barred galaxies tend to be more baryonic dominated in the inner parts  than unbarred galaxies. This baryonic concentration is present but less strong for larger apertures, such that there is no significant difference in the total stellar mass fraction.}
\label{fig:m200overmstar2kpc}
\end{figure}

To further explain how the baryonic component is related to the dark matter, we investigate the local content of baryons in barred and unbarred galaxies at different scales.  Fig. \ref{fig:m200overmstar2kpc}  shows the median stellar mass fraction, $M_{*}/(f_{\rm b}M_{\rm tot})$ at different apertures, as a function of $M_{200}$ (top panel) and $M_{*}$ (bottom panel).
The stellar mass fraction is normalised by $f_{\rm b}$, that is the universal baryonic fraction and  $M_{\rm tot}$ is the mass from the stellar, gas and dark matter components inside $r\leq2\,\kpc$ (the faintest and thinnest line) and  $r\leq 5 \,\rm kpc$ (the thickest solid line) at  $z=0$.   On average,  the stellar mass fractions increase with increasing $M_{200}$ and $M_{*}$  for both barred  (lines with circles) and unbarred galaxies (lines with triangles)  and for all apertures.  Something interesting is that, for a given stellar (halo) mass, the inner parts ($ r\leq 2\, \kpc$) of the barred galaxies are slightly more dominated by baryons ($M_{*}(r<2\,{\rm kpc})/(f_{\rm b}M_{\rm tot})(r<2\,\kpc)\sim 4$) than unbarred galaxies,  which instead have overall stellar mass fractions slightly smaller than $3$ in their interior. At apertures of $5\,\kpc$, we find a similar pattern. The global stellar mass fraction ($M_{*}/(f_{\rm b}M_{200})$) also show no substantial difference between halos with a barred galaxy and those without a bar.  For instance, for haloes with $M_{200}=10^{11.6-11.9}\Msun$, the ratio of the stellar mass fractions between barred and unbarred is $1.19$ at $2$ kpc, $1.25$ at $5$ kpc and $1.09$ at $r_{200}$.
According to the inset plot of the top panel of Fig. \ref{fig:m200overmstar2kpc}, which displays the stellar mass-halo mass relation of barred and unbarred galaxies and the entire halo population in TNG50, this conclusion is verified. The inset figure shows that for a given halo mass, there is no significant variation in stellar mass. A similar result can be seen in the inset plot of the bottom panel, which displays the global stellar mass fraction as a function of stellar mass.

It is worth mentioning that we find similar behaviour in the local and global $M_{*}/f_{\rm b}M_{\rm tot}$ ratios at $z\leq 1$ as a function of halo mass and stellar mass at $z=[2,4]$.
This is also consistent with \cite*{izquierdo2022}, who, using a sample of disc galaxies in TNG100 which contains more massive unbarred galaxies at $z=0$, found compatible results in the stellar-dark matter mass ratios and found this difference in baryons at different radii prevails even before the bar formation time. This is in broad agreement with recent findings from \cite*{fragkoudi2021} in Auriga barred galaxies, who find that the contribution of the stellar component is dominant in the total rotation curve  since  $z=0.5$


\section{Summary and discussion}
\label{sec:summary}
In this paper, we analyse the properties of barred galaxies predicted by the $\Lambda$CDM magneto-cosmological hydrodynamical simulation TNG50 from the IllustrisTNG project.  The simulation evolves a comoving region of 51.7 $\rm cMpc$ on a side with an initial number of particles and gas cells of $2\times2160^3$   and reproduces fairly well the cosmological evolution of the main properties of disc galaxies \citep{pillepich2019,nelson2019b}.
The resolution of the simulation allows us to explore bar formation and evolution in a cosmological context across time.  Our study focuses on massive galaxies (with more than  $M_{*}\gsim 10^{10} \Msun$) having a dominant disc component  ($D/T\geq0.5$), as obtained through a kinematic decomposition \citep{genel2015}.

 To identify barred galaxies, we Fourier decompose the face-on stellar surface density and calculate the second term, $\Ato$, as a function of the cylindrical radius. We define the extent of a bar as the radius at which $\Ato$ reaches  $\rm max(0.15, min(\Ato))$ in the range where the phase is constant.  We consider as barred galaxies all disc galaxies with $\Amax\geq0.2$ and $\rbar>r_{\rm min}$, where $r_{\rm min}$ is $1.38$ times the proper softening length, and for which the bar structure is present at least for a period of time that corresponds to two outputs in the simulation. We have also compared our bar identification method to a subtracted sample from Zana et al., submitted, which do a kinematic decomposition study of TNG50 galaxies, finding good agreement at all redshifts

Our results include the following:
 \begin{itemize}

\item The first stable bars appear as early as $z=4$, and  both the bar and disc scale lengths grow at comparable rates throughout time (Fig. \ref{fig:rbarhdisc}).  However, bar lengths  are less correlated with stellar mass than the disc scale lengths (Fig. \ref{fig:rbarhdiscrelations}). We also compare the predicted correlation between  bar size and  stellar mass  with the local observational constraints of   \cite{gadotti2011} \& \cite{erwin2018}.  The results show a weaker correlation than the observed ones, but still consistent with observations.

\item The bar fraction mildly evolves with redshift  (Fig. \ref{fig:barfractionz}), and is always above $\sim 0.4$,  except at $z=0$ (0.30) and $z=4$ (0.28). While we cannot make an apples-to-apples comparison with available observations and other simulation results, since both observational  and simulated samples have different selection criteria, we find that the predicted bar fraction recovered at $z=0$ is within the range reported by different observational results  \citep[e.g.][]{masters2012,cervantes2017,gavazzi2015,erwin2018} and other theoretical studies \citep[e.g.][]{algorry2017,peschken2019,rosasguevara2020,zhao2020}. We find that the mild evolution of the bar fraction with redshift could be reconciled with observational results  \citep{sheth2008,melvin2014} if only long bars are included. This is because at higher redshifts bars are smaller, generally below the angular resolution limit of current observational facilities. Indeed, when only long bars are analysed, the predicted bar fraction rapidly declines with increasing redshift, as observed by \citet{sheth2008,melvin2014}. Consistent with observations, we find that the bar fraction is stellar mass-dependent, being larger for more massive galaxies ($0.65$ at $z=0$ and $0.7$ at $z=3$ for $M_{*}\geq10^{10.5}\Msun$).

\item We explore several internal properties of barred galaxies and find that they have lower gas fractions (Fig. \ref{fig:gasfracmstar}), are more massive and have an older stellar population (Fig. \ref{fig:agespassembly}) than unbarred galaxies at all redshifts analysed. Our results are compatible with observational studies of the local Universe that show the low gas fractions of barred galaxies  such as \cite{masters2012} who combine data on morphologies of the galaxy zoo project with HI data from the ALFALFA survey, and \cite{cervantes2017} who analyse the data from the SSDS DR7 and ALFALFA survey. \cite{gavazzi2015} using the H$_{\alpha}$ follow-up survey from ALFALFA show the low star formation activity of massive, strongly-barred galaxies. Moreover, we find barred galaxies feature a younger stellar population along the bar (Fig. \ref{fig:gasdensitymaps}). This is consistent with a scenario where a younger stellar population is trapped on more elongated orbits  shaping a thinner part of the bar  and also explains the V-shape signature found in the star formation histories perpendicular to the bar major axis of nine barred galaxies in the local universe analysed  by \cite{neumann2020}. The strongest differences in gas fractions, mass and stellar population age between barred and unbarred galaxies are at $z=0$ and are consistent with the early assembly of barred galaxies (see also \citetalias{rosasguevara2020} and \citealt{izquierdo2022}).

\item We find that the haloes of barred galaxies formed earlier than the ones of unbarred galaxies for a given halo mass (Fig. \ref{fig:timeassembly}). Furthermore, at fixed halo mass, the haloes hosting barred galaxies are also slightly more concentrated and spin less rapidly than the host haloes of unbarred galaxies (Fig. \ref{fig:m200concentration}).  We also found similar or smaller differences in spin and concentration for the counterpart halos from the dark matter-only simulation between barred and unbarred galaxies.  These results are consistent with numerical results from idealised simulations suggesting that spinning haloes promote bar formation (e.g., \citealt{saha2013}), but hinder  bar growth in strength and size (e.g., \citealt{long2014}).

\item The global and local stellar mass fractions at all redshifts show that barred galaxies are more baryon dominated than unbarred galaxies in the inner parts of the galaxy, with weakly or no significant differences at a global scale for haloes with mass smaller than $10^{12.2}\Msun$ and $z=0$  (Fig. \ref{fig:m200overmstar2kpc}). We find similar differences for $z<2$.  Our results are consistent with the ones of \citealt{fragkoudi2021} who, using the Auriga suite, report that barred galaxies are more baryon dominated in the inner regions than unbarred galaxies since $z=0.5$.  Similar trends will be shown in  \cite*{izquierdo2022} even during the period of bar formation  using  TNG100 and TNG50 disc galaxies.
\end{itemize}

Combining our findings together, the formation and  evolution of a bar could  be a potential  tracer of assembly history of galaxies because of the early and fast growth of galaxies that eventually formed a bar. Early disc assembly \citep{rosasguevara2020}, together with a hampered growth of the bulge \citep{bonoli2016,saha2018}, seem to lead to the favourable conditions for global disc instabilities and bar formation (\citealt{zana2018c}, \citealt{izquierdo2022}). Furthermore, we speculate that the host haloes of barred galaxies have more time to accrete gas and efficiently convert it into stars, as their haloes are spinning less rapidly and are more concentrated than those of unbarred galaxies, suggesting more effective gas accretion and star formation \citep{correa2020}. On the contrary, the host haloes of unbarred galaxies  formed later, having less time to accrete gas. Besides, their host halo is less concentrated and spinning more rapidly and this, combined with feedback processes, might limit the star formation efficiency.
To what extent the contribution of the feedback processes affects the formation and evolution of bars in comparison to the cosmological nature of the  host halo is something to study in more detail in future works.
On the other hand, it is clear that feedback processes could influence the evolution of a bar. This has been extensively discussed by  \cite{zana2018c}, who analyse a suite of zoom-in cosmological hydrodynamical simulations of Milky Way haloes, which essentially differ from each other in the feedback process modelling. The authors have found that feedback processes can influence the time of bar formation and the strength and size of a bar. This is also recently confirmed by \cite{fragkoudi2021} comparing barred galaxies from the Auriga, Eagle and Illustris simulations. The authors show that the global and local stellar mass fractions were different among the simulations and influence the pattern speed of the bars.  However, there is a caveat regarding the impact of the different hydrodynamic schemes on the bar properties.
Our results highlight, in agreement with the aforementioned studies, that the properties of bars can provide information on the history of the host  galaxy and halo and can be used as an additional diagnostic of the subgrid physics of galaxy formation.

\section*{Acknowledgements}
\addcontentsline{toc}{section}{Acknowledgements}
The authors thank the referee for the constructive comments of the manuscript that improved the clarity of the paper. The authors thank Annalisa Pillepich for early discussions and early access to the TNG50 data prior to its public release. YRG acknowledges the support of the``Juan de la Cierva Incorporation'' Fellowship (IJC2019-041131-I) and the European Research Council through grant number ERC-StG/716151. D.I.V acknowledges financial support provided under the European Union’s H2020 ERC Consolidator Grant ``Binary Massive Black Hole Astrophysics''
(B Massive, Grant Agreement: 818691) and INFN H45J18000450006. MV acknowledges support through NASA ATP grants 16-ATP16-0167, 19-ATP19-0019, 19-ATP19-0020, 19-ATP19-0167, and NSF grants AST-1814053, AST-1814259,  AST-1909831 and AST-2007355. We thank
contributors to SciPy\footnote{http://www.scipy.org}, Matplotlib\footnote{https://matplotlib.org}, and the Python\footnote{http://www.python.org} programming language.


\section*{Data Availability}
The data from the TNG050 simulations can be found on the websites: \url{https://www.tng-project.org} \citep{nelson2019a}. The catalogues of bars will be made public in~\url{https://www.tng-project.org/data/docs/specifications/\#sec5j}.







\appendix

\section{Method of two-component decomposition of surface face-on density profile}
\label{app:decomp}
In this section, we describe the method to decompose the surface brightness profiles of the disc sample and find the length scale of the disc. 
Surface density profiles are computed in a face-on view in concentric annuli of 0.1 kpc in width and centred on the position of the minimum potential of the stellar component. 

We use the particle swarm optimization (PSO) code PSOBacco \footnote{ \url{https://github.com/hantke/pso_bacco}} \citep{arico2021}  to find the minimum $\chi^{2}$ of a fit  of the sum of an exponential profile and a Sersic profile,  that in terms of the stellar mass is given by 

\begin{equation}
\Sigma(R) =\Sigma_{\rm d,0}\, {\rm exp} \bigg (R/\hdisc \bigg ) + \Sigma_{\rm b,e} \,{\rm exp}\bigg ( -b_{n} \bigg [r/r_{\rm b,eff}^{1/n} -1  \bigg ] \bigg )
\label{eq:sigma}
\end{equation}
where $\Sigma_{\rm d,0}$ is the central surface density  of the disc component, $\hdisc$ is the disc scale length,  $r_{b,\rm eff}$  the effective radius that encloses half of the stellar mass of the predicted one of the Sersic profile,  $\Sigma_{\rm b,e} $ corresponds to the surface density at $r_{b,\rm eff}$, and $n$ is the Sersic index.  The quantity $b_{n}$ depends on the complete gamma function, such that $\Gamma(2n)= 2\gamma(2n,b_{n})$. 

We perform the method of fitting two components in the region at radii smaller than $r_{\rm fit} = \rm max (\rm log_{10}(\Sigma(r)/\Sigma_{\rm max})\geq -2.6,10\,\kpc)$. This condition ensures fitting both high-redshift discs, which are more compact (smaller disc scale lengths for a given stellar mass), and their analogues at low redshifts whose discs are less compact (larger disc scale lengths for a given stellar mass).

\begin{figure}
\begin{tabular}{cc}
\includegraphics[width=0.5\columnwidth]{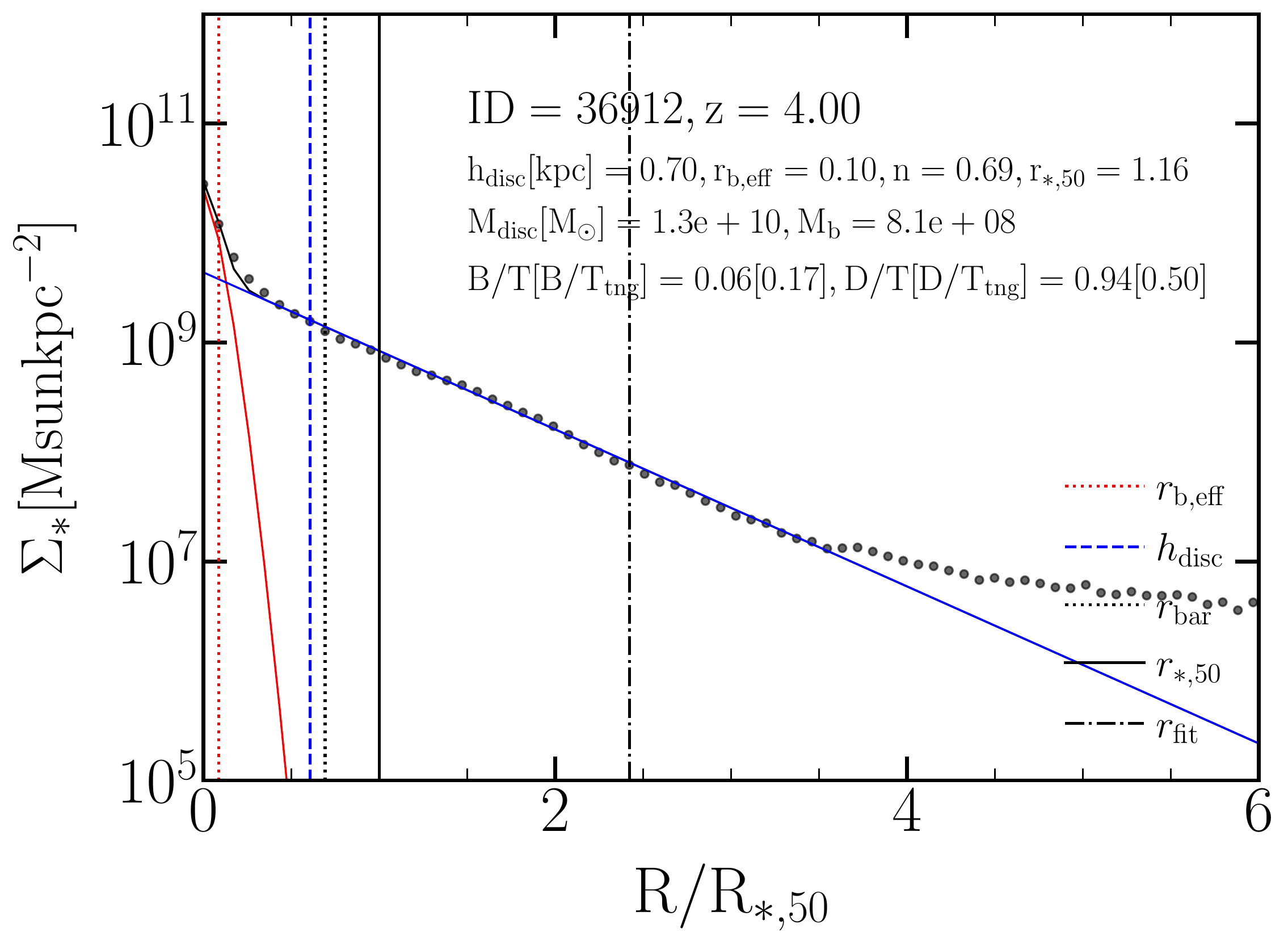} &
\includegraphics[width=0.5\columnwidth]{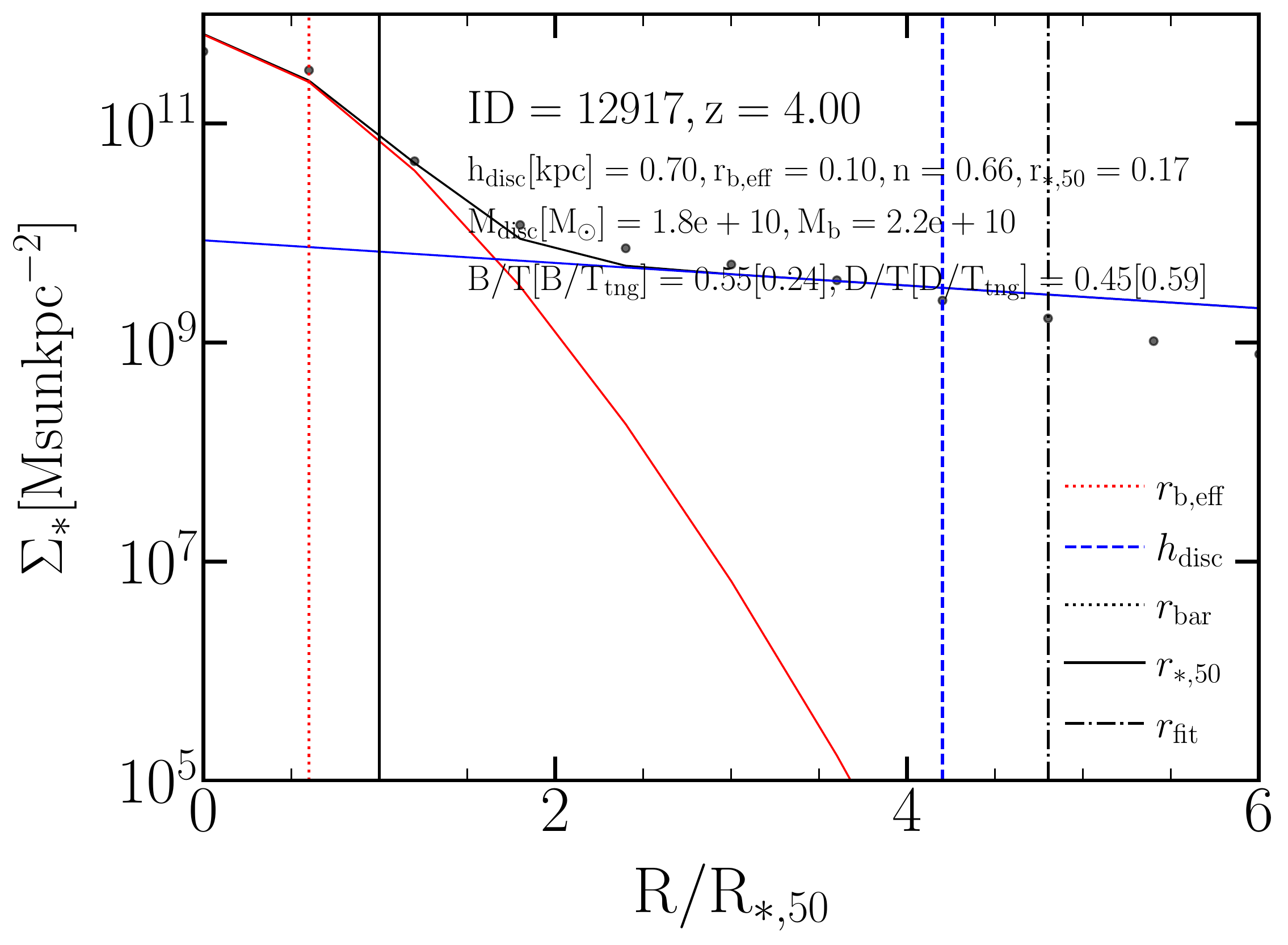}\\
\includegraphics[width=0.5\columnwidth]{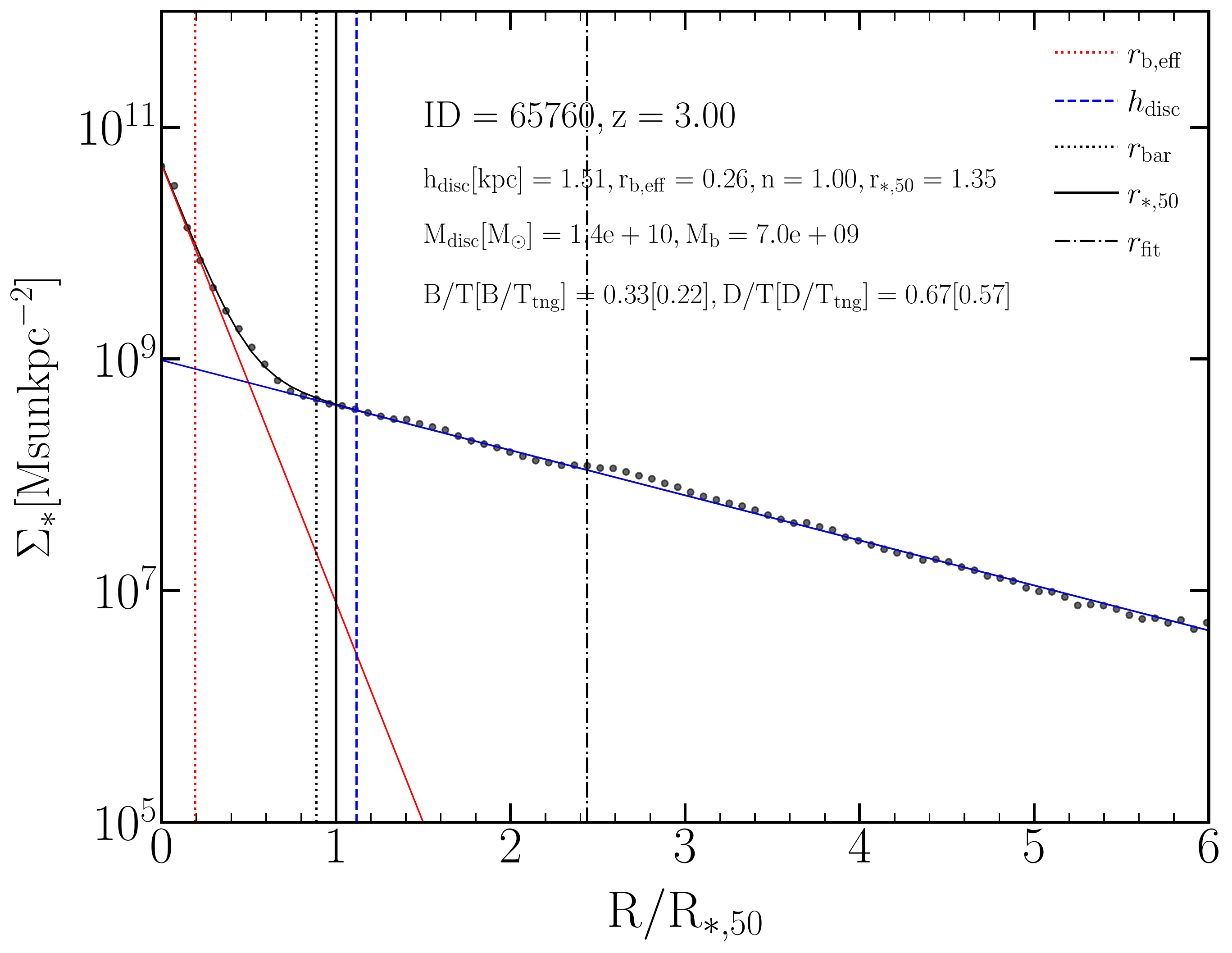} &
\includegraphics[width=0.5\columnwidth]{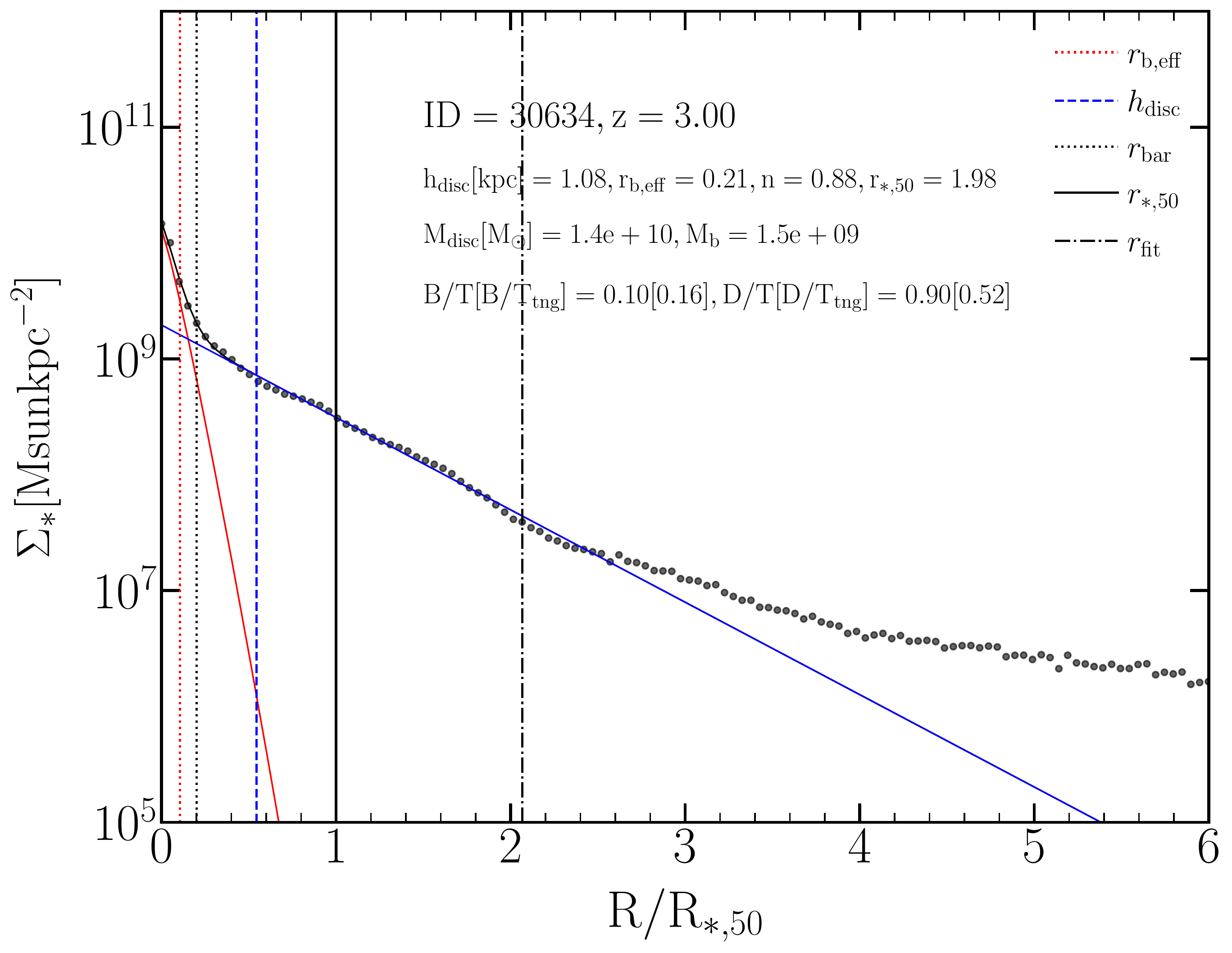}    \\
\includegraphics[width=0.5\columnwidth]{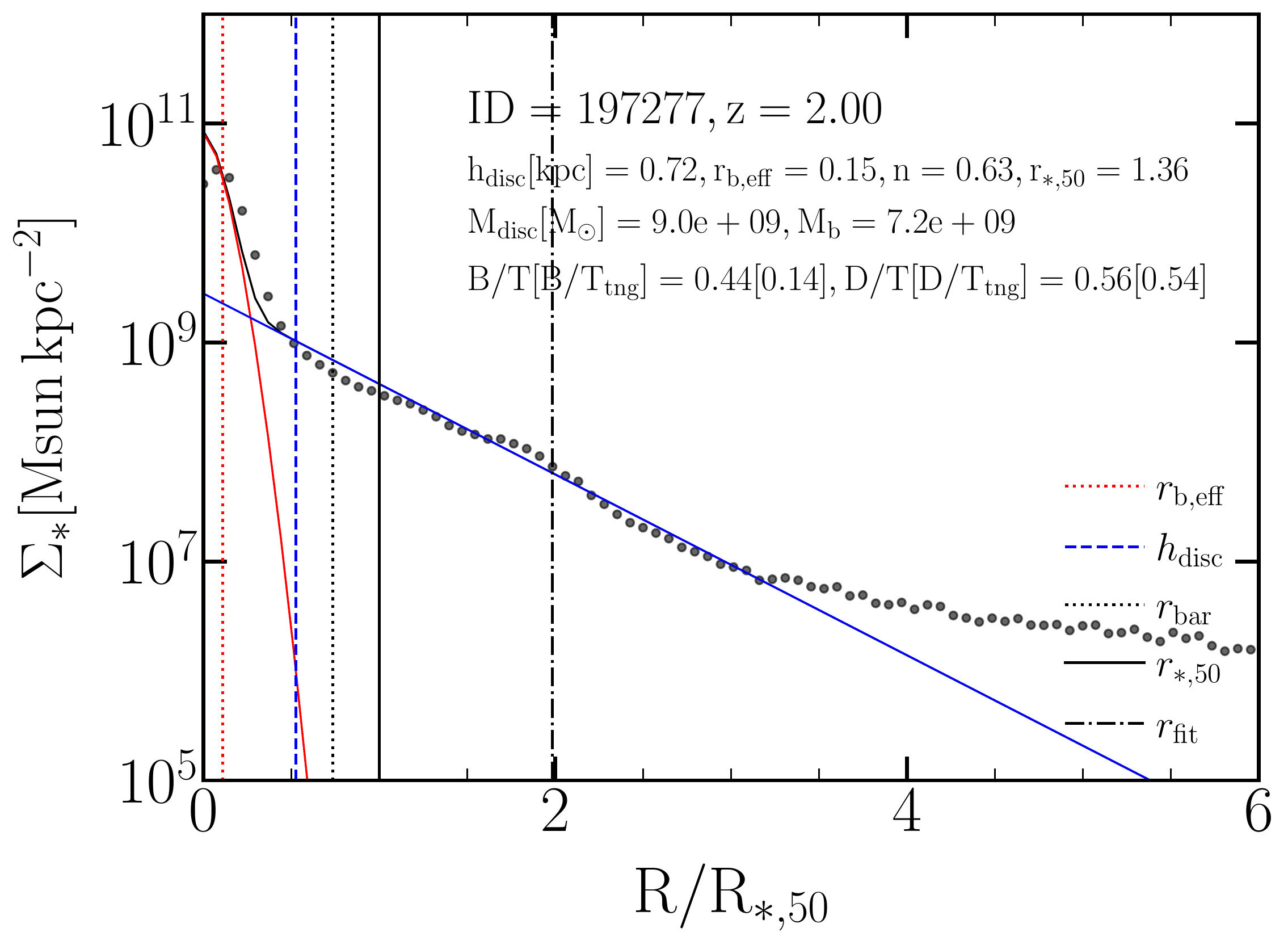}  &
\includegraphics[width=0.5\columnwidth]{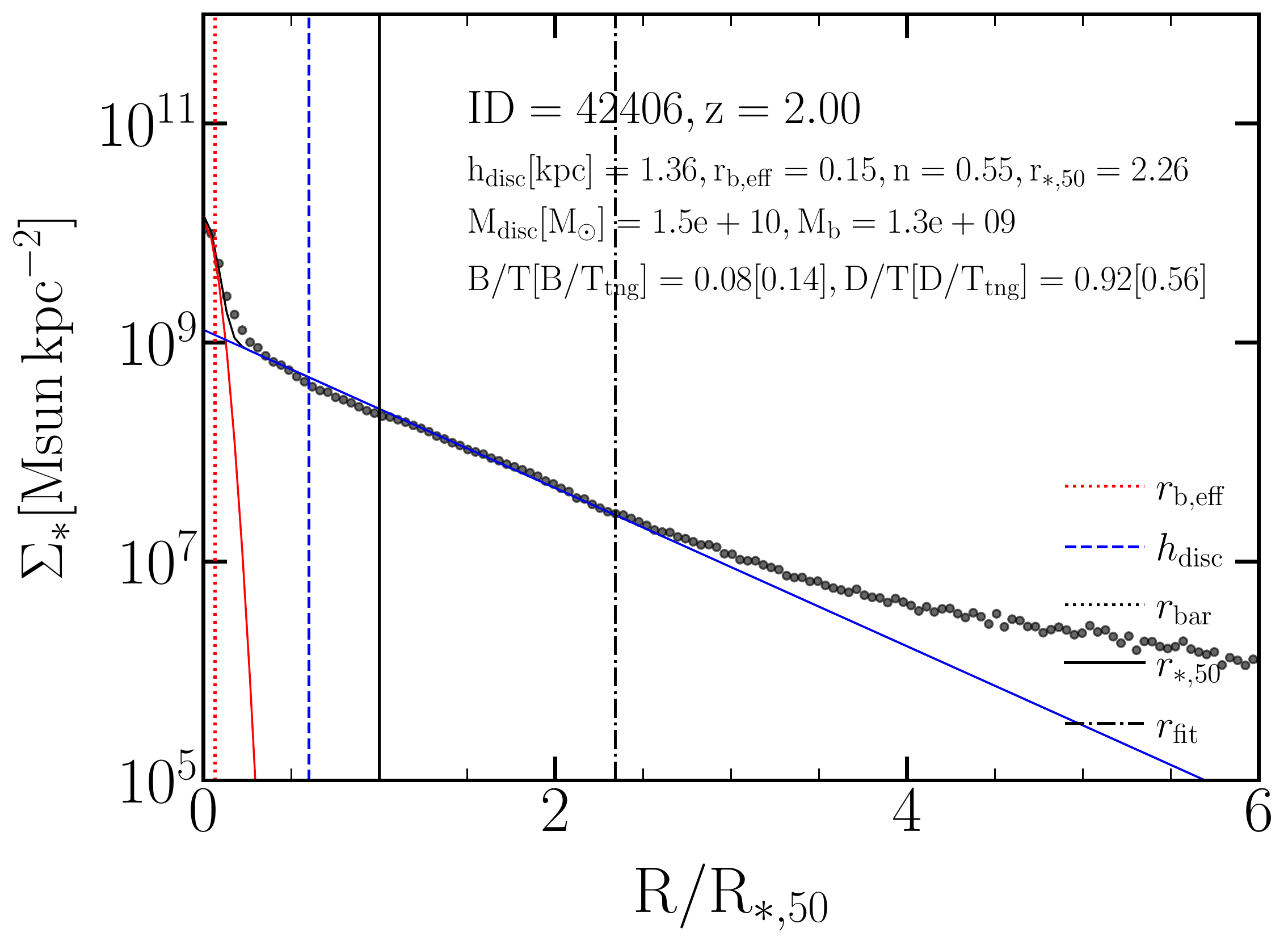}  \\
\includegraphics[width=0.5\columnwidth]{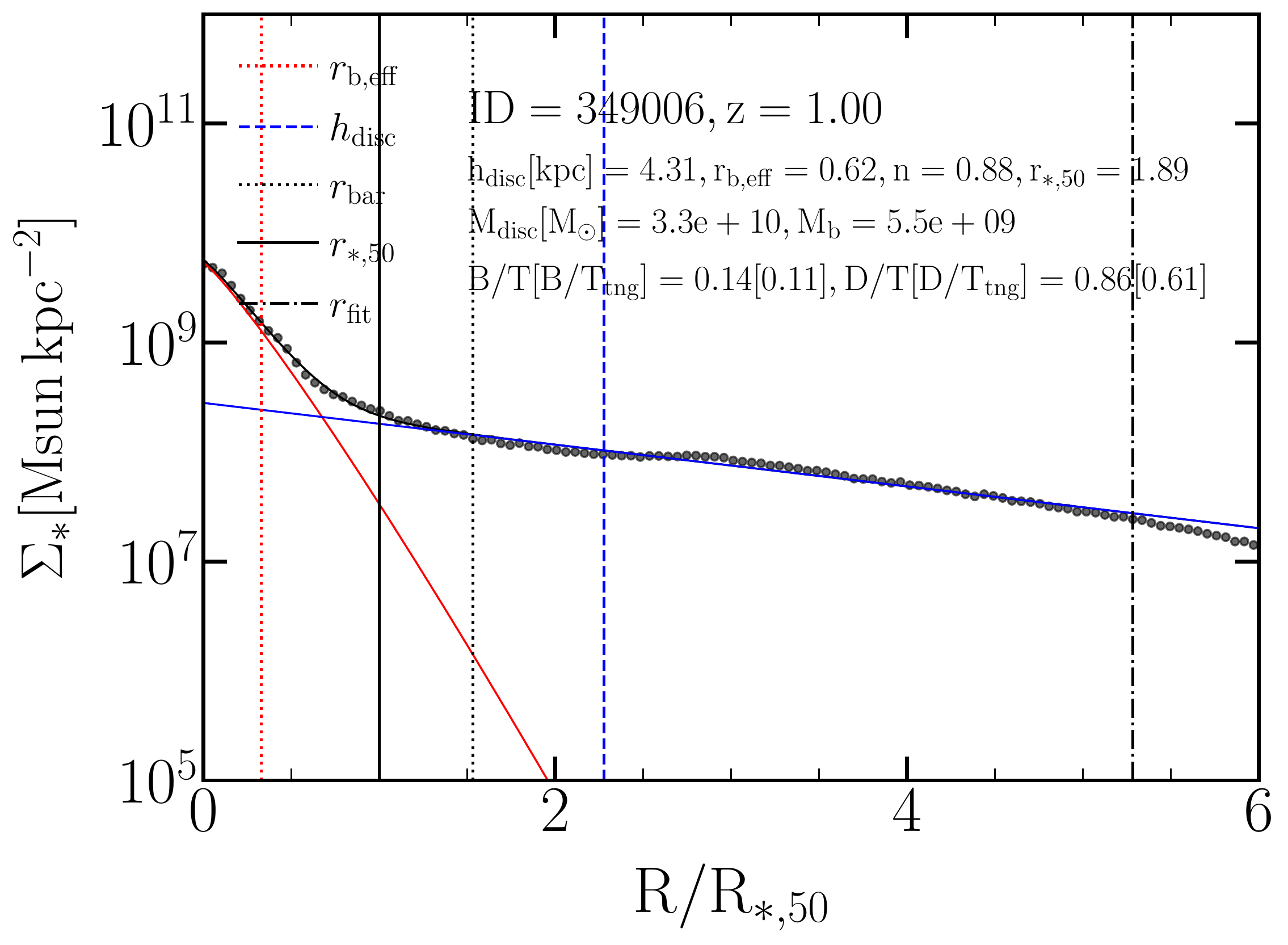} &
\includegraphics[width=0.5\columnwidth]{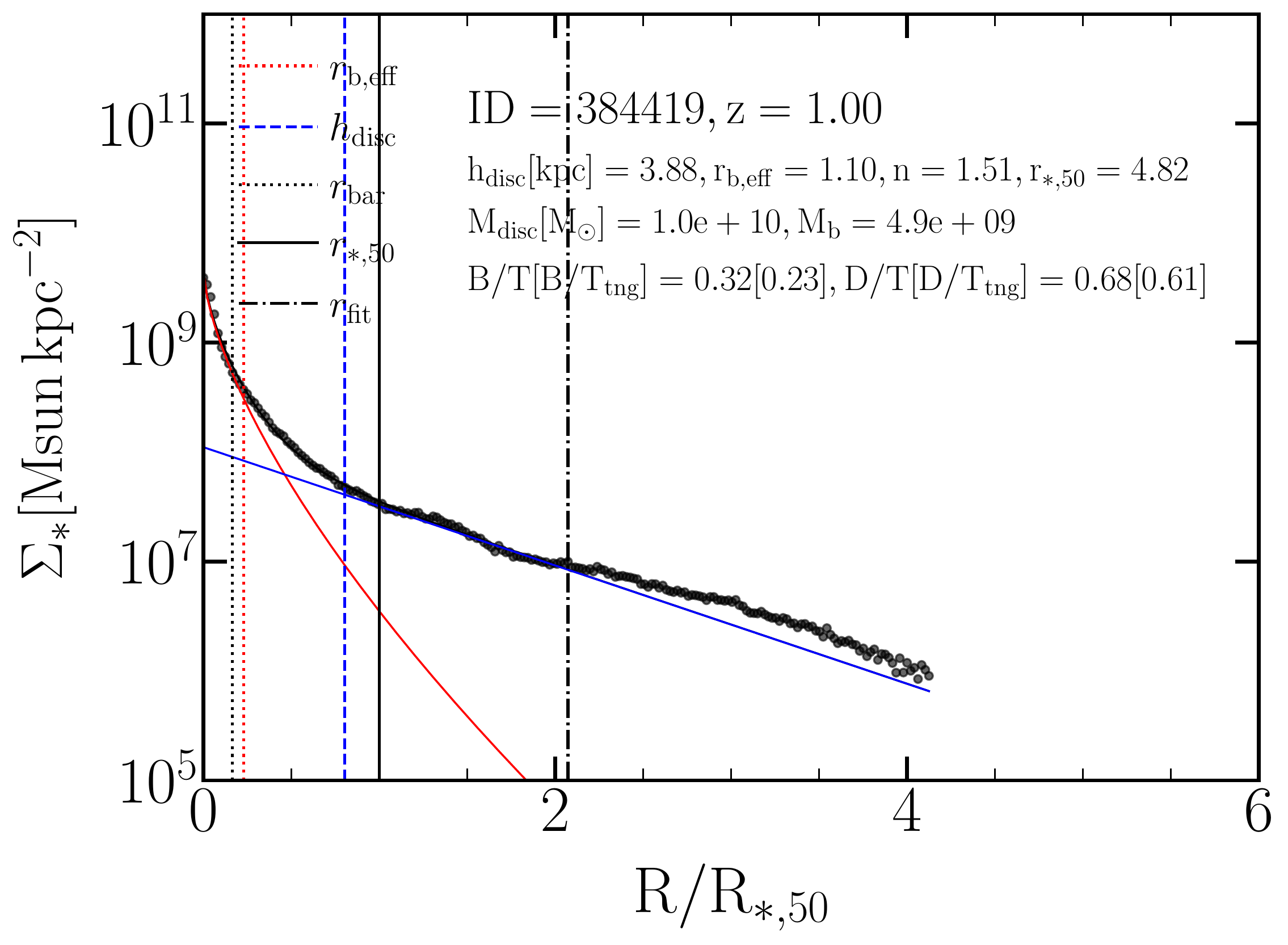} \\
\includegraphics[width=0.5\columnwidth]{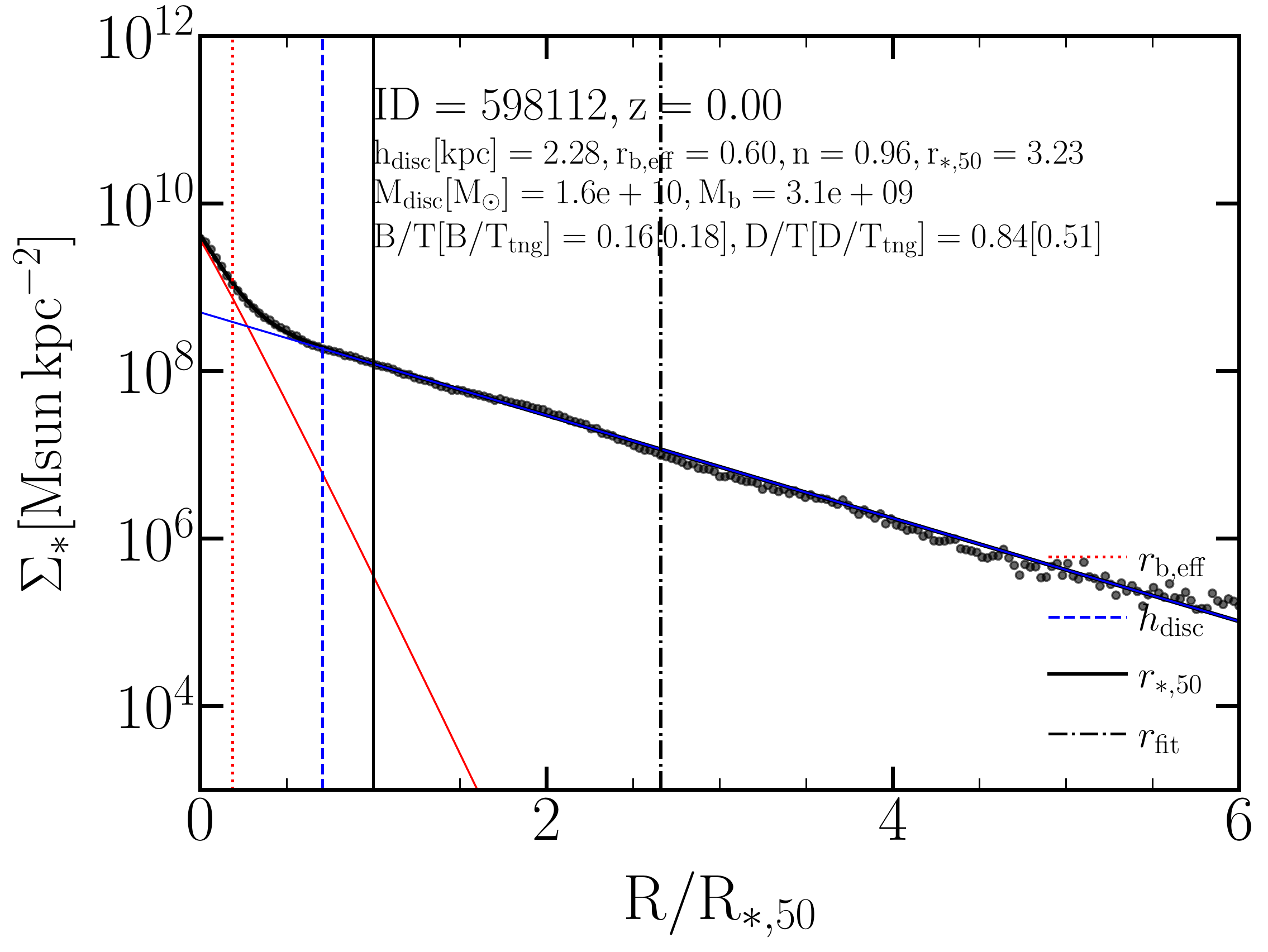} &
\includegraphics[width=0.5\columnwidth]{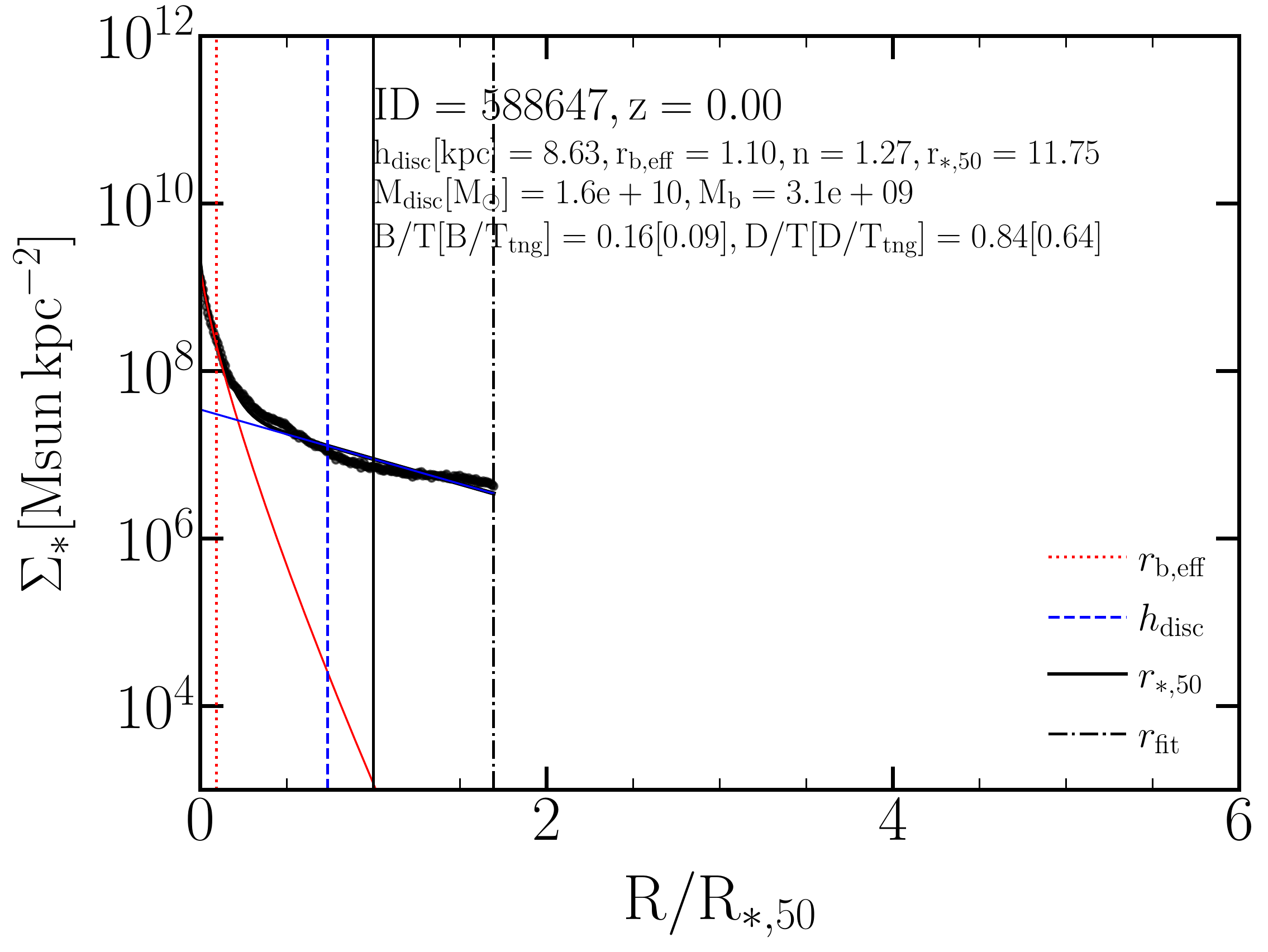}\\

\end{tabular}
\caption{Examples of face-on stellar surface density profiles in terms of  $R/R_{*,50}$ where $R_{*,50}$ is the half stellar mass radius, for barred (left column) and unbarred (right column) galaxies at $z=4,3,2,1,0$. The  profiles (black lines) were
obtained by fitting simultaneously a Sersic (red lines) and an exponential profile (blue lines). Horizontal  dashed lines correspond to the radius at which the fit is done. Dotted horizontal lines  correspond to the bar length (black), effective radius of the bulge (red), and scale length of the disc (blue).}
\label{fig:fitexamples}
\end{figure}

Figure \ref{fig:fitexamples} shows the surface density profiles and their fits for galaxies that have a  bar (left column)  and without a bar (right column) in terms of the stellar half mass radius at $z=4,3,2,1,0$. The total fitted profile is indicated as a black line. Blue and red lines correspond to the disc and bulge components, respectively. As a reference, coloured horizontal lines correspond to the scale-length of a disc,$\hdisc$, the effective radius of the bulge, $r_{\rm b,eff}$, and the length of the bar in case of barred galaxies. The dashed horizontal lines correspond to the radius at which the fit is done.

\section{Impact of galaxy size estimation on the galaxy scaling relations}
\label{app:sizes}

\begin{figure}
\includegraphics[width=\columnwidth]{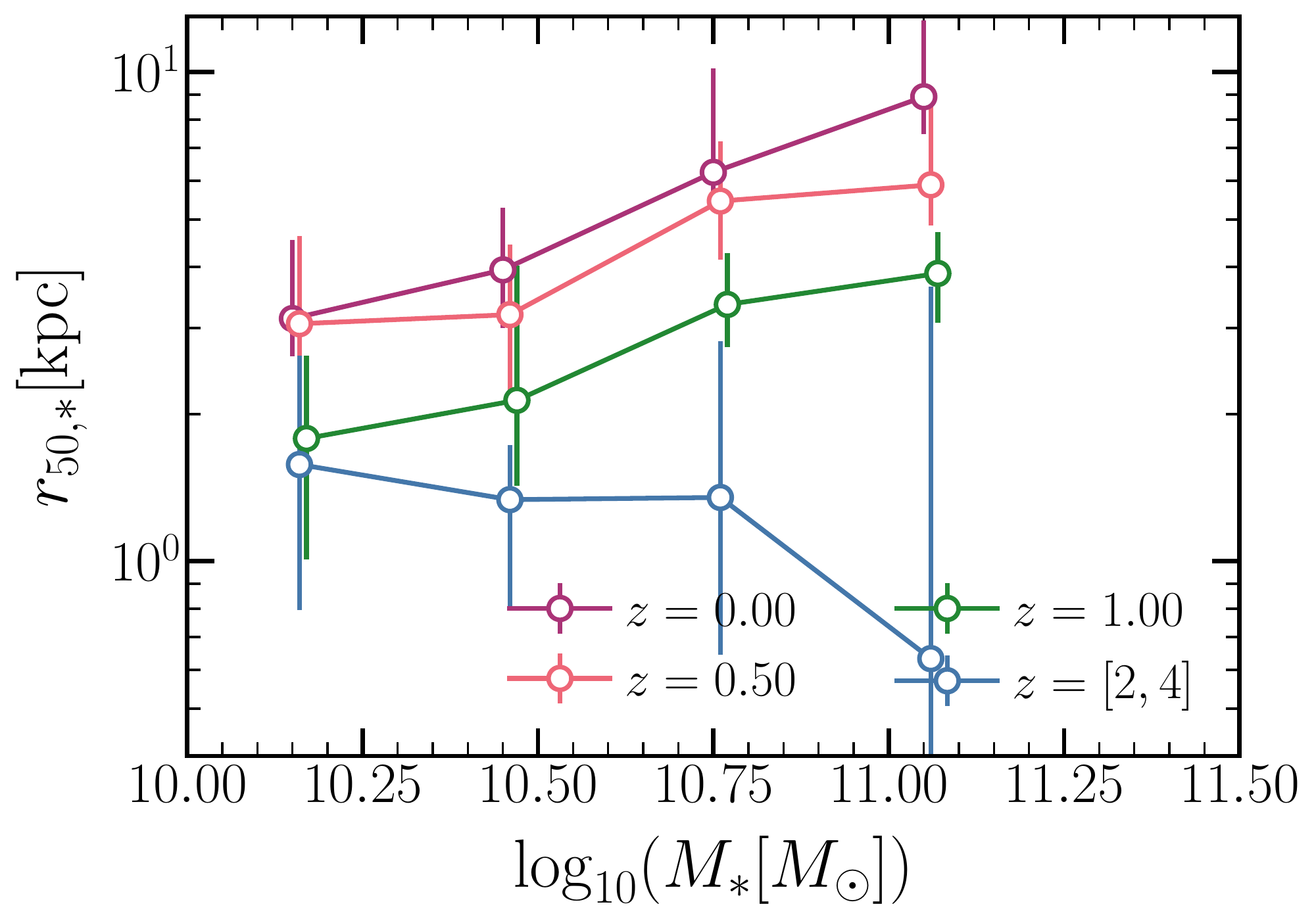}
\caption{ Stellar half mass radius (left column)  as a function of stellar mass. Solid lines with circles correspond to the median relation in barred galaxies and dashed lines with triangles to unbarred galaxies. Error bars correspond to the 20$^{\rm th}$ and $80^{\rm th}$ percentiles of the distribution in each sample.  There is an evolution in general with $\rhalf$. Barred galaxies are smaller in comparison with unbarred galaxies at the same observed time.}  \label{fig:rbarhdiscrelations2}
\end{figure}

In this section we present the relation between galaxy size  and stellar mass using the half-stellar mass radius as a proxy of galaxy size instead of the disc scale length as in Section~\ref{sec:results} to show that our results are robust and do not depend on the method used to calculate the disc scale length.  Fig~\ref{fig:rbarhdiscrelations2} shows this relation for barred galaxies at different redshifts as in Fig~\ref{fig:rbarhdiscrelations}. We see that the mild positive relation between galaxy sizes and their stellar mass prevails for barred galaxies.

\begin{figure}
\includegraphics[width=1.\columnwidth]{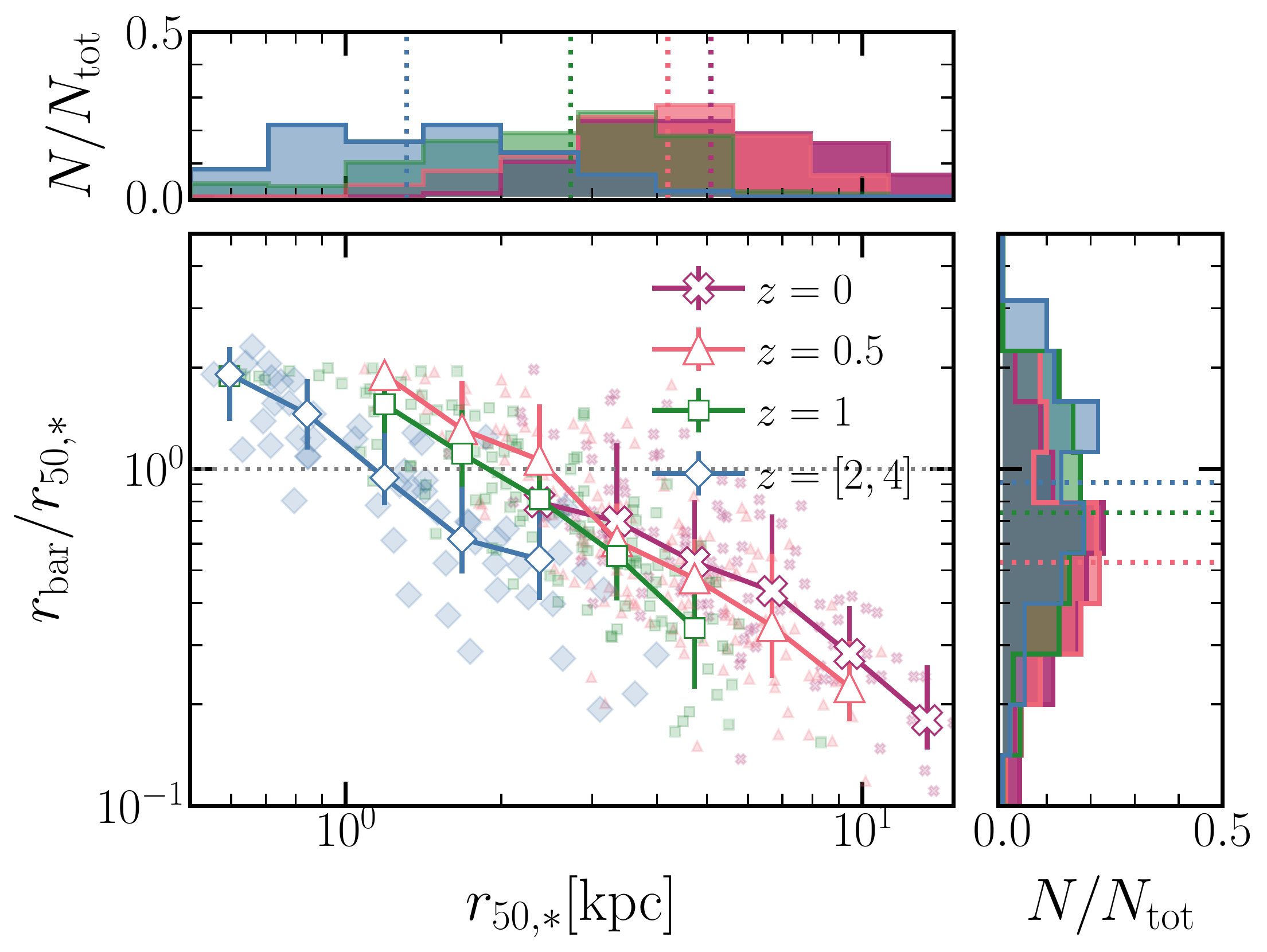}  
\caption{The bar extent relative to the stellar half mass radius, $ \rbar/r_{\rm 50,*}$, in the barred TNG50 galaxies as a function of $r_{\rm 50,*}$. Different markers and colours represent different redshifts as indicated in the legend. The median relation is shown in solid lines with markers for each redshift and only in disc length bins with more than 5 galaxies and error bars represent the $20^{\rm th}$ and $80^{\rm th}$ percentiles of the distribution. Panels along the margins show the distributions of the stellar half mass radius and  $\rbar/r_{\rm 50,*}$. Dotted lines represent the median of each distribution.}
\label{fig:rbarhdisc2}

\end{figure}

We also present the relation between the bar sizes relative to the galaxy size as a function of galaxy sizes in Fig~\ref{fig:rbarhdisc2}, where we show that the evolution of this relation is similar to the one found when using the disc scale length instead in Fig~\ref{fig:rbarhdisc}.

\bsp	
\label{lastpage}
\end{document}